\input harvmac.tex
\input amssym.tex
\input labeldefs.tmp



\font\teneurm=eurm10 \font\seveneurm=eurm7 \font\fiveeurm=eurm5

\newfam\eurmfam

\textfont\eurmfam=\teneurm \scriptfont\eurmfam=\seveneurm

\scriptscriptfont\eurmfam=\fiveeurm

 \font\teneusm=eusm10 \font\seveneusm=eusm7 \font\fiveeusm=eusm5

\newfam\eusmfam

\textfont\eusmfam=\teneusm \scriptfont\eusmfam=\seveneusm

\scriptscriptfont\eusmfam=\fiveeusm

\def\eusm#1{{\fam\eusmfam\relax#1}}

\font\tencmmib=cmmib10 \skewchar\tencmmib='177

\font\sevencmmib=cmmib7 \skewchar\sevencmmib='177

\font\fivecmmib=cmmib5 \skewchar\fivecmmib='177

\newfam\cmmibfam

\textfont\cmmibfam=\tencmmib \scriptfont\cmmibfam=\sevencmmib

\scriptscriptfont\cmmibfam=\fivecmmib

\def\cmmib#1{{\fam\cmmibfam\relax#1}}



\writedefs


\lref\AtiyahF{M.F.~Atiyah, ``Floer homology,'' Progr. Math.
Birkhauser 133 (1995) 105.}

\lref\Atiyah{M.F.~Atiyah, ``The Geometry and Physics of Knots,''
Cambridge Univ. Press, 1990.}

\lref\Rasmussen{J.~Rasmussen, ``Floer homology and knot
complements," math.GT/0306378.}

\lref\OShf{P.~Ozsvath, Z.~Szabo, ``Holomorphic disks and
topological invariants for closed three-manifolds,''
math.SG/0101206.}

\lref\OShfk{P.~Ozsvath, Z.~Szabo, ``Holomorphic disks and knot
invariants," math.GT/0209056.}

\lref\OSlens{P.~Ozsvath, Z.~Szabo, ``On knot Floer homology and
lens space surgeries,'' math.GT/0303017.}

\lref\OSappl{P.~Ozsvath, Z.~Szabo, ``Holomorphic disks and
three-manifold invariants: properties and applications,''
math.SG/0105202.}

\lref\OSreview{P.~Ozsvath, Z.~Szabo, ``Heegaard diagrams and
holomorphic disks,'' math.GT/0403029.}

\lref\OShfl{P.~Ozsvath, Z.~Szabo, ``Holomorphic disks and link
invariants,'' math.GT/0512286.}

\lref\MengT{G.~Meng, C.~Taubes, ``$SW=$ Milnor Torsion,'' Math.
Res. Lett. {\bf 3} (1996) 661.}

\lref\GSV{S.~Gukov, A.~Schwarz and C.~Vafa, ``Khovanov-Rozansky
homology and topological strings,'' hep-th/0412243.}

\lref\WittenJones{ E.~Witten, ``Quantum Field Theory And The Jones
Polynomial,'' Commun.\ Math.\ Phys.\  {\bf 121}, 351 (1989).}

\lref\HOMFLY{P.~Freyd, D.~Yetter, J.~Hoste, W.~Lickorish,
K.~Millett, A.~Oceanu, ``A New Polynomial Invariant of Knots and
Links,'' Bull. Amer. Math. Soc. {\bf 12} (1985) 239.}

\lref\Wittencsstring{ E.~Witten, ``Chern-Simons gauge theory as a
string theory,'' Prog.\ Math.\  {\bf 133} (1995) 637,
hep-th/9207094.}

\lref\OV{ H.~Ooguri, C.~Vafa, ``Knot Invariants and Topological
Strings,'' Nucl.Phys. {\bf B577} (2000) 419.}

\lref\Khovanov{M.~ Khovanov, ``A categorification of the Jones
polynomial,'' math.QA/9908171.}

\lref\Khovanovii{M.~ Khovanov, ``Categorifications of the colored
Jones polynomial,'' math.QA/0302060.}

\lref\Khovanoviii{M.~ Khovanov, ``$sl(3)$ link homology I,''
math.QA/0304375.}

\lref\Khovanoviv{M.~ Khovanov, ``An invariant of tangle
cobordisms,'' math.QA/0207264.}

\lref\Khovanovtangles{M.~Khovanov, ``A functor-valued invariant of
tangles,'' Algebr. Geom. Topol. {\bf 2} (2002) 665,
math.QA/0103190.}

\lref\DBN{D.~Bar-Natan, ``On Khovanov's categorification of the
Jones polynomial,'' math.QA/0201043.}

\lref\DBNnews{D.~Bar-Natan, ``Some Khovanov-Rozansky
Computations'' \semi
{http://www.math.toronto.edu/~drorbn/Misc/KhovanovRozansky/index.html}}

\lref\Jacobsson{M.~Jacobsson, ``An invariant of link cobordisms
from Khovanov's homology theory,'' math.GT/0206303.}

\lref\RKhovanov{M.~Khovanov, L.~Rozansky, ``Matrix factorizations
and link homology,'' math.QA/0401268.}

\lref\Rfoam{L.~Rozansky, ``Topological A-models on seamed Riemann
surfaces,'' hep-th/0305205.}

\lref\KRfoam{M.~Khovanov and L.~Rozansky, ``Topological
Landau-Ginzburg models on a world-sheet foam,'' hep-th/0404189.}

\lref\GopakumarV{R.~Gopakumar and C.~Vafa, ``On the gauge
theory/geometry correspondence,'' Adv.\ Theor.\ Math.\ Phys.\
{\bf 3} (1999) 1415, hep-th/9811131.}

\lref\GViii{R.~Gopakumar and C.~Vafa, ``M-theory and topological
strings. I,II,'' hep-th/9809187; hep-th/9812127.}

\lref\KKV{S.~Katz, A.~Klemm and C.~Vafa, ``M-theory, topological
strings and spinning black holes,'' Adv.\ Theor.\ Math.\ Phys.\
{\bf 3} (1999) 1445, hep-th/9910181.}

\lref\mirbook{``Mirror Symmetry'' (Clay Mathematics Monographs, V.
1), K.~Hori et.al. ed, American Mathematical Society, 2003.}

\lref\HV{K.~Hori, C.~Vafa, ``Mirror Symmetry,'' hep-th/0002222;
K.~Hori, A.~Iqbal, C.~Vafa, ``D-Branes And Mirror Symmetry,''
hep-th/0005247.}

\lref\AKV{M.~Aganagic, A.~Klemm, C.~Vafa, ``Disk Instantons,
Mirror Symmetry and the Duality Web,'' hep-th/0105045.}

\lref\AKMV{M.~Aganagic, A.~Klemm, M.~Marino and C.~Vafa, ``Matrix
model as a mirror of Chern-Simons theory,'' JHEP {\bf 0402}, 010
(2004), hep-th/0211098.}

\lref\AAHV{B.~Acharya, M.~Aganagic, K.~Hori and C.~Vafa,
``Orientifolds, mirror symmetry and superpotentials,''
hep-th/0202208.}

\lref\LMV{J.~M.~F.~Labastida, M.~Marino and C.~Vafa, ``Knots,
links and branes at large N,'' JHEP {\bf 0011}, 007 (2000),
hep-th/0010102.}

\lref\LMtorus{J.~M.~F.~Labastida and M.~Marino, ``Polynomial
invariants for torus knots and topological strings,'' Commun.\
Math.\ Phys.\  {\bf 217} (2001) 423, hep-th/0004196.}

\lref\LMqa{J.~M.~F.~Labastida and M.~Marino, ``A new point of view
in the theory of knot and link invariants,'' math.qa/0104180.}

\lref\HSTa{S.~Hosono, M.-H.~Saito, A.~Takahashi, ``Holomorphic
Anomaly Equation and BPS State Counting of Rational Elliptic
Surface,'' Adv.Theor.Math.Phys. {\bf 3} (1999) 177.}

\lref\HSTb{S.~Hosono, M.-H.~Saito, A.~Takahashi, ``Relative
Lefschetz Action and BPS State Counting,'' Internat. Math. Res.
Notices, (2001), No. 15, 783.}

\lref\Kprivate{M.~Khovanov, private communication.}

\lref\Taubes{ C.~Taubes, ``Lagrangians for the Gopakumar-Vafa
conjecture,'' math.DG/0201219.}

\lref\Wittenams{E.~Witten, ``Dynamics of Quantum Field Theory,''
{\it Quantum Fields and Strings: A Course for Mathematicians} (P.
Deligne, {\it et.al.} eds.), vol. 2, AMS Providence, RI, (1999)
pp. 1313-1325.}

\lref\LVW{W.~Lerche, C.~Vafa and N.~P.~Warner, ``Chiral Rings In
N=2 Superconformal Theories,'' Nucl.\ Phys.\ B {\bf 324}, 427
(1989).}

\lref\HMoore{J.~A.~Harvey and G.~W.~Moore, ``On the algebras of
BPS states,'' Commun.\ Math.\ Phys.\  {\bf 197} (1998) 489,
hep-th/9609017.}

\lref\IqbalV{T.~J.~Hollowood, A.~Iqbal and C.~Vafa, ``Matrix
models, geometric engineering and elliptic genera,''
hep-th/0310272.}

\lref\Schwarz{A.~Schwarz, ``New topological invariants arising in
the theory of quantized fields,'' Baku International Topological
Conf., Abstracts (part II) (1987).}

\lref\SchwarzS{A.~Schwarz and I.~Shapiro, ``Some remarks on
Gopakumar-Vafa invariants,'' hep-th/0412119.}

\lref\Aspinwallrev{ P.~S.~Aspinwall, ``D-branes on Calabi-Yau
manifolds,'' hep-th/0403166.}

\lref\MNOP{D.~Maulik, N.~Nekrasov, A.~Okounkov, R.~Pandharipande,
``Gromov-Witten theory and Donaldson-Thomas theory, I,''
math.AG/0312059.}

\lref\Katz{S.~Katz, ``Gromov-Witten, Gopakumar-Vafa, and
Donaldson-Thomas invariants of Calabi-Yau threefolds,''
math.ag/0408266.}

\lref\FultonH{W.~Fulton, J.~Harris, ``Representation Theory: A
First Course,'' Springer-Verlag 1991.}

\lref\Kontsevich{M.~Kontsevich, unpublished.}

\lref\KapustinLi{A.~Kapustin and Y.~Li, ``D-branes in
Landau-Ginzburg models and algebraic geometry,'' JHEP {\bf 0312}
(2003) 005, hep-th/0210296.}

\lref\KapustinLii{A.~Kapustin and Y.~Li, ``Topological correlators
in Landau-Ginzburg models with boundaries,'' Adv.\ Theor.\ Math.\
Phys.\  {\bf 7} (2004) 727, hep-th/0305136.}

\lref\KapustinLiii{A.~Kapustin and Y.~Li, ``D-branes in
topological minimal models: The Landau-Ginzburg approach,'' JHEP
{\bf 0407} (2004) 045, hep-th/0306001.}

\lref\Brunneriii{ I.~Brunner, M.~Herbst, W.~Lerche and J.~Walcher,
``Matrix factorizations and mirror symmetry: The cubic curve,''
hep-th/0408243.}

\lref\Brunner{I.~Brunner, M.~Herbst, W.~Lerche and B.~Scheuner,
``Landau-Ginzburg realization of open string TFT,''
hep-th/0305133.}

\lref\HLLii{M.~Herbst, C.~I.~Lazaroiu and W.~Lerche, ``D-brane
effective action and tachyon condensation in topological minimal
models,'' hep-th/0405138.}

\lref\HLL{M.~Herbst, C.~I.~Lazaroiu and W.~Lerche,
``Superpotentials, A(infinity) relations and WDVV equations for
open topological strings,'' hep-th/0402110. }

\lref\LercheJW{W.~Lerche and J.~Walcher, ``Boundary rings and N =
2 coset models,'' Nucl.\ Phys.\ B {\bf 625} (2002) 97,
hep-th/0011107.}

\lref\HoriJW{K.~Hori and J.~Walcher, ``F-term equations near
Gepner points,'' hep-th/0404196.}

\lref\Orlov{D.~Orlov, ``Triangulated Categories of Singularities
and D-Branes in Landau-Ginzburg Orbifold,'' math.AG/0302304.}

\lref\Emanuelii{S.~K.~Ashok, E.~Dell'Aquila, D.~E.~Diaconescu and
B.~Florea, ``Obstructed D-branes in Landau-Ginzburg orbifolds,''
hep-th/0404167.}

\lref\Emanuel{ S.~K.~Ashok, E.~Dell'Aquila and D.~E.~Diaconescu,
``Fractional branes in Landau-Ginzburg orbifolds,''
hep-th/0401135.}

\lref\Guadagnini{E~.Guadagnini, ``The Link Invariants of the
Chern-Simons Field Theory: New Developments in Topological Quantum
Field Theory,'' Walter de Gruyter Inc., 1997.}

\lref\MOY{H.~Murakami, T.~Ohtsuki, S.~Yamada, ``HOMFLY polynomial
via an invariant of colored plane graphs,'' Enseign. Math. {\bf
44} (1998) 325.}

\lref\Shumakovitch{A.~Shumakovitch, {\it KhoHo} --- a program for
computing and studying Khovanov homology,
{http://www.geometrie.ch/KhoHo}}

\lref\DBNtangles{D.~Bar-Natan, ``Khovanov's Homology for Tangles
and Cobordisms,'' math.GT/0410495.}

\lref\Lee{E.S.~Lee, ``The support of the Khovanov's invariants for
alternating knots,'' math.GT/0201105.}

\lref\HTi{A.~Hanany and D.~Tong,``Vortices, instantons and
branes,'' JHEP {\bf 0307}, 037 (2003), hep-th/0306150.}

\lref\HTii{A.~Hanany and D.~Tong, ``Vortex strings and
four-dimensional gauge dynamics,'' JHEP {\bf 0404}, 066 (2004),
hep-th/0403158.}

\lref\Bradlowi{S.~Bradlow, ``Vortices in Holomorphic Line Bundles
over Closed Kahler Manifolds,'' Commun. Math. Phys. {\bf 135} (1990) 1.}

\lref\BGPvortices{S.~Bradlow, O.~Garcia-Prada, ``Non-abelian
monopoles and vortices,'' math.AG/9602010.}

\lref\OTi{C.~Okonek, A.~Teleman, ``The Coupled Seiberg-Witten
Equations, vortices, and Moduli spaces of stable pairs,''
alg-geom/9505012.}

\lref\OTii{C.~Okonek, A.~Teleman, ``Quaternionic Monopoles,''
alg-geom/9505029.}

\lref\OTiii{C.~Okonek, A.~Teleman, ``Recent Developments in
Seiberg-Witten Theory and Complex Geometry,'' alg-geom/9612015.}

\lref\Gornik{B.~Gornik, ``Note on {K}hovanov link cohomology,''
math.QA/0402266.}

\lref\DGR{ N.~Dunfield, S.~Gukov, J.~Rasmussen, ``The
Superpolynomial for Knot Homologies,'' math.GT/0505662.}

\lref\Fukaya{K.~Fukaya, ``Floer homology for 3-manifolds with
boundary,'' in {\it Topology, Geometry, and Field Theory}, World
Sci. Publishing, River Edge, NJ, 1994.}

\lref\KMbook{P.B.~Kronheimer, T.S.~Mrowka, ``Floer homology for
Seiberg-Witten monopoles,'' in preparation.}

\lref\HKhovanov{R.~Huerfano, M.~Khovanov, ``A category for the
adjoint representation,'' math.QA/0002060.}

\lref\SeidelSmith{P.~Seidel, I.~Smith, ``A link invariant from the
symplectic geometry of nilpotent slices,'' math.SG/0405089.}

\lref\CFrenkel{L.~Crane, I.~Frenkel, ``Four-dimensional
topological quantum field theory, Hopf categories, and the
canonical bases,'' J. Math. Phys. {\bf 35} (1994) 5136.}

\lref\Fischer{J.~Fischer, ``2-categories and 2-knots,'' Duke Math.
J. {\bf 75} (1994) 493.}

\lref\Seibergetal{P.~C.~Argyres, M.~R.~Plesser and N.~Seiberg,
``The Moduli Space of N=2 SUSY {QCD} and Duality in N=1 SUSY
{QCD},'' Nucl.\ Phys.\ B {\bf 471} (1996) 159, hep-th/9603042.}

\lref\DGNV{R.~Dijkgraaf, S.~Gukov, A.~Neitzke and C.~Vafa,
``Topological M-theory as unification of form theories of
gravity,'' hep-th/0411073.}

\lref\JaffeT{A.~Jaffe, C.~Taubes, ``Vortices and monopoles,''
Birkhauser, Boston MA, 1980.}

\lref\HoriH{A.~Hanany and K.~Hori, ``Branes and N = 2 theories in
two dimensions,'' Nucl.\ Phys.\ B {\bf 513} (1998) 119,
hep-th/9707192.}

\lref\WittenDonaldson{E.~Witten, ``Topological Quantum Field
Theory,'' Commun.\ Math.\ Phys.\  {\bf 117} (1988) 353.}

\lref\Wittenmonopoles{E.~Witten, ``Monopoles and four manifolds,''
Math.\ Res.\ Lett.\  {\bf 1} (1994) 769, hep-th/9411102.}

\lref\Wittensigma{E.~Witten, ``Topological Sigma Models,''
Commun.\ Math.\ Phys.\  {\bf 118} (1988) 411.}

\lref\Wittenmirror{E.~Witten, ``Mirror manifolds and topological
field theory,'' hep-th/9112056.}

\lref\BaezD{J.~C.~Baez and J.~Dolan, ``Higher dimensional algebra
and topological quantum field theory,'' J.\ Math.\ Phys.\  {\bf
36} (1995) 6073, q-alg/9503002.}

\lref\Kontsevichhom{M.~Kontsevich, ``Homological Algebra of Mirror
Symmetry,'' alg-geom/9411018.}

\lref\BondalK{A.I.~Bondal, M.M.~Kapranov, ``Framed triangulated
categories,'' Math. USSR-Sb.  {\bf 70}  (1991) 93.}

\lref\Rastorus{J.~Rasmussen, ``Khovanov's invariant for closed
surfaces,'' math.GT/050252.}

\lref\CSS{J.S.~Carter, M.~Saito, S.~Satoh, ``Ribbon-moves for
2-knots with 1-handles attached and Khovanov-Jacobsson numbers,''
math.GT/0407493.}

\lref\LMnonab{J.~M.~F.~Labastida and M.~Marino, ``NonAbelian
monopoles on four manifolds,'' Nucl.\ Phys.\ B {\bf 448} (1995)
373, hep-th/9504010.}

\lref\PTyurin{V.~Pidstrigach, A.~Tyurin, ``Localisation of the
Donaldson's invariants along Seiberg-Witten classes,''
dg-ga/9507004.}

\lref\Taubeslett{C.~Taubes, ``The Seiberg-Witten and Gromov
invariants,'' Math. Res. Lett. {\bf 2}  (1995) 221.}

\lref\Taubesbook{C.~Taubes, ``Seiberg Witten and Gromov invariants
for symplectic 4-manifolds,''
International Press, Somerville, MA, 2000.}

\lref\Leung{N.~C.~Y.~Leung, ``Topological quantum field theory for
Calabi-Yau threefolds and G(2) manifolds,'' Adv.\ Theor.\ Math.\
Phys.\  {\bf 6} (2003) 575, math.DG/0208124.}

\lref\LeungXW{N.~C.~Y.~Leung, X.~Wang, ``Intersection theory of
coassociative submanifolds in G(2)-manifolds and Seiberg-Witten
invariants,'' math.DG/0401419.}

\lref\ASalur{S.~Akbulut and S.~Salur, ``Calibrated manifolds and
gauge theory,'' math.GT/0402368; ``Associative submanifolds of a
G2 manifold,'' math.GT/0412032.}

\lref\WMoore{ G.~Moore, E.~Witten, ``Integration over the u-plane
in Donaldson theory,'' Adv. Theor. Math. Phys. {\bf 1} (1998) 298,
hep-th/9709193.}

\lref\Dorey{N.~Dorey, ``The BPS Spectra of Two-Dimensional
Supersymmetric Gauge Theories with Twisted Mass Terms,'' JHEP {\bf
9811} (1998) 005, hep-th/9806056.}

\lref\DHT{N.~Dorey, T.J.~Hollowood, D.~Tong, ``The BPS Spectra of
Gauge Theories in Two and Four Dimensions,'' JHEP {\bf 9905}
(1999) 006, hep-th/9902134.}

\lref\SWi{N.~Seiberg and E.~Witten, ``Electric - magnetic duality,
monopole condensation, and confinement in N=2 supersymmetric
Yang-Mills theory,'' Nucl.\ Phys.\ B {\bf 426} (1994) 19
[Erratum-ibid.\ B {\bf 430} (1994) 485], hep-th/9407087.}

\lref\SWii{N.~Seiberg and E.~Witten, ``Monopoles, duality and
chiral symmetry breaking in N=2 supersymmetric QCD,'' Nucl.\
Phys.\ B {\bf 431} (1994) 484, hep-th/9408099.}

\lref\ALabastida{M.~Alvarez, J.M.F.~Labastida, ``Topological
Matter in Four Dimensions,'' Nucl.Phys. {\bf B437} (1995) 356,
hep-th/9404115.}

\lref\WessBagger{J.~Wess, J.~Bagger, ``Supersymmetry and
Supergravity,'' 2nd edition, Princeton University Press, 1992.}

\lref\LLectures{J.~M.~F.~Labastida and C.~Lozano, ``Lectures on
topological quantum field theory,'' hep-th/9709192.}

\lref\Wphases{E.~Witten, ``Phases of N = 2 theories in two
dimensions,'' Nucl.\ Phys.\ B {\bf 403} (1993) 159,
hep-th/9301042.}

\lref\Wgrassmannian{E.~Witten, ``The Verlinde algebra and the
cohomology of the Grassmannian,'' hep-th/9312104.}

\lref\BS{M.~Bershadsky and V.~Sadov, ``Theory of Kahler gravity,''
Int.\ J.\ Mod.\ Phys.\ A {\bf 11}, 4689 (1996), hep-th/9410011.}

\lref\AKMVvertex{M.~Aganagic, A.~Klemm, M.~Marino and C.~Vafa,
``The topological vertex,'' hep-th/0305132.}

\lref\Apol{S.~Gukov, ``Three-dimensional quantum gravity,
Chern-Simons theory, and the  A-polynomial,'' hep-th/0306165.}

\lref\BCOVa{M.~Bershadsky, S.~Cecotti, H.~Ooguri and C.~Vafa,
``Holomorphic anomalies in topological field theories,'' Nucl.\
Phys.\ B {\bf 405}, 279 (1993).}

\lref\BCOV{M.~Bershadsky, S.~Cecotti, H.~Ooguri and C.~Vafa,
``Kodaira-Spencer theory of gravity and exact results for quantum
string amplitudes,'' Commun.\ Math.\ Phys.\  {\bf 165}, 311
(1994).}

\lref\BJSV{M.~Bershadsky, A.~Johansen, V.~Sadov and C.~Vafa,
``Topological Reduction of 4-d SYM to 2-d Sigma Models,'' Nucl.\
Phys.\ B {\bf 448} (1995) 166, hep-th/9501096.}

\lref\VafaW{C.~Vafa and E.~Witten, ``A Strong coupling test of S
duality,'' Nucl.\ Phys.\ B {\bf 431} (1994) 3, hep-th/9408074.}

\lref\Manolescu{C.~Manolescu, ``Nilpotent slices, Hilbert schemes,
and the Jones polynomial,'' math.SG/0411015.}

\lref\Manolescusln{C.~Manolescu, ``Link homology theories from
symplectic geometry,'' math.SG/0601629.}

\lref\Bigelow{S.~Bigelow, ``A homological definition of the Jones
polynomial,'' Geom. Topol. Monogr. {\bf 4} (2002) 29,
math.GT/0201221.}

\lref\Beilinson{A.A.~Beilinson, V.G.~Drinfeld, ``Quantization of
Hitchin's fibrations and Langlands' program,'' Math. Phys. Stud.
{\bf 19}, Kluwer Acad. Publ. (1996) 3.}

\lref\DBZias{D.~Ben-Zvi, ``Geometric Langlands as a Quantization
of the Hitchin System,'' lecture at the Princeton workshop on the
Langlands correspondence and physics.}

\lref\CCGLS{D.~Cooper, M.~Culler, H.~Gillet, D.D.~Long, P.B.
Shalen, ``Plane curves associated to character varieties of
3-manifolds,'' Invent. Math. {\bf 118} (1994) 47.}

\lref\Stavros{S.~Garoufalidis, J.~Geronimo, ``Asymptotics of
$q$-difference Equations,'' math.QA/0405331; S.~Garoufalidis,
``Difference and differential equations for the colored Jones
function,'' math.GT/0306229.}

\lref\KWitten{A.~Kapustin, E.~Witten, ``Electric-Magnetic Duality
and the Geometric Langlands Program,'' hep-th/0604151.}

\lref\BTnovel{M.~Blau, G.~Thompson, Nucl.\ Phys.\ B {\bf 492}
(1997) 545; Phys.\ Lett.\ B {\bf 415} (1997) 242.}

\lref\BTbranes{M.~Blau and G.~Thompson, ``Aspects of N(T) >= 2
topological gauge theories and D-branes,'' Nucl.\ Phys.\ B {\bf
492} (1997) 545, hep-th/9612143.}

\lref\BTeucl{M.~Blau and G.~Thompson, ``Euclidean SYM theories by
time reduction and special holonomy  manifolds,'' Phys.\ Lett.\ B
{\bf 415} (1997) 242, hep-th/9706225.}

\lref\BTonRW{M.~Blau and G.~Thompson, ``On the relationship
between the Rozansky-Witten and the three-dimensional
Seiberg-Witten invariants,'' Adv.\ Theor.\ Math.\ Phys.\  {\bf 5}
(2002) 483, hep-th/0006244.}

\lref\GMext{B.~Geyer, D.~Mülsch, ``$N_T=4$ equivariant extension
of the 3D topological model of Blau and Thompson,'' Nucl.Phys.
{\bf B616} (2001) 476, hep-th/0108042.}

\lref\MMsympl{M.~Marino and G.~W.~Moore, ``Donaldson invariants
for non-simply connected manifolds,'' Commun.\ Math.\ Phys.\  {\bf
203} (1999) 249, hep-th/9804104.}

\lref\MMthreed{M.~Marino and G.~W.~Moore, ``3-manifold topology
and the Donaldson-Witten partition function,'' Nucl.\ Phys.\ B
{\bf 547} (1999) 569, hep-th/9811214.}

\lref\MMcoulomb{M.~Marino and G.~W.~Moore,
``Integrating over the Coulomb branch in N = 2 gauge theory,''
Nucl.\ Phys.\ Proc.\ Suppl.\  {\bf 68} (1998) 336, hep-th/9712062.}

\lref\MMrank{M.~Marino and G.~W.~Moore,
``The Donaldson-Witten function for gauge groups of rank larger than one,''
Commun.\ Math.\ Phys.\  {\bf 199} (1998) 25, hep-th/9802185.}

\lref\MMPeradze{M.~Marino, G.~W.~Moore and G.~Peradze,
``Superconformal invariance and the geography of four-manifolds,''
Commun.\ Math.\ Phys.\  {\bf 205} (1999) 691, hep-th/9812055.}

\lref\Issues{A.~Losev, N.~Nekrasov and S.~L.~Shatashvili,
``Issues in topological gauge theory,''
Nucl.\ Phys.\ B {\bf 534} (1998) 549, hep-th/9711108.}

\lref\RW{L.~Rozansky and E.~Witten, ``Hyper-Kaehler geometry and
invariants of three-manifolds,'' Selecta Math.\  {\bf 3} (1997)
401, hep-th/9612216.}

\lref\Simpson{C.~Simpson, ``Nonabelian Hodge theory,'' Proc. I.C.M., Kyoto 1990,
Springer-Verlag, 1991, pp. 198-230.}

\lref\IwasakiU{K.~Iwasaki, T.~Uehara, ``Periodic Solutions
to Painleve VI and Dynamical System on Cubic Surface,'' math.AG/0512583.}

\lref\IISaito{M.~Inaba, K.~Iwasaki and M.-H.~Saito,
``Dynamics of the sixth Painleve equation,''
Proceedings of Conference Internationale Theories Asymptotiques
et Equations de Painleve, Seminaires et Congres, Soc. Math. France, math.AG/0501007.}

\lref\Iwasaki{K.~Iwasaki, ``An area-preserving action of the modular group
on cubic surfaces and the Painleve VI equation,'' Comm. Math. Phys. {\bf 242} (2003) 185.}

\lref\Iwasakid{K.~Iwasaki, ``A modular group action on cubic surfaces and the monodromy
of the Painleve VI equation,'' Proc. Japan Acad. Ser. A Math. Sci. {\bf 78} (2002) 131.}

\lref\Sakai{H.~Sakai, ``Rational surfaces associated with affine root systems
and geometry of the Painleve equations,'' Comm. Math. Phys. {\bf 220}  (2001) 165.}

\lref\HitSchles{N.~Hitchin, ``Geometrical aspects of Schlesinger's equation,''
J. Geom. Phys. {\bf 23}  (1997) 287.}

\lref\Lin{X.S.~Lin, ``A knot invariant via representation spaces,''
J. Diff. Geom. {\bf 35} (1992) 337.}

\lref\Herald{C.~Herald, ``Flat connections, the Alexander invariant,
and Casson's invariant,'' Comm. Anal. Geom. {\bf 5}  (1997) 93.}

\lref\CollinSteer{O.~Collin, B.~Steer, ``Instanton Floer homology for knots
via 3-orbifolds,'' J. Diff. Geom. {\bf 51} (1999) 149.}

\lref\Collin{O.~Collin, ``Floer Homology for Knots and $SU(2)$-Representations
for Knot Complements and Cyclic Branched Covers,'' Canad. J. Math. {\bf 52} (2000) 293.}

\lref\Klassen{E.~Klassen, ``Representations of knot groups in $SU(2)$,''
Trans. Amer. Math. Soc. {\bf 326} (1991) 795.}

\lref\BHKK{H.~Boden, C.~Herald, P.~Kirk, E.~Klassen,
``Gauge Theoretic Invariants of Dehn Surgeries on Knots,''
Geom. Topol. {\bf 5} (2001) 143, math.GT/9908020.}

\lref\WLi{W.~Li, ``Knot and link invariants and moduli space
of parabolic bundles,'' Commun. Contemp. Math. {\bf 3} (2001) 501.}

\lref\DonaldsonMT{S.K.~Donaldson,
``Topological field theories and formulae of Casson and Meng-Taubes,''
Geom. Topol. Monogr.{\bf 2} (1999) 87.}

\lref\Bradlow{S.~Bradlow, ``Special metrics and stability for holomorphic bundles
with global sections,'' J. Diff. Geom. {\bf 33} (1991) 169.}

\lref\OSadjunction{P.~Ozsvath, Z.~Szabo, ``Higher type adjunction inequalities
in Seiberg-Witten theory,'' J. Diff. Geom. {\bf 55} (2000) 385.}

\lref\MrowkaOY{T.~Mrowka, P.~Ozsvath, B.~Yu, ``Seiberg-Witten monopoles on Seifert
fibered spaces,'' Comm. Anal. Geom. {\bf 5} (1997) 685.}

\lref\MorganST{J.~Morgan, Z.~Szabo, C.~Taubes, ``A product formula for the Seiberg-Witten
invariants and the generalized Thom conjecture,''  J. Diff. Geom. {\bf 44} (1996) 706.}

\lref\MarkSW{T.~Mark, ``Torsion, TQFT, and Seiberg-Witten invariants of 3-manifolds,''
Geom.Topol. {\bf 6} (2002) 27.}

\lref\MunozWang{V.~Munoz, B.-L.~Wang, ``Seiberg-Witten-Floer homology of a surface
times a circle for non-torsion ${\rm spin}\sp {\Bbb C}$ structures,''
Math. Nachr. {\bf 278}  (2005) 844.}

\lref\NielsenOlesen{H.~B.~Nielsen and P.~Olesen,
``Vortex-Line Models For Dual Strings,''  Nucl.\ Phys.\ B {\bf 61} (1973) 45}

\lref\KMi{P.~Kronheimer, T.~Mrowka, ``Gauge theory for embedded surfaces. I,''
Topology {\bf 32} (1993) 773.}

\lref\KMii{P.~Kronheimer, T.~Mrowka, ``Gauge theory for embedded surfaces. II,''
Topology {\bf 34} (1995) 37.}

\lref\KMiandii{P.~Kronheimer, T.~Mrowka, ``Gauge theory for embedded surfaces, I, II,''
Topology {\bf 32} (1993) 773; Topology {\bf 34} (1995) 37.}

\lref\KMstructure{P.~Kronheimer, T.~Mrowka, ``Embedded surfaces and the structure
of Donaldson's polynomial invariants,'' J. Diff. Geom. {\bf 41} (1995) 573.}

\lref\WittenAbelian{ E.~Witten, ``On S duality in Abelian gauge theory,''
Selecta Math.\  {\bf 1} (1995) 383, hep-th/9505186.}

\lref\VerlindeAbelian{E.~Verlinde, ``Global aspects of electric - magnetic duality,''
Nucl.\ Phys.\ B {\bf 455} (1995) 211, hep-th/9506011.}

\lref\Gottsche{L.~G\"{o}ttsche, ``The Betti numbers of the Hilbert scheme of points
on a smooth projective surface,''  Math. Ann. {\bf 286} (1990) 193.}

\lref\Braverman{A.~Braverman,
``Instanton counting via affine Lie algebras I:
Equivariant J-functions of (affine) flag manifolds and Whittaker vectors,''
math.AG/0401409.}

\lref\BEtingof{A.~Braverman, P.~Etingof,
``Instanton counting via affine Lie algebras II: from Whittaker
vectors to the Seiberg-Witten prepotential,'' math.AG/0409441.}


\lref\Seidel{P.~Seidel, ``Lagrangian two-spheres can be symplectically knotted,''
J. Diff. Geom. {\bf 52} (1999) 145.}

\lref\SThomas{P.~Seidel, R.~Thomas, ``Braid group actions on derived categories
of coherent sheaves,'' Duke Math. J. {\bf 108} (2001) 37.}

\lref\DeligneBraid{P.~Deligne, ``Action du groupe des tresses sur une categorie,''
Invent. Math. {\bf 128} (1997) 159.}

\lref\Bielawski{R.~Bielawski, ``Hyper-Kähler structures and group actions,''
J. London Math. Soc. {\bf 55} (1997) 400.}

\lref\Kronheimeri{P.~Kronheimer, ``Instantons and the Geometry
of the Nilpotent Variety,'' J. Diff. Geom. {\bf 32}  (1990) 473.}

\lref\Kronheimerii{P.~Kronheimer, ``A Hyper-Kahlerian Structure on
Coadjoint Orbits of a Semisimple Complex Group,'' J. London Math.
Soc. {\bf 42} (1990) 193.}

\lref\ChrissG{N.~Chriss, V.~Ginzburg, ``Representation theory and
complex geometry,'' Birkhauser Boston, Inc., Boston, MA, 1997.}

\lref\Bezrukavnikov{R.~Bezrukavnikov,
``Noncommutative Counterparts of the Springer Resolution,'' math.RT/0604445.}

\lref\Bridgelandi{T.~Bridgeland,
``T-structures on some local Calabi-Yau varieties,'' math.AG/0502050.}

\lref\Bridgelandii{T.~Bridgeland,
``Stability conditions and Kleinian singularities,'' math.AG/0508257.}

\lref\Bridgelandiii{T.~Bridgeland,
``Stability conditions on a non-compact Calabi-Yau threefold,'' math.AG/0509048.}

\lref\IshiiUehara{A.~Ishii, H.~Uehara,
``Autoequivalences of derived categories on the minimal resolutions
of $A_n$-singularities on surfaces,'' math.AG/0409151.}

\lref\KVesserot{M.~Kapranov, E.~Vasserot,
``Kleinian singularities, derived categories and Hall algebras,''
math.AG/9812016.}

\lref\Landweber{G.~Landweber, ``Singular instantons with SO(3) symmetry,''
math.DG/0503611.}

\lref\Witteninst{E.~Witten,
``Some Exact Multi - Instanton Solutions Of Classical Yang-Mills Theory,''
Phys.\ Rev.\ Lett.\  {\bf 38} (1977) 121.}

\lref\Nekrasov{N.~Nekrasov, ``Seiberg-Witten Prepotential From
Instanton Counting,'' Adv. Theor. Math. Phys. {\bf 7} (2004) 831.}

\lref\BodenY{H.U.~ Boden, K.~ Yokogawa, ``Moduli Spaces of Parabolic
Higgs Bundles and Parabolic K(D) Pairs over Smooth Curves: I,'' J.
Math. {\bf 7} (1996) 573.}

\lref\GPGM{O.~García-Prada, P.B.~Gothen, V.~Munoz,
``Betti numbers of the moduli space of rank 3 parabolic Higgs bundles,''
math.AG/0411242.}

\lref\Kirwan{F.~Kirwan, ``Cohomology of quotients in symplectic and
algebraic geometry,'' Mathematical Notes {\bf 31}, Princeton University Press, 1984.}

\lref\WittenSolutions{E.~Witten,
``Solutions of four-dimensional field theories via M-theory,''
Nucl.\ Phys.\ B {\bf 500} (1997) 3, hep-th/9703166.}

\lref\CMcGovern{D.~Collingwood, W.~McGovern,
``Nilpotent Orbits in Semisimple Lie Algebras,''
Van Nostrand Reinhold Math. Series, New York, 1993.}

\lref\LusztigS{G.~Lusztig, N.~Spaltenstein, ``Induced Unipotent
Classes,'' J. London Math. Soc. {\bf 19} (1979) 41.}

\lref\Lusztigiv{G.~Lusztig, ``A Class of Representations of a Weyl Group,''
Proc. Kon. Nederl. Akad. A {\bf 82} (1979) 323.}

\lref\Hirai{T.~Hirai, ``On Richardson classes of unipotent elements in
semisimple algebraic groups,''  Proc. Japan Acad. Ser. A Math. Sci. {\bf 57} (1981) 367;
``Structure of unipotent orbits and Fourier transform of unipotent orbital
integrals for semisimple Lie groups,''
Lectures on harmonic analysis on Lie groups and related topics,
Lectures in Math.{\bf 14}, Tokyo, 1982.}

\lref\Spaltenstein{N.~Spaltenstein, ``Classes Unipotentes et
Sous-groupes de Borel,'' Lecture Notes in Math. 946, Springer,
Berlin-New York, 1982.}

\lref\Hesselink{W.~Hesselink, ``Polarizations in the classical groups,''
Math. Z. {\bf 160} (1978) 217.}

\lref\Lusztigi{G.~Lusztig, ``Intersection cohomology complexes on a reductive group,''
Invent. Math. {\bf 75} (1984) 205.}

\lref\Lusztigii{G.~Lusztig, ``Notes on Unipotent Classes,'' Asian J.
Math. {\bf 1} (1997) 194.}

\lref\Lusztigiii{G.~Lusztig, ``Characters of Reductive Groups over a
Finite Field.'' Annals of Mathematics Studies, 107. Princeton
University Press, Princeton, NJ, 1984.}

\lref\Achar{P.~Achar, ``An order-reversing duality map for conjugacy classes
in Lusztig's canonical quotient,'' Transform. Groups {\bf 8} (2003) 107.}

\lref\ASage{P.~Achar, D.~Sage, ``On special pieces, the Springer correspondence,
and unipotent characters,'' math.RT/0606075.}

\lref\Richardson{R.~Richardson, ``Conjugacy classes of parabolic subgroups
in semisimple algebraic groups,'' Bull. London Math. Soc. {\bf 6} ( 1974 ) 21.}

\lref\Fu{B.~Fu, ``Symplectic Resolutions for Nilpotent Orbits,''
Invent. Math. {\bf 151} (2003) 167.}

\lref\Namikawa{Y.~Namikawa, ``Birational Geometry of symplectic resolutions
of nilpotent orbits I-II'', math.AG/0404072, math.AG/0408274.}

\lref\KLmap{D.Kazhdan, G.~Lusztig, ``Fixed Point Varieties on Affine Flag Manifolds,''
Isr. J. Math. {\bf 62} (1988) 129.}

\lref\SpaltensteinKL{N.~Spaltenstein, ``Order Relations on Conjugacy
Classes and the Kazhdan-Lusztig Map,'' Math. Ann. {\bf 292} (1992)
281.}

\lref\Ramified{S.~Gukov, E.~Witten, ``Gauge Theory, Ramification,
and the Geometric Langlands Program,'' hep-th/0612073.}

\lref\Nakajimai{H.~Nakajima, ...}

\lref\KSwann{P.~Kobak, A.~Swann, ``Classical nilpotent orbits as hyperKahler quotients,''
Int. J. Math. {\bf 7} (1996), 193.}

\lref\BBMacPherson{W.~Borho, J.-L.~Brylinski, R.~MacPherson,
``Nilpotent Orbits, Primitive Ideals, and Characteristic Classes,''
{\it A Geometric Perspective in Ring Theory. Progress in Mathematics,}
78. Birkhauser Boston, Inc., Boston, MA, 1989.}

\lref\dCKac{C.~De Concini, V.~Kac, ``Representations of Quantum
Groups at Roots of $1$: Reduction to the Exceptional Case,'' {\it
Infinite analysis, Part A, B (Kyoto, 1991)}, Adv. Ser. Math. Phys.
{\bf 16} (1992) 141.}

\lref\Maldacena{J.~M.~Maldacena,
``The Large N Limit of Superconformal Field Theories and Supergravity,''
Adv.\ Theor.\ Math.\ Phys.\  {\bf 2} (1998) 231, hep-th/9711200.}

\lref\KnotHom{S.~Gukov,
``Surface Operators and Knot Homologies,'' arXiv:0706.2369 [hep-th].}

\lref\Wild{E.~Witten,
``Gauge Theory and Wild Ramification,'' arXiv:0710.0631 [hep-th].}

\lref\Hitchin{N.~Hitchin,
``The Self-Duality Equations on a Riemann Surface,''
Proc. London Math. Soc. (3) {\bf 55} (1987) 59.}

\lref\Vogan{D.~Vogan, ``Singular unitary representations,''
{\it Non-commutative Harmonic Analysis and Lie groups,}
J.~Carmona and M.~Vergne, eds., Lecture Notes in Mathematics {\bf 880},
Springer-Verlag, Berlin-Heidelberg-New York (1981) 506.}

\lref\Vogansing{D.~Vogan,
``Three-dimensional subgroups and unitary representations,''
Challenges for the 21st century (Singapore, 2000),
World Sci. Publ., River Edge, NJ (2001) 213.}

\lref\BravermanJ{A.~Braverman, A.~Joseph,
``The Minimal Realization from Deformation Theory,''
J. Alg. {\bf 205} (1998) 13.}

\lref\toappear{S.~Gukov, E.~Witten, ``D-branes and Quantization,'' to appear.}

\lref\quantizationnote{{\TeX}note ``quantization.tex''}

\lref\rigidnote{{\TeX}note ``rigid.tex''}

\lref\KSaulina{A.~Kapustin, N.~Saulina, ``The Algebra of Wilson-'t
Hooft Operators,''  arXiv:0710.2097 [hep-th].}

\lref\Freedman{O.~DeWolfe, D.~Freedman, H.~Ooguri,
``Holography and Defect Conformal Field Theories,''
Phys. Rev. {\bf D66} (2002) 025009, hep-th/0111135.}

\lref\Wbaryons{E.~Witten, ``Baryons and Branes in Anti de Sitter
Apace,'' JHEP {\bf 9807} (1998) 006, arXiv:hep-th/9805112.}

\lref\BergmanGS{O.~Bergman, E.~G.~Gimon, and S.~Sugimoto,
``Orientifolds, RR Torsion, and K-theory,'' JHEP {\bf 0105} (2001)
047, arXiv:hep-th/0103183.}

\lref\LLMii{H.~Lin, J.~M.~Maldacena, ``Fivebranes from Gauge
Gheory,'' Phys.\ Rev.\  D {\bf 74} (2006) 084014, arXiv:hep-th/0509235.}

\lref\LLMi{H.~Lin, O.~Lunin, J.~M.~Maldacena, ``Bubbling AdS Space
and 1/2 BPS Geometries,'' JHEP {\bf 0410} (2004) 025,
arXiv:hep-th/0409174.}

\lref\GomisM{J.~Gomis, S.~Matsuura, ``Bubbling Surface Operators and
S-duality,'' JHEP {\bf 0706} (2007) 025 arXiv:0704.1657 [hep-th].}

\lref\Brylinski{R.~Brylinski, ``Geometric Quantization of Real
Minimal Nilpotent Orbits,''
 Diff. Geom. Appl. {\bf 9} (1998) 5.}

\lref\KSavin{D.~Kazhdan, G.~Savin, ``The smallest representation of simply laced groups,''
Festschrift in honor of I. I. Piatetski-Shapiro on the occasion of his sixtieth birthday,
Part I (Ramat Aviv, 1989),  209. Israel Math. Conf. Proc., 2, Weizmann, Jerusalem, 1990.}

\def\boxit#1{\vbox{\hrule\hbox{\vrule\kern8pt
\vbox{\hbox{\kern8pt}\hbox{\vbox{#1}}\hbox{\kern8pt}}
\kern8pt\vrule}\hrule}}
\def\mathboxit#1{\vbox{\hrule\hbox{\vrule\kern8pt\vbox{\kern8pt
\hbox{$\displaystyle #1$}\kern8pt}\kern8pt\vrule}\hrule}}


\let\includefigures=\iftrue
\newfam\black
\includefigures
\input epsf
\def\figin{\epsfcheck\figin}\def\figins{\epsfcheck\figins}
\def\epsfcheck{\ifx\epsfbox\UnDeFiNeD
\message{(NO epsf.tex, FIGURES WILL BE IGNORED)}
\gdef\figin##1{\vskip2in}\gdef\figins##1{\hskip.5in}
\else\message{(FIGURES WILL BE INCLUDED)}%
\gdef\figin##1{##1}\gdef\figins##1{##1}\fi}
\def\DefWarn#1{}
\def\figinsert{\goodbreak\midinsert}
\def\ifig#1#2#3{\DefWarn#1\xdef#1{fig.~\the\figno}
\writedef{#1\leftbracket fig.\noexpand~\the\figno}%
\figinsert\figin{\centerline{#3}}\medskip\centerline{\vbox{\baselineskip12pt
\advance\hsize by -1truein\noindent\footnotefont{\bf
Fig.~\the\figno:} #2}}
\bigskip\endinsert\global\advance\figno by1}
\else
\def\ifig#1#2#3{\xdef#1{fig.~\the\figno}
\writedef{#1\leftbracket fig.\noexpand~\the\figno}%
\global\advance\figno by1} \fi
\def\N{{\cal N}}

\newdimen\tableauside\tableauside=1.6ex
\newdimen\tableaurule\tableaurule=0.6pt
\newdimen\tableaustep
\def\phantomhrule#1{\hbox{\vbox to0pt{\hrule height\tableaurule width#1\vss}}}
\def\phantomvrule#1{\vbox{\hbox to0pt{\vrule width\tableaurule height#1\hss}}}
\def\sqr{\vbox{%
  \phantomhrule\tableaustep
  \hbox{\phantomvrule\tableaustep\kern\tableaustep\phantomvrule\tableaustep}%
  \hbox{\vbox{\phantomhrule\tableauside}\kern-\tableaurule}}}
\def\squares#1{\hbox{\count0=#1\noindent\loop\sqr
  \advance\count0 by-1 \ifnum\count0>0\repeat}}
\def\tableau#1{\vcenter{\offinterlineskip
  \tableaustep=\tableauside\advance\tableaustep by-\tableaurule
  \kern\normallineskip\hbox
    {\kern\normallineskip\vbox
      {\gettableau#1 0 }%
     \kern\normallineskip\kern\tableaurule}%
  \kern\normallineskip\kern\tableaurule}}
\def\gettableau#1 {\ifnum#1=0\let\next=\null\else
  \squares{#1}\let\next=\gettableau\fi\next}

\tableauside=1.6ex \tableaurule=0.6pt


\def\IB{\relax\hbox{$\inbar\kern-.3em{\rm B}$}}
\def\IC{\relax\hbox{$\inbar\kern-.3em{\rm C}$}}
\def\IQ{\relax\hbox{$\inbar\kern-.3em{\rm Q}$}}
\def\ID{\relax\hbox{$\inbar\kern-.3em{\rm D}$}}
\def\IE{\relax\hbox{$\inbar\kern-.3em{\rm E}$}}
\def\IF{\relax\hbox{$\inbar\kern-.3em{\rm F}$}}
\def\IG{\relax\hbox{$\inbar\kern-.3em{\rm G}$}}
\def\IGa{\relax\hbox{${\rm I}\kern-.18em\Gamma$}}
\def\IH{\relax{\rm I\kern-.18em H}}
\def\IK{\relax{\rm I\kern-.18em K}}
\def\IL{\relax{\rm I\kern-.18em L}}
\def\IP{\relax{\rm I\kern-.18em P}}
\def\IR{\relax{\rm I\kern-.18em R}}
\def\Z{\Bbb{Z}}

\def\II{\relax{\rm I\kern-.18em I}}

\def\S{{\bf S}}

\def\R{{\Bbb R}}
\def\C{{\Bbb C}}
\def\RP{{\Bbb {RP}}}

\def\CA {{\cal A}}
\def\CB {{\cal B}}

\def\CF {{\cal F}}

\def\CH {{\cal H}}
\def\CI {{\cal I}}

\def\CN {{\cal N}}

\def\CQ {{\cal Q}}
\def\CR {{\cal R}}

\def\CZ {{\cal Z}}


\def\p{\partial}

\def\tilde{\widetilde}
\def\hat{\widehat}
\def\bar{\overline}


\def\Tr{{\rm Tr}}

\def\p{\partial}

\def\inbar{\,\vrule height1.5ex width.4pt depth0pt}

\def\a{\alpha}
\def\b{\beta}

\def\g{\gamma}

\def\la{\lambda}
\def\th{\theta}

\def\bar{\overline}

\def\det{{\rm det}}

\def\Tr{{\rm Tr}}

\def\IH{{\bf H}}

\def\Fl{{\CF {\kern -1.2pt \ell} }}

\def\leadsto{\rightsquigarrow}

\def\example#1{\bgroup\narrower\footnotefont\baselineskip\footskip\bigbreak
\hrule\medskip\nobreak\noindent {\bf Example}. {\it
#1\/}\par\nobreak}
\def\endexample{\medskip\nobreak\hrule\bigbreak\egroup}

\def\btimes{~{{{\lower1pt\hbox{$\square$}} \kern-7.6pt \times}}~}
\def\TT{{\Bbb{T}}}
\def\LL{{\Bbb{L}}}
\def\Weyl{{\cal W}}

\def\C{{\Bbb{C}}}
\def\R{{\Bbb{R}}}

\def\sp{{\rm sp}}


\def\dual#1{{^L\negthinspace #1}}

\def\LG{\dual{G}}
\def\Gad{G_{{\rm ad}}}
\def\Gsc{G_{{\rm sc}}}


\def\G{G}


\Title{\vbox{\baselineskip12pt
}}
{\vbox{ \centerline{Rigid Surface Operators} }}
\centerline{Sergei Gukov$^{a,b}$\footnote{$^{\dagger}$}{On leave from
California Institute of Technology.} and Edward Witten$^{c}$}
\medskip
\medskip
\centerline{$^a$ \it{Department of Physics, University of California}}
\centerline{\it{Santa Barbara, CA 93106}}
\medskip
\centerline{$^b$ \it{School of Mathematics, Institute for Advanced Study}}
\centerline{\it{Princeton, New Jersey 08540}}
\medskip
\centerline{$^c$ \it{School of Natural Sciences, Institute for Advanced Study}}
\centerline{\it{Princeton, New Jersey 08540}}
\noindent
\vskip 20pt {\bf \centerline{Abstract}}
\noindent
Surface operators in gauge theory are analogous to Wilson
and 't Hooft line operators except that they are supported on
a two-dimensional surface rather than a one-dimensional curve.
In a previous paper, we constructed a certain class of half-BPS
surface operators in ${\cal N}=4$ super Yang-Mills theory,
and determined how they transform under $S$-duality.
Those surface operators depend on a relatively large
number of freely adjustable parameters.
In the present paper, we consider the opposite case of half-BPS surface
operators that are ``rigid'' in the sense that they do not depend on
any parameters at all. We present some simple constructions of rigid
half-BPS surface operators and attempt to determine how they
transform under duality. This attempt is only partially successful,
suggesting that our constructions are not the whole story.
The partial match suggests interesting connections with quantization.
We discuss some possible refinements and some string theory
constructions which might lead to a more complete picture.

\medskip
\Date{April 2008}

\listtoc\writetoc


\newsec{Introduction}\seclab\intro

The familiar examples of non-local operators in four-dimensional gauge theory
include line operators, such as Wilson and 't Hooft operators,
supported on a one-dimensional curve $L$ in the space-time manifold $M$.
While a Wilson operator labeled by a representation $R$ of the gauge group $G$
can be defined by modifying the measure in the path integral,
namely by inserting a factor
\eqn\wilson{W_R (L) = \Tr_R ~{\rm Hol}_{L} (A) = \Tr_R \left( P \!
\exp \oint_L A \right), }
an 't Hooft operator is defined by modifying the space of fields
over which one performs the path integral.

Similarly, a surface operator in four-dimensional gauge theory is an
operator supported on a two-dimensional submanifold $D \subset M$ in
the space-time manifold $M$. Although in this paper we mainly take
$M=\R^4$ and $D=\R^2$, the constructions are local and one might
consider more general space-time four-manifolds $M$ and embedded
surfaces $D$.   In general, surface operators do not admit a simple
``electric'' description analogous to the definition of Wilson
lines, and should be defined, like 't Hooft operators, by modifying
the domain of integration in the path integral, that is by requiring
the gauge field $A$ (and, possibly, other fields) to have prescribed
singularities along $D$.

Four-dimensional gauge theories admit surface operators, and in the
supersymmetric case, they often admit supersymmetric surface
operators, that is, surface operators that preserve some of the
supersymmetry.  In this paper, we consider   $\CN=4$ super
Yang-Mills theory in four dimensions, the maximally supersymmetric
case.  This theory has many remarkable properties, including
electric-magnetic duality, and has been extensively studied in the
context of string dualities, in particular in the AdS/CFT
correspondence \Maldacena. It also has a rich spectrum of non-local
operators, including supersymmetric Wilson and 't Hooft operators
which play an important role in many applications, as well as
supersymmetric surface operators and domain walls.

A half-supersymmetric or half-BPS Wilson operator is determined by
discrete data, namely the choice of a representation of the gauge
group $G$. Similarly, a half-BPS 't Hooft operator is determined by
discrete data. In contrast, the half-BPS surface operators that we
constructed in previous work \Ramified\ depend on freely adjustable
parameters, typically quite a few of them. Much of their interest
actually comes from the dependence on these parameters.

As will become clear, the problem of describing all half-BPS surface
operators in ${\cal N}=4$ super Yang-Mills theory is rather
involved. In this paper, we will consider  the opposite case from
what was considered in \Ramified: surface operators that depend on
no continuously variable parameters at all. We call these {\it
rigid} surface operators.

It is purely for simplicity that we consider only maximally
supersymmetric or half-BPS surface operators.  In the case of line
operators, in addition to the half-BPS Wilson and 't Hooft
operators, there are many more ${1 \over 4}$-BPS line operators;
their analysis is very interesting but is much more complex than the
half-BPS case, as shown in \KSaulina. Surface operators with reduced
supersymmetry are probably also interesting, but harder to study.

In addition to being rigid, the surface operators that we consider
here are in a certain sense minimal or irreducible.  They do not
have any extra fields supported on the surface.  This notion is
clarified in section \minimal; in the meanwhile, we simply remark that
our surface operators are related to individual orbits of the gauge
group $G$ (or rather its complexification), and this leads to
minimality.

Finally, rigid surface operators are probably automatically
conformally invariant.  They must be scale-invariant, or a scale
transformation would introduce a free parameter.  In local quantum
field theory, scale invariance usually implies conformally
invariance.  Our constructions will be manifestly conformally
invariant at the classical level.  Quantum conformal invariance can
probably be argued along the lines of \Freedman, and is manifest for
some of our surface operators in the string theory construction of
section \holog.  ${\cal N}=4$ super Yang-Mills theory also has
(non-rigid) half-BPS surface operators that are not conformally
invariant \Wild.

\def\neg{\negthinspace}

\medskip\noindent{{\it Organization Of The Paper}}\medskip

In section \rigidsurf, after a brief review of the surface operators
considered in \Ramified, we describe two constructions of rigid
surface operators. Some further refinements leading to additional
rigid surface operators are described in section \additional.

$S$-duality must transform rigid surface operators of ${\cal N}=4$
super Yang-Mills theory with gauge group $G$ to similar operators in
the same theory with the dual gauge group $^L\neg G$. Aiming to
understand this, we describe in section \invariants\ some properties
of surface operators that are computable and should  be invariant under
electric-magnetic duality or should transform in a known way.

In section \dualfor, we attempt to use this information to
determine, in examples, how our surface operators transform under
duality.  In doing this, we concentrate on orthogonal and
symplectic gauge groups of small rank. A simple example involving
unitary groups is also discussed in section \rigdual. We omit
exceptional groups, which are more complicated. It is especially
interesting to consider the dual pairs of groups $G=SO(2n+1)$ and
$\LG=Sp(2n)$, whose Lie algebras are not isomorphic. In carrying
out this analysis, we do find some interesting examples of what
appear to be dual pairs of surface operators, but we are not able
to get a complete duality conjecture. It is quite likely that our
constructions of rigid surface operators are in need of some
further refinement. There may be a relation to the construction in
.

The remainder of the paper is devoted to some attempts at a more
systematic understanding. In section \families, we try to be more
systematic, at least for certain families of rigid surface
operators, in orthogonal and symplectic gauge groups of any rank.
In section \specialsurf, we argue that the mathematical theory of
special unipotent conjugacy classes \refs{\Lusztigiv,\Lusztigii}
provides the right framework for a duality conjecture for a
certain family of surface operators. In section \ssfamilies, we
make analogous proposals for other families of surface operators.
This discussion is somewhat similar to a relation between
conjugacy classes defined in \Lusztigiii, section 13.3.
 In
section \quant, we make a general conjecture about how the conjugacy
class associated with a rigid surface operator transforms under
duality.  Finally, in section \holog, we describe string theory
constructions of some of the rigid surface operators of section 2.

The paper contains two appendices.
In Appendix A, we describe rigid nilpotent orbits for exceptional groups
which, together with the material of section \rigidsurf,
can be used to study rigid surface operators in super Yang-Mills theories
with exceptional gauge groups.
In Appendix B, we review the root systems and matrix realizations
of the Lie algebras $\frak{so} (2N+1)$ and $\frak{sp} (2N)$.
In particular, we identify the invariant polynomials of the Higgs
field in dual theories with gauge groups $G=SO(2n+1)$ and $\LG=Sp(2n)$
which play an important role in identifying dual pairs of rigid surface operators.


\newsec{Rigid Surface Operators}\seclab\rigidsurf

\subsec{Review}\subseclab\review

To keep this paper self-contained, we begin with a brief review of
the surface operators  constructed in \Ramified. We consider $\N=4$
super Yang-Mills theory on $\R^4$, with coordinates
$x^0,x^1,x^2,x^3$. The support $D$ of the surface operator will be a
copy of $\R^2$ at $x^2=x^3=0$. The supersymmetry preserved by the
surface operator is $(4,4)$ supersymmetry in the two-dimensional
sense.  We recall that the vector multiplet of $(4,4)$ supersymmetry
in two dimensions consists of a gauge field and four scalars in the
adjoint representation (plus fermions). Accordingly, components
$A_0,A_1$ of the four-dimensional gauge field plus four of the six
scalars of ${\cal N}=4$ super Yang-Mills theory transform in a
vector multiplet of two-dimensional $(4,4)$ supersymmetry.  The
``normal'' components $A_2$ and $A_3$ of the gauge field transform
in a hypermultiplet of the unbroken supersymmetry, along with two of
the scalars.  It is convenient to denote those two scalars as
$\phi_2$ and $\phi_3$.

Surface operators were defined in \Ramified\ by postulating a
suitable singular behavior of the hypermultiplets, that is the
fields $A_2,A_3,\phi_2,\phi_3$, at $x^2=x^3=0$.  Of course, the
singularity must be chosen to be compatible with supersymmetry. The
condition for supersymmetry is that $A=A_2 dx^2+A_3 dx^3$ and
$\phi=\phi_2 dx^2+\phi_3dx^3$ must obey certain equations that are
known as Hitchin's equations \Hitchin.  Hitchin's equations are
equations in the $x^2-x^3$ plane that can be written as follows:
\eqn\hitchineqs{\eqalign{ & F_A - \phi \wedge \phi = 0 \cr & d_A
\phi = 0,\quad d_A \star \phi = 0. }}
Originally, these equations were obtained in \Hitchin\ as the
dimensional reduction of the self-dual Yang-Mills equations from
four to two dimensions; $\phi$ simply arises as the components of
the gauge field in the two hidden dimensions.  (This approach is
natural if one considers $\N=4$ super Yang-Mills theory to arise by
dimensional reduction from ten dimensions.)  This interpretation of
Hitchin's equations makes it clear they are associated with unbroken
supersymmetry.

To define a supersymmetric surface operator, one picks a solution of
Hitchin's equations with a singularity along $D$, and one requires
that quantization of ${\cal N}=4$ super Yang-Mills theory should be
carried out for fields with precisely this kind of singularity.  For
the surface operator to be superconformal, the singularity must be
scale-invariant.  In addition, it is natural to look for surface
operators that are invariant under rotations of the $x^2-x^3$ plane.
If we set $x^2+ix^3=re^{i\theta}$, then the most general possible
rotation-invariant ansatz is
\eqn\abcabsatz{\eqalign{ A & = a(r) d \th + f(r) {dr \over r} \cr
\phi & = b(r) {dr \over r} - c (r) d \th. }}
Setting $f(r)=0$ by a gauge transformation and introducing a new
variable $s = - \ln r$, we can write the supersymmetry equations
\hitchineqs\ in the form of Nahm's equations:
\eqn\nahmeqs{\eqalign{ & {da \over ds} = [b,c] \cr & {db \over ds} =
[c,a] \cr & {dc \over ds} = [a,b]. }}
A conformally invariant solution is invariant under scalings of $r$
and therefore is independent of $s$.  (As we discuss later,
solutions that are not quite conformally invariant can also be used
to construct conformally invariant surface operators.) So the most
general conformally invariant solution is obtained by setting
$a,b,c$ to constant elements $\alpha,\beta,\gamma$ of the Lie
algebra $\frak{g}$ of $G$. The equations imply that $\alpha$,
$\beta,$ and $\gamma$ must commute, so we can conjugate them to the
Lie algebra $\frak t$ of a maximal torus $\Bbb{T}$ of $G$.  The
resulting singular solution of Hitchin's equations then takes the
simple form \eqn\norto{\eqalign{A&=\alpha\,d\theta\cr \phi&=\beta
{dr\over r}-\gamma\,d\theta.}} Hitchin's equations with a
singularity of this form were first studied mathematically in
\Simpson.

Roughly speaking, surface operators were defined in \Ramified\ by
requiring that the fields have a singularity of this kind, with
specified values\foot{If instead of specifying the values of
$\alpha,\beta,$ and $\gamma$, we treat them as dynamical fields, we
get a non-minimal surface operator, in the sense of section \minimal.}
of $\alpha,\beta,$ and $\gamma$.  More exactly, to study ${\cal
N}=4$ super Yang-Mills theory in the presence of the surface
operator, one performs the path integral (or one quantizes) in a
space of fields that take the form given in eqn. \norto\
{\it modulo terms that are less singular than $1/r$.}

There are two important caveats. First, it turns out that one can
add an additional parameter $\eta$, also $\frak t$-valued. $\eta$ is
a sort of two-dimensional theta angle and plays an important role
because it transforms into $\alpha$ under duality. (For rigid
surface operators, $\eta$ at most has only a discrete analog.)
Second, to quantize in the presence of the singularity described in
\norto, one should divide only by gauge transformations that, along
the locus $D$ of the singularity, take values in the subgroup of $G$
that commutes with $\alpha,\beta,$ and $\gamma$ (and $\eta$).
Generically, this subgroup is the maximal torus $\Bbb{T}$. But in
general, it may be any subgroup $\Bbb{L}$ of $G$ that contains
$\Bbb{T}$.  Such a subgroup is called a Levi subgroup.  In studying
a surface operator of this type, we regard the choice of $\Bbb{L}$
as part of the definition. Having chosen $\Bbb{L}$, we pick
$\alpha,\beta,\gamma$, and $\eta$ to be an $\Bbb{L}$-regular
quadruple, meaning that the subgroup of $G$ that commutes with all
four of them is precisely $\Bbb{L}$.  Then, to calculate Yang-Mills
observables in the presence of the surface operator, we perform a
path integral over fields with the indicated type of singularity,
dividing by gauge transformations that along $D$ are
$\Bbb{L}$-valued.  This gives a surface operator that varies
smoothly with $\alpha,\beta,\gamma,\eta$ as long as those parameters
form an $\Bbb{L}$-regular quadruple.  But when the parameters are
varied so that the unbroken group becomes a larger group
${{\Bbb{L}}}'$, a singularity emerges.  In a sense, the residue of
this singularity is a surface operator that can be constructed in
the same way, but starting with $\Bbb{L}'$ rather than $\Bbb{L}$.
One of the main ideas in \Ramified\ was to study the monodromies in
the space of $\Bbb{L}$-regular parameters.

\subsec{Limit For $\alpha,\beta,\gamma\to 0$}\subseclab\limit

As a preliminary to discussing rigid surface operators, we will
consider what happens to the above construction in the limit that
$\alpha,\beta,\gamma\to 0$.  To keep things simple, we begin with
the case $G=SU(2)$.  For more detail on the following, see
\Ramified, section 3.3.

The naive idea is that the singularity of $A$ and $\phi$ is linear
in $\alpha,\beta$, and $\gamma$, so that if we set
$\alpha,\beta,\gamma$ to zero, there is no singularity and no
surface operator.   However, as we have already noted, the
definition of the surface operator is that $A$ and $\phi$ have
singularities proportional to $\alpha,\beta,\gamma$ modulo terms
that are less singular than $1/r$. Generically, for
$\alpha,\beta,\gamma\to 0$, we should not conclude that $A$ and
$\phi$ are nonsingular, but only that they are less singular than
$1/r$.  In fact, Hitchin's equations do have a rotationally
symmetric solution that is singular at $r=0$ but less singular than
$1/r$. The Nahm equations \nahmeqs\ are solved with
\eqn\nahmsol{ a = - {t_1 \over s+1/ f} \quad,\quad b = - {t_2 \over
s+1/f} \quad,\quad c = - {t_3 \over s+1/f }}
where $t_1$, $t_2$, and $t_3$ are elements of the Lie algebra
$\frak g$, which satisfy the usual $\frak{su} (2)$ commutation relations,
$[t_1,t_2] = t_3$, {\it etc}.  Moreover, $f$ is an arbitrary
non-negative constant. Since we are taking $G=SU(2)$, the matrices
$t_i$, if nonzero, correspond to the two-dimensional representation
of $SU(2)$.

Because of the factor of $-1/s=1/\ln r$,
this solution is less singular at $r=0$ than the solutions
considered before in which $a,b,c$ are set to commuting constants
$\alpha,\beta,\gamma$.  A surface operator with nonzero
$\alpha,\beta,\gamma$ converges for $\alpha,\beta,\gamma\to 0$ to
one that is characterized by the statement that the singularity at
$r=0$ looks like the solution of eqn. \nahmsol, for some $f$.
(We also allow the limiting case $f=0$, in which there is no
singularity.) Any choice of $f$ would spoil conformal invariance.
But it is not natural to make a choice of $f$, because the
derivative of $A$ and $\phi$ with respect to $f$ is
square-integrable. So the surface operator that we get from the
ansatz \nahmsol, with $f$ allowed to fluctuate, is actually
conformally invariant.

A convenient way to describe this surface operator is to say that
the fields behave near $r=0$ as
\eqn\nahmsolut{\eqalign{
A&={t_1\,d\theta\over \ln r}+\dots \cr
\phi&={t_2\,dr\over r\ln r}-{t_3\,d\theta\over \ln r}+\dots,}}
where the ellipses refer to terms that are less singular (at most of
order $1/r\,\ln^2 r$)  at $r=0$.

Concretely, a generic field with the singularity determined by
$\alpha,\beta,\gamma$ has (in a basis in which $\alpha,\beta,\gamma$
are diagonal) off-diagonal terms that are singular, but less
singular than $1/r$.  For $\alpha,\beta,\gamma\to 0$, a sequence of
such solutions can converge to the one given in eqn. \nahmsol.  Such
a sequence can also converge to a non-singular solution
(corresponding to $f=0$), but that is non-generic.

\bigskip\noindent{\it The Monodromy}

\def\CA{{\cal A}}
\def\CF{{\cal F}}
The following considerations give a useful picture of what is
happening.  The complex-valued flat connection $\CA=A+i\phi$ is
invariant under part of the supersymmetry preserved by the surface
operator. Hence the conjugacy class of the monodromy
\eqn\dolf{U=P\exp\left(-\int_{\ell} \CA\right)}
is a supersymmetric observable. Here $\ell$ is a contour surrounding
the singularity. Hitchin's equations imply that the curvature
of $\CA,$ namely $\CF=d\CA+\CA\wedge \CA$, is equal to zero.
So if Hitchin's equations are obeyed, then the conjugacy class of $U$
is invariant under deformations of $\ell$.
Of course, $U$ is an element of $G_\C$, the complexification of $G$.

In general, in quantum theory, the fields fluctuate and Hitchin's
equations are only obeyed near the singularity (where they are
imposed as a boundary condition). However, the conjugacy class of
$U$ is independent of $\ell$ as an observable in a suitable chiral
algebra, defined using some of the supersymmetries, since $\CF$
vanishes in that chiral algebra. Alternatively, one can simply
define the conjugacy class of $U$ for the limiting case that $\ell$
is a small loop surrounding the singularity.

So let us compute the conjugacy class of $U$ for the surface
operators that were described above.  For a generic surface operator
with parameters $\alpha,\beta,\gamma$, we set $\xi=\alpha-i\gamma$.
Then $\CA=\xi d\theta$, and the monodromy is hence
\eqn\doof{U=\exp(-2\pi\xi).}
This is independent of the choice of $\ell$.

On the other hand, for the solution \nahmsol, we find
$\CA=-d\theta (t_1-it_3)/(s+1/f)$.
If we take $\ell$ to be the circle $s=s_1$,
the monodromy comes out to be
\eqn\oof{U'=\exp(-2\pi (t_1-it_3)/(s_1+1/f)).}
At first sight, it is not obvious that the
conjugacy class of $U'$ is independent of $s_1$, as it should be.
What saves the day is that $t_1-it_3$ is nilpotent, because of the
commutation relation
\eqn\zelgo{ [it_2,t_1-it_3]=t_1-it_3.}
In a form of the two-dimensional representation of $SU(2)$,
with $t_2$ being diagonal, $t_1-it_3$ is lower triangular.
Thus $U'$ takes the form
\eqn\nordo{U'=\left(\matrix{1& 0 \cr w & 1\cr}\right),}
for some $w$.

The conjugacy class of $U'$ is independent of $w$, as long as $w$ is
nonzero, because $w$ can be changed by conjugating $U'$ by a
diagonal matrix.  Now let us reconsider the monodromy \doof\ of the
surface operator with $\alpha,\gamma\not=0$.  If $\xi\not=0$, then
$\xi$ can be diagonalized with eigenvalues $\pm \xi_0$.  $U$ can
also be diagonalized, with eigenvalues $\exp(\pm 2\pi\xi_0)$:
\eqn\ordo{U=\left(\matrix{\exp(-2\pi \xi_0) & 0 \cr
0 & \exp(2\pi \xi_0)\cr}\right).}
As long as $\xi_0\not=0$, this matrix is conjugate to
\eqn\ordo{U_w=\left(\matrix{\exp(-2\pi \xi_0) & 0 \cr
w& \exp(2\pi \xi_0)\cr}\right),}
so it does not matter if $w$ is zero or not.
In fact, $U$ can be transformed to $U_w$ by conjugation
by a lower triangular matrix
\eqn\rdo{\left(\matrix{1 & 0 \cr * & 1\cr}\right).}
But if $\xi_0=0$, then of course, the conjugacy class of $U_w$ does
depend on whether $w$ vanishes or not.

Let $ \frak C_\xi$ be the conjugacy class in $SL(2,\C)$ that
contains the element $U=\exp(-2\pi \xi)$, with generic $\xi$.
Then $\frak C_\xi$ is of complex dimension two.  Indeed, $U$ commutes
only with a one-parameter subgroup of diagonal matrices, so its
orbit in the three-dimensional group $SL(2,\C)$ is two-dimensional.
Similarly, the lower triangular matrix $U'$ commutes only with the
one-parameter group of lower-triangular matrices, so it lies in a
two-dimensional conjugacy class $\frak C'$.  The limit of the
conjugacy class $\frak C_\xi$ for $\xi\to 0$ is $\frak C'$
(or more precisely its closure, as we note in a moment).
It is not the conjugacy class $\frak C_0$ of the identity element of $SL(2,\C)$,
as we would expect if we naively set $\xi_0=0$ in the expression
\ordo\ for $U$.

In fact, the conjugacy class $\frak C_\xi$ can be defined by the
equation \eqn\dor{\Tr\,U=\exp(-2\pi\xi_0)+\exp(2\pi \xi_0).} The
limit of this equation for $\xi_0=0$ is \eqn\zor{\Tr\,U=2,} which
is obeyed by $U'$.  In fact, the equation $\Tr\,U=2$ defines a union
of two conjugacy classes: one conjugacy class $\frak C'$ that
contains  $U'$, and a second class $\frak C_0$ that consists of a
single element, the identity element of $SL(2,\C)$.

This gives us a new perspective on why the surface operator defined
by generic values of $\alpha,\beta,\gamma$ can have for a limit the
surface operator associated with the solution \nahmsol\ of Nahm's
equations.  The former surface operator is associated with monodromy
in the class $\frak C_\xi$.  The latter one is associated with
monodromy that is generically in the class $\frak C'$, but can also
be in the class $\frak C_0$, corresponding to trivial monodromy, in
the special case $f=0$.  The limit of $\frak C_\xi$ for $\xi\to 0$
is the union of $\frak C'$ and $\frak C_0$.  This is why the limit
of the generic surface operator can be the one associated with
Nahm's equations.

The conjugacy class $\frak C'$ is not closed in $SL(2,\C)$, because
the matrix $U'$ of eqn. \nordo\ jumps from being in the class $\frak
C'$ to the class $\frak C_0$ when $w$ becomes 0.  The closure of
$\frak C'$ therefore includes the point $\frak C_0$.  When we say
that the monodromy associated with a given surface operator is in
the conjugacy class $\frak C'$, we will always mean that it is
generically in that conjugacy class and in general is in the closure
of the stated conjugacy class.

An element of a complex Lie group -- $SL(2,\C)$ in our example --
is called semisimple if it can be diagonalized (or conjugated to
a maximal torus).  As in our example, the conjugacy class of a
semisimple element is always closed. We call this a semisimple
conjugacy class. By contrast, an element $U$ is called unipotent
if, in any finite-dimensional representation, it takes the form
$U=\exp(n)$, where $n$ is nilpotent. In our above example, $U'$ is
unipotent. The conjugacy class of a unipotent element is called a
unipotent conjugacy class.  As in our above example, a unipotent
class of positive dimension is never closed; its closure always
contains the class $\frak C_0$ of the identity element of $G_\C$.
In general, for a group of higher rank, the closure of a unipotent
conjugacy class is a union of many (but only finitely many)
conjugacy classes.

If a surface operator is associated with a semisimple or unipotent
conjugacy class, we call it a semisimple or unipotent surface
operator.


\bigskip\noindent{\it Counting Dimensions}

In our above example, the conjugacy class $\frak C_0$ consists
of a single point, while $\frak C'$ has complex dimension 2
or real dimension 4.
Let us understand this from the point of view of Hitchin's equations.
To get trivial monodromy, we must set $f=0$ in \nahmsol.
This involves adjusting one real parameter. In addition,
at $f=0$, the solution reduces to $A=\phi=0$, which is invariant
under global $SU(2)$ gauge rotations. In fixing the gauge
invariance, one is then free to make global $SU(2)$ gauge rotations
on the other fields, away from the support of the surface operator.
As the real dimension of $SU(2)$ is 3, the real codimension of the
locus (in a family of solutions of Hitchin's equations,
or a family of fields in the path integral) at which the monodromy
is trivial rather than being conjugate to $U'$ is $1+3=4$.

\def\M{{\cal M}}
Now suppose that we compactify ${\cal N}=4$ super Yang-Mills theory
from four dimensions to two dimensions on a Riemann surface $C$, the
four-manifold being then $\R^2\times C$. It is possible to make a
topological twist so that supersymmetry is preserved; Hitchin's
equations for the pair $(A,\phi)$ are the condition for unbroken
supersymmetry \BJSV.  Let $\M_H$ be the moduli space of solutions of
Hitchin's equations.  It is a hyper-Kahler manifold.  In one complex
structure, it parametrizes, up to conjugation, homomorphisms from
the fundamental group of $C$ to $G_\C$, the complexification of $G$.
Concretely, if $C$ has genus $g$, and $V_i$, $W_j$, $i,j=1,\dots,g$
are the monodromies around a complete set of $A$-cycles and
$B$-cycles, then such a flat connection corresponds to a solution of
the equation \eqn\gofer{V_1W_1V_1{}^{-1}W_1{}^{-1}\cdots
V_gW_gV_g{}^{-1}W_g{}^{-1}=1,} modulo conjugation by an element of
$G$.  The complex dimension of the solution space is thus
$2(g-1){\rm dim}\,G$.  (The coefficient of ${\rm dim}\, G$ is
obtained by counting the $2g$ group elements $V_i$ and $W_j$, and
subtracting 1 for the equation and 1 for dividing by conjugation.)

Now include a surface operator, supported on $D=\R^2\times p$ for
$p$ a point in $C$.  We suppose that the surface operator is
associated with a conjugacy class $\frak C$, which in our above
examples is $\frak C_\xi$ or $\frak C'$. Let $n$ be the complex
dimension of $\frak C$.  The equation for the monodromies becomes
\eqn\zofer{V_1W_1V_1{}^{-1}W_1{}^{-1}\cdots
V_gW_gV_g{}^{-1}W_g{}^{-1}=U,} where $U$ may be any element of the
class $\frak C$ (or in general of its closure).  Since $U$ takes
values in an $n$-dimensional space, the dimension of the moduli
space becomes $2(g-1){\rm dim}\,G+n$.

For instance, if $\frak C'$ is the unipotent conjugacy class
described above, then $n=2$ and including the surface operator
increases the complex dimension of the moduli space by 2.


\bigskip\noindent{\it More General Conjugacy Classes}

For $G=SU(2)$, the unipotent surface operator that we have described
above is not essentially new, in the sense that it is the limit of a
semisimple surface operator with parameters $\alpha,\beta,\gamma$ as
the parameters go to zero. However, the same construction can be
applied for other groups $G$ and in general does give essentially
new surface operators. In fact, the construction that we have
explained above can be directly adapted to give a surface operator
for any unipotent conjugacy class $\frak C\subset G_\C$.

Unipotent elements $U$ of $G_\C$ correspond naturally to nilpotent
elements $n$ of the Lie algebra $\frak g_\C$ of $G_\C$, via
$U=\exp(n)$. It is convenient to think in terms of the Lie algebra.
A natural source of nilpotent elements of $G_\C$ comes by picking an
embedding of Lie algebras $\rho:\frak{sl}(2,\C)\to \frak g_\C$.
Then the raising (or lowering) operator for this embedding gives us
a nilpotent element $n\in \frak g_\C$.

Conversely, the Jacobson-Morozov theorem states that every nilpotent
element $n\in \frak g_\C$ is the raising operator for some
$\frak{sl}(2,\C)$ embedding.  In fact, up to conjugacy, every
nilpotent element is the raising operator of some unitary embedding
\eqn\jmhom{\rho:\frak{su}(2)\to \frak g}
of the real Lie algebra of $SU(2)$ to that of the compact form of $G$.
We pause to explain this theorem for $G=SU(N)$.
(A similar verification can be made for the other
classical groups $SO(N)$ and $Sp(2N)$.)
Every nilpotent element $n$ of $\frak{sl}(N,\C)$
can be put in Jordan canonical form.
In this form, $n$ is block diagonal with off-diagonal
blocks vanishing, as shown here
\eqn\blocdi{\left(\matrix{* &
* &
*&0&0&0\cr
*&*&*&0&0&0\cr
*&*&*&0&0&0\cr 0&0&0&*&*&0\cr 0&0&0&*&*&0\cr 0&0&0&0&0&*\cr}\right).}
In this examples, the blocks have sizes $\la_1=3,$ $\la_2=2,$
$\la_3=1$. Moreover, in Jordan canonical form, each diagonal block
is a ``principal nilpotent element'' with 1's just above the main
diagonal and all other matrix elements vanishing:
 \eqn\fredo{n=\left(\matrix{0&1&
0&0&\cdots & 0\cr
 0 & 0 & 1 & 0 &\cdots & 0 \cr  &&&\ddots &&\cr
  0& 0&0&0&\cdots &1\cr
  0&0 &0 & 0 & \cdots &
  0\cr}\right).}
In general, the sizes of the blocks are $\la_1,\la_2,\dots,\la_k$,
where $\la_1+\la_2+\dots+\la_k=N$, and we may as well assume
$\la_1\geq \la_2\geq \la_3\geq \dots\geq \la_k$.
On the other hand, up to isomorphism, there
is one irreducible representation of $SU(2)$ for each positive
integer dimension.  If we choose the $SU(2)$ embedding that
corresponds to the decomposition $N=\la_1+\la_2+\dots+\la_k$, then the
raising operator is conjugate to a matrix in Jordan canonical form
with blocks of the indicated size.

An important special case is the case that $\rho:\frak{su}(2)\to
\frak{su}(N)$ is an irreducible representation.  Then its raising
operator is simply an $N\times N$ matrix of the form in \fredo, up
to conjugacy. Such an element is called a principal nilpotent
element of $\frak{su}(N)$.

Now it is clear how to make a surface operator associated with any
unipotent conjugacy class  $\frak C\subset G_\C$.  We pick an
$SU(2)$ embedding $\rho:{\frak su}(2)\to \frak g$, and define the
surface operator using eqn. \nahmsol, where $t_1,t_2,$ and $t_3$ are
now the images of the standard $SU(2)$ generators under the chosen
embedding.

The classification of $\frak{su}(2)$ embeddings in $\frak{su}(N)$
has a close analog for orthogonal and symplectic groups.  We need
only to know a few facts. Irreducible representations of
$\frak{su}(2)$ are real or pseudoreal  according to whether their
dimension is odd or even. (A real representation admits an invariant
symmetric bilinear form, and a pseudoreal one admits an invariant
antisymmetric bilinear form.) In addition, if $R$ is a real or
pseudoreal representation (it admits an invariant quadratic form
that is either symmetric or antisymmetric), then the direct sum
$R\oplus R$ can be endowed with an invariant quadratic form that is
either symmetric or antisymmetric, as one prefers.

A homomorphism $\rho:\frak{su}(2)\to \frak{so}(N)$ is the same as an
$N$-dimensional real representation of $\frak{su}(2)$, or in other
words an $N$-dimensional representation that admits an invariant
symmetric form.  If $\rho$ is given by a decomposition
$N=\la_1+\la_2+\dots+\la_k$, then the condition, in view of the facts
cited in the last paragraph, is that the $\la_i$ each either are odd
or occur with even multiplicity.

A homomorphism $\rho:\frak{su}(2)\to\frak{sp}(2N)$ is the same as a
$2N$-dimensional pseudoreal representation of $\frak{su}(2)$. If
$\rho$ is given by a decomposition $N=\la_1+\la_2+\dots+\la_k$, then
the condition is that the $\la_i$ either are even or occur with even
multiplicity.

A decomposition $N=\la_1+\la_2+\dots+\la_k$ is called a partition of $N$,
and the $\la_i$ are called parts.
To summarize the above, for $G$ of  type $A$, $B$, $C$, or $D$, we
have the following classification of nilpotent orbits in terms of
partitions (see {\it e.g.} \CMcGovern, section 5):

\medskip

\item{$(A_{N}):$} partitions of $N+1$, $\sum \la_i = N+1$;

\item{$(B_{N}):$} partitions of $2N+1$, $\sum \la_i = 2N+1$, with
a constraint that the multiplicity of every even part $\la_i$ is
even;

\item{$(C_{N}):$} partitions of $2N$, $\sum \la_i = 2N$, with a
constraint that the multiplicity of every odd part $\la_i$ is even;

\item{$(D_{N}):$} partitions of $2N$, $\sum \la_i = 2N$, with a
constraint that the multiplicity of every even part $\la_i$ is even.
(Moreover, though this will not be important in the present paper,
partitions with all $\la_i$ even correspond to {\it two} nilpotent
orbits.)

\medskip
\noindent In what follows, we denote the nilpotent orbit associated
with a partition $\la$ by $\frak c_{\la}$, and the corresponding
unipotent conjugacy class by $\frak C_{\la}$.


\subsec{Searching For Rigid Surface Operators}\subseclab\searching

For any $G$ and any $\rho:\frak{su}(2)\to \frak g$, the above
construction gives a surface operator.  But generically it is not
rigid. For example, $G_\C=SL(N,\C)$ has no rigid conjugacy classes
at all, except the central elements.  Surface operators associated
with central classes have been considered in \Ramified\ and will be
described in section \centertopology. They are rigid, but they are
not good illustrations of the ideas of the present paper as they are
too special.  Let us explain why $SL(N,\C)$ has no other rigid
conjugacy classes.

We consider first the semisimple case.  Consider a semisimple
element of $SL(N,\C)$, say $U={\rm diag}(u_1,u_2,\dots,u_N)$,
with $u_i\in \C^*$.
Now let us try to vary the $u_i$ in such a way that the
conjugacy class $\frak C_U$ containing $U$ varies smoothly.
In doing this, we must preserve the condition
\eqn\zongo{u_1 u_2\cdots u_N=1,}
so as to remain in $SL(N,\C)$.
Also, regardless of whether $u_i=u_j$
or $u_i\not=u_j$ for some $i,j$, when we vary the
$u_i$, we want to preserve these conditions,
so that the subgroup of $SL(N,\C)$ that commutes with $U$ does not jump.
As long as $U$ is not central, so that the $u_i$ are not all equal,
these conditions allow us to vary at least one parameter.
So semisimple conjugacy classes in $SU(N)$ are never rigid.

Now let us consider unipotent conjugacy classes.  The basic case
in a sense is the principal unipotent conjugacy class.  This is
the class of an element  $U=\exp(n)$ (or equally well $U=1+n$),
where $n$ is a principal nilpotent element of the Lie algebra,  of
the form in \fredo. For $G_\C=SL(2,\C)$, we have seen in detail
that this conjugacy class is the limit of a semisimple conjugacy
class $\Tr\,U=\exp(-2\pi\xi_0)+\exp(2\pi\xi_0)$ for $\xi_0\to 0$.
So this conjugacy class is not rigid.  Similarly, for any $N$,
a principal nilpotent element \fredo\ can be deformed to the
following family:
 \eqn\bredo{\tilde n=\left(\matrix{0&1&
0&0&\cdots & 0\cr
 0 & 0 & 1 & 0 &\cdots & 0 \cr  &&&\ddots &&\cr
  0& 0&0&0&\cdots &1\cr
  a_N&a_{N-1} &a_{N-2} & a_{N-3} & \cdots &
  0\cr}\right).}
(The lower right matrix element of $\tilde n$ is set to zero to
ensure that $\Tr\,\tilde n=0$.) Any element of $\frak{sl}(N,\C)$ of
this form is regular, meaning that the subgroup of $SL(N,\C)$ that
commutes with $\tilde n$ has complex dimension $n-1$ (the dimension
of a maximal torus). The coefficients $a_k$ can be interpreted as
$\Tr\,\tilde n^k$, $k=2,\dots,N$, the Casimir invariants of this
group.  A generic regular conjugacy class in the Lie algebra is
specified by giving the values of the Casimir invariants; the
regular nilpotent element of eqn. \fredo\ is what we get
(generically) if we set the Casimir invariants to zero.  The
deformation from $U=\exp(n)$ to $\tilde U=\exp(\tilde n)$ shows that
the conjugacy class of $U$ is not rigid and in fact it can be
deformed to a generic regular semisimple conjugacy class.  This
means that, just as we explained in detail for $SU(2)$, a surface
operator constructed using an irreducible embedding
$\rho:\frak{su}(2)\to \frak{su}(N)$ is a limit for
$\alpha,\beta,\gamma\to 0$ of the surface operator constructed with
the general ansatz \norto.

In general, any element of $SL(N,\C)$ can be put in the
block-diagonal form
\eqn\blotcdi{\left(\matrix{* &
* &
*&0&0&0\cr
*&*&*&0&0&0\cr
*&*&*&0&0&0\cr 0&0&0&*&*&0\cr 0&0&0&*&*&0\cr 0&0&0&0&0&*\cr}\right)}
where now each diagonal block, say of size $k\times k$, is the
product of a scalar ``eigenvalue'' $u\in\C^*$ and a principal
unipotent element of $GL(k,\C)$.  Such a conjugacy class is not
rigid if $k>1$ (for any block), since then we can make in that block
the argument of the last paragraph.  If the blocks are all $1\times 1$ blocks,
we are back in the case, treated first, that $U$ is diagonalizable.

To summarize, we have shown that there are no noncentral rigid
conjugacy classes in $SL(N,\C)$. To find rigid (noncentral) surface
operators, we will have to look farther.


\bigskip\noindent{\it Some Examples}

However, complex semisimple Lie groups other than $SL(N,\C)$
{\it do} have rigid surface operators.

Let us first give some simple examples.  For $G=Sp(2N)$, we consider
the $\frak{su}(2)$ embedding corresponding to the decomposition
\eqn\gret{2N=2+1+1+\dots +1.}
The corresponding partition is $\la = [2,1,1,\ldots,1]$ which we
also write as $\la = [2,1^{2N-2}]$. The Lie algebra of $Sp(2N)$
consists of symmetric matrices $n_{ij}$. The raising operator of an
$\frak{su}(2)$ embedding associated to the decomposition \gret\ is a
rank 1 matrix of the form $n_{ij}=b_ib_j$, for some vector $b_i$.
The conjugacy class $\frak C_n$ of an element $U = \exp(n)$ for such
a $n$ is parametrized by $b$ up to $b\to -b$, and so has complex
dimension $2N$. Indeed, its closure (obtained by allowing $b=0$) is
simply
\eqn\hoto{\bar{\frak C}_n = \C^{2N}/\Z_2.}
Not coincidentally, this is a hyper-Kahler orbifold.
The orbit of any element of a complex semisimple Lie algebra is always
hyper-Kahler, as it can be realized as a moduli space of solutions
of Nahm's equations \Kronheimeri.

The conjugacy class $\frak C_n$ is rigid, if $N>1$, simply because
it has the smallest dimension of any non-central conjugacy class in
$G_\C=Sp(2N,\C)$. To see that $\frak C_n$ cannot be deformed to a
semisimple conjugacy class, note that a non-central semisimple
conjugacy class in $Sp(2N,\C)$ of smallest dimension is the
conjugacy class of the element ${\rm diag}(u,u^{-1},1,1,\dots,1)$. A
small calculation shows that the conjugacy class of this element is
of complex dimension $2(2N-1)$, and this exceeds $2N$ if $N>1$.

For $N=1$, the conjugacy class $\frak C_n$ is equivalent to the regular
unipotent conjugacy class in $SL(2,\C)$ that we analyzed earlier, and is not rigid.
This is related to the fact that for $N=1$, the hyper-Kahler
orbifold in eqn. \hoto\ can be blown up or deformed
(while for $N>1$, this hyper-Kahler singularity has no moduli).

For $G=SO(N)$, an example of a rigid unipotent conjugacy class can be
constructed similarly. The Lie algebra $\frak{so}(N)$ consists of
antisymmetric matrices $a_{ij}$. A minimal (nonzero) nilpotent
element of the Lie algebra $\frak{so}(N)$ corresponds to the
decomposition $N=2+2+1+1+\dots +1$.  An element of the Lie algebra
corresponding to such a decomposition takes the form
$a_{ij}=b_ic_j-b_jc_i$, where $b$ and $c$ are vectors obeying
$b\cdot b=b\cdot c=c\cdot c=0$ (and modulo an action of $SL(2,\C)$
on the pair $b,c$).  The conjugacy class $\frak C_a$ of $\exp(a)$
has dimension $2N-6$. For $N>4$, this is the least dimension of any
non-central conjugacy class in $SO(N,\C)$, so again this is a rigid conjugacy class.

Rigid unipotent conjugacy classes or rigid nilpotent orbits
also exist in exceptional groups (see Appendix A).
A (noncentral) unipotent conjugacy class of minimal dimension in
a complex semisimple Lie group is always rigid, except for $A_N$.
In the table, we indicate the dimensions of these minimal conjugacy classes.

\vskip 0.8cm \vbox{ \centerline{\vbox{ \hbox{\vbox{\offinterlineskip
\def\tablespace{height7pt&\omit&&\omit&&\omit&&\omit&&\omit
&&\omit&&\omit&&\omit&&\omit&&\omit&\cr}
\def\tablerule{\tablespace\noalign{\hrule}\tablespace}

\hrule\halign{&\vrule#&\strut\hskip0.2cm\hfill #\hfill\hskip0.2cm\cr
\tablespace & Type && $A_N$ && $B_N$ && $C_{N}$ && $D_{N}$ && $E_6$
&& $E_7$ && $E_8$ && $G_2$ && $F_4$ &\cr
\tablerule & $\dim (\frak C_{{\rm min}})$ && $2N$ && $4N-4$ && $2N$
&& $4N-6$ && $22$ && $34$ && $58$ && $6$ && $16$ &\cr
\tablespace}\hrule}}}}
%
%
} \vskip 0.5cm


\bigskip\noindent{\it Computing The Dimension Of A Unipotent Orbit}

As in the examples just described, it is convenient to be able to
compute the dimension of a unipotent conjugacy class in $G_\C$,
or equivalently of a nilpotent orbit in $\frak{g}_\C$.
So we pause to explain how to do this.

Let $d$ be the complex dimension of $G_\C$, and let $s$ be the
complex dimension of the subgroup $\G_{\C}^n \subset G_\C$ of elements that
commute with a given $n\in \frak{g}_\C$.  The dimension of the orbit
of $n$ (or of $\exp(n)$) is $d-s$.  So it suffices to compute $s$.

The element $n$ is the raising operator for some embedding
$\rho:\frak{su}(2)\to \frak g$.
We decompose $\frak g$ in irreducible representations
${\cal R}_i$ of $\frak{su}(2)$:
\eqn\tofo{\frak g=\oplus_{i=1}^s{\cal R}_i.}
The subspace of $\frak g$ that commutes
with the raising operator $n$ is precisely the space of highest
weight vectors for the action of $\frak{su}(2)$.  Each irreducible
summand ${\cal R}_i $ has a one-dimensional space of highest weight
vectors.  So the subspace of $\frak g$ that commutes with $n$ is of
dimension equal to $s$, the number of summands in \tofo.

For example, one can use this method to compute the dimensions of
the minimal unipotent conjugacy classes in $SO(N,\C)$ or $Sp(2N,\C)$.
We leave this to the reader. For another important example,
we re-examine the regular unipotent orbit of $SL(N,\C)$.
This corresponds to an
irreducible $N$-dimensional representation of $\frak{su}(2)$, and
the summands in \tofo\ are of dimension $3,5,7,\dots,2N-1$.  There
are $N-1$ summands. This shows that the subgroup of $SL(N,\C)$ that
commutes with a principal unipotent element has dimension $N-1$.
(Indeed, for $n$ as in \fredo, this subgroup is generated by the
matrices $n,n^2,\dots,n^{N-1}$.)  The number $N-1$ equals the
dimension of the maximal torus, showing that a principal unipotent
orbit has the same dimension as a generic semisimple orbit
(to which it can be deformed, as we have already discussed).


\bigskip\noindent{\it Strongly Rigid Orbits In Orthogonal And Symplectic Groups}

We will now introduce some useful terminology.  We will say that an
orbit in a Lie algebra (resp. a conjugacy class in a group) is
strongly rigid if its dimension is less than the dimension of any
nearby orbit (resp. conjugacy class). Strongly rigid orbits are
rigid in a very robust way. For suitable $G$, there are also rigid
conjugacy classes that are not strongly rigid; this more delicate
phenomenon is described momentarily.

A nilpotent element $n\in \frak g_{\C}$ is strongly rigid if and
only if the corresponding unipotent group element $U=\exp(n)$ is
strongly rigid. So as long as we focus on unipotent conjugacy
classes, we can equally well work in the group or the Lie algebra.

An equivalent definition is that $U\in G_\C$ (or $n\in \frak
g_{\C}$) is strongly rigid if the dimension of its centralizer is
greater than the dimension of the centralizer of any nearby element
of $G_\C$ (or of $\frak g_{\C}$).  In due course, we will also
consider a weaker notion that applies to group elements (but not to
elements of a Lie algebra): $U\in G_\C$ is rigid (but not strongly
rigid) if its centralizer includes as a proper subgroup the
centralizer of any nearby element of $G_\C$.  Thus any nearby
element has a centralizer that is strictly smaller than that of $U$.
(We also use the term weakly rigid to describe an element that is
rigid but not strongly rigid.) For unipotent orbits, there is no
difference between rigid and strongly rigid.

At the end of section \limit, we explained how to classify unipotent
orbits in $SO(N)$ or $Sp(2N)$ in terms of partitions.  For a
unipotent orbit to be strongly rigid, the partition must obey two
conditions.  We here explain why the conditions are necessary,
referring to \CMcGovern, section 7.3, for a proof that they are sufficient.

The first condition reflects the fact that the identity orbit of
$SO(2)$ is not rigid.  Indeed, $SO(2)$ is abelian, so every orbit
consists of only one point.  The identity orbit is rigid in any
other orthogonal or symplectic group.

Let us begin with $G=SO(N)$. Consider a partition
$N=\la_1+\la_2+\dots+\la_k$ in which one of the parts, say $\la^*$,
occurs with multiplicity $r>1$. Let $\rho:\frak{su}(2) \to
\frak{so}(N)$ be a corresponding homomorphism. The subgroup of $G$
that commutes with $\rho$ and acts only on the summands of dimension
$\la^*$ is $\G^*=SO(r)$ if $\la^*$ is odd, and $\G^*=Sp(r)$ if
$\la^*$ is even. (We recall that if $\la^*$ is even, then $r$ is
always also even.) Let $n$ be the raising operator of $\rho$ and
$U=\exp(n)$ the corresponding unipotent element. If $\lambda^*=2$
and $r=2$, then because of the exceptional property of $SO(2)$ just
noted, we can modify $U$ by multiplying it by an element of $\G^*$,
without changing the dimension of its orbit.

So a partition of $N$ in which an odd part occurs with multiplicity
2 does not lead to a strongly rigid orbit in $SO(N)$. For example,
for $G=SO(9)$, the  orbit labeled by the partition $\la =
[3,2,2,1,1]$ is not strongly rigid, since the odd number 1 appears
with multiplicity 2. The same reasoning shows that a partition of
$2N$ in which an even part occurs with multiplicity 2 does not lead
to a strongly rigid orbit in $Sp(2N)$.

%
%
%

Now we consider the second constraint required in order for a
unipotent orbit to be rigid.  In terms of partitions, this
constraint occurs if there are gaps in the sequence of the $\la_i$.
To be precise, arranging the $\la_i$ so that $\la_1\geq \la_2\geq
\dots\geq \la_k$, the condition is that all positive integers that
are less than $\lambda_1$ do occur in this sequence with positive
multiplicity.

A partition with a gap does not lead to a strongly rigid orbit. We
will discuss the case that the gap separates two parts $\lambda_j$,
$\lambda_{j+1}$ with $\lambda_j\geq \lambda_{j+1}+2$. (The other
case with a gap is the case that $\lambda_k\geq 2$; it can be
treated similarly, replacing the numbers $\lambda_j$ and
$\lambda_{j+1}$ in the following construction with $\lambda_k$ and
0.)   For odd $\la_j, \la_{j+1}$, a deformation showing that such an
orbit is not strongly rigid can be constructed in a subspace
involving only the two blocks of size $\la_j$ and $\la_{j+1}$, as
shown here for $\la_j=3,$ $\la_{j+1}=1$:
\eqn\welf{\left(\matrix{
*&*&*&0\cr
*&*&*&0\cr
*&*&*&0\cr
0&0&0&*\cr }\right). }
So we can replace $N$ by $N'=\la_j + \la_{j+1}$ and
$SO(N)$ by $SO(N')$.  (If $\la_j$ and $\la_{j+1}$ are not odd,
the corresponding blocks occur with multiplicity at least 2
and we have to keep 2 blocks of the relevant dimension
in making the construction of the next paragraph.)

So we are reduced to the case that $G=SO(N)$ with a decomposition
$N=m + m'$, with $m \geq m'+2$. We write $U'$ for a unipotent
element of $SO(N)$ associated with this embedding. It is the product
of principal unipotent elements  in the two blocks. (Each is
associated with irreducible $\frak{su}(2)$ embedding in that block.)
The conjugacy class of $U'$ can be deformed to a non-unipotent (but
also not semisimple) conjugacy class of the following type. We
consider an $SO(N)$ matrix $U$ that is the direct sum of three
blocks: a generic semisimple $2\times 2$ block, a principal
unipotent $(m-2)\times (m-2)$ block, and a principal unipotent
$m'\times m'$ block. Thus $U$ looks something like
\eqn\tolg{U=\left(\matrix{
*&*&0&0\cr
*&*&0&0\cr
0&0&\times&0\cr
0&0&0&\times\cr}\right),}
where the upper left $2\times 2$ block is a generic element of $SO(2)$
\eqn\libo{\left(\matrix{a  & b \cr -b & a\cr}\right),~~a^2+b^2=1,}
and the diagonal elements denoted $\times$ in eqn. \tolg\
represent principal unipotent elements of $SO(m-2)$
and $SO(m')$, respectively. A family of matrices conjugate to $U$
for some $a,b$ can as $a\to 1$, $b\to 0$ approach $U'$. This is very
similar to the relation between  \nordo\ and \doof\ in the
$SL(2,\C)$ example that we studied in detail (and in fact, if we set
$m=3$, $m'=1$, and use the fact that $SL(2,\C)$ is a double cover of
$SO(3,\C)$, the previous example becomes a special case of the
present discussion).
The conjugacy class of $U$ has the same dimension as that of $U'$, as one can
verify by computing the dimension of the subgroups of $SO(N)$ that
commute with $U$ or $U'$, using the method\foot{To be more exact,
one can use this method to compute the dimension of the centralizer
of $U'$ of equivalently the dimension of its conjugacy class.
The dimension of the centralizer of $U$ equals the sum of 1 -- coming
from the fact that $U$ commutes with an $SO(2)$ that is embedded
as the upper left block in $SO(N)$ -- plus the dimension of
the conjugacy class of a unipotent element of $SO(N-2)$
associated with the decomposition $N-2=(m-2)+m'$.
This dimension can be computed using eqn. \tofo,
and finally one shows that $U$ and $U'$
have centralizers of the same dimension.} of eqn. \tofo.

The conditions that we have just described, taken together,
completely characterize rigid nilpotent orbits  for orthogonal and
symplectic gauge groups (\CMcGovern, section 7.3). In the following
table, we list the rigid nilpotent orbits in classical groups of
small rank. In the table, a partition corresponding to a
decomposition $N=\la_1+\la_2+\dots+\la_k$ is denoted simply
$[\la_1,\la_2,\dots,\la_k]$. In the table, we do not include the
orbit of the identity element, though it is rigid for all $G$. (It
corresponds to the partition $[1,1,\ldots, 1]$.)
\bigskip
\centerline{\vbox{\offinterlineskip
\def\tablerule{\noalign{\hrule}}
\halign to 3.5truein{\tabskip=1em plus 2em#\hfil&\vrule height12pt
depth5pt#&#\hfil&\vrule height12pt depth5pt#&#\hfil\tabskip=0pt\cr
\hfil ~~$G$ \hfil&&\hfil rigid nilpotent orbit
$\frak c_{\la}$ \hfil&&\hfil $\dim (\frak c_{\la})$ \hfil\cr
\tablerule
~~$B_2$  && $[2,2,1]$ && 4 \cr
~~$C_2$   && $[2,1,1]$  && 4 \cr
~~$B_3$  && $[2,2,1,1,1]$  && 8 \cr
~~$C_3$   && $[2,1,1,1,1]$ && 6 \cr
~~$B_4$  && $[2^4,1]$  && 16 \cr
&& $[2,2,1^5]$ && 12 \cr
~~$C_4$   && $[2,2,2,1,1]$ && 18 \cr
&& $[2,1^6]$ && 8 \cr
~~$D_4$   && $[3,2,2,1]$ && 16 \cr
&& $[2,2,1^4]$ && 10 \cr
~~$\ldots$ && $\ldots$ && $\ldots$ \cr
}}}\bigskip
\noindent


\subsec{Strongly Rigid Semisimple Orbits}
\subseclab\rigidsemisimple

For what we have just described, it is equivalent to consider a
nilpotent element $n$ of the Lie algebra $\frak g_{\C}$ or
a unipotent element $U=\exp(n)$ of the group $G_{\C}$.
Indeed, $n$ is a strongly rigid element of the Lie algebra
if and only if $U$ is a strongly rigid element of the group.

A strongly rigid element $n\in\frak g_{\C}$ is always nilpotent,
for the following reason. First of all, if $t$ is a non-zero complex
number, $n$ and $tn$ always have orbits of the same dimension.
On the other hand, if $n$ is not nilpotent, it has nonzero Casimir
invariants, which differ from those of $tn$ (if $t$ is close to but
not equal to 1), showing that $tn$ is not conjugate to $n$.
So the orbit of $n$, if $n$ is not nilpotent, can always be deformed
to a nearby orbit of the same dimension, namely the orbit of $tn$.

However, it is possible for a semisimple conjugacy class in the
group $G$ or $G_\C$ (as opposed to an orbit in the Lie algebra) to
be strongly rigid. This does not occur for $G=SU(N)$, as we
explained in section \searching. But if $G$ is any other simple Lie
group, there are strongly rigid semisimple conjugacy classes in $G$.
For example, for $G=SO(N)$, a strongly rigid conjugacy class
contains an element of the form
\eqn\sssimple{S_i = {\rm diag} \big( +1,+1 \ldots,+1, \underbrace{-1,-1,\dots,-1,-1}_{2i} \big) }
where the subscript $i$ refers to the total number of pairs of
$-1$'s, and we require $i>1$. The subgroup $\G^{S_i}$ of $SO(N)$ that
commutes with $S_i$ is a double cover of $SO(N-2i)\times SO(2i)$.
The double cover in question might be denoted as $S(O(N-2i)\times
O(2i))$. Indeed, $S_i$ commutes with a block diagonal matrix
\eqn\delf{\left(\matrix{A&0\cr 0&B\cr}\right),} where $A\in O(2i)$,
$B\in O(N-2i)$; and such a matrix is in $SO(N)$ if $\det A\,\det
B=1$.

For $i>1$, the conjugacy class of $S_i$ is strongly rigid, since
perturbing the eigenvalues of $S_i$ away from $\pm 1$ causes the
dimension of the centralizer to become smaller.  After such a
perturbation,  the orthogonal groups $SO(N-2i)$ and $SO(2i)$ are
replaced by unitary groups or products of unitary and orthogonal
groups of lower dimension.

The case $i=1$ is special, because $SO(2)$ is abelian. We do not
change the dimension of the conjugacy class of $S_1$ if we deform it
so that the lower right $2\times 2$ block changes from ${\rm
diag}(-1,-1)$ to a generic element
\eqn\zolf{\left(\matrix{a&b\cr -b&a\cr}\right),~~a^2+b^2=1}
of $SO(2)$.  Hence, the conjugacy class of $S_1$ is not strongly
rigid. It actually is our first example of a group element that is
weakly rigid but not strongly rigid. If we deform the lower right
block of $S_1$ as in \zolf, its centralizer is reduced from
$S(O(N-2)\times O(2))$ to $SO(N-2)\times SO(2)$. The centralizer of
the nearby conjugacy class is smaller, but has the same dimension.
It is of index 2 in the centralizer of $S_1$. We discuss this more
fully in section \additional.

There is a similar story for $G=Sp(2N)$.  A rigid element is again
conjugate to the element $S_i$ of eqn. \sssimple.  For $S_i$ to be
non-central, we need $1\leq i\leq N-1$.   Again, if we deform
$S_i$ so that its eigenvalues are not $\pm 1$, then the dimension
of its centralizer becomes less and the dimension of its conjugacy class
increases. So these elements are strongly rigid.

It is not hard to show that these are the only rigid semisimple
elements in $SO(N)$ or $Sp(2N)$. If $S$ has a pair of eigenvalues
$u,u^{-1}$ that do not equal $1$ or $-1$, then one can
vary $u$ without changing the centralizer of $S$.
(If there are several eigenvalue pairs all equal to $u,u^{-1}$,
then one must vary these pairs while preserving their equality.)

As one can see in the above examples, if $S$ is a rigid semisimple
element of $G$, then the subgroup $\G^S$ of $G$ that commutes with $S$
has the same rank as $G$, though of course its dimension is smaller
(unless $S$ is central).  A further study of the above examples
shows that the Dynkin diagram of $\G^S$ can always be obtained from
the extended Dynkin diagram of  $G$ by removing one node. (Extended
Dynkin diagrams of the simple Lie groups are shown in the figure below.)
Finally, the order of $S$ in the adjoint form of $G$ divides the
Coxeter label (or Kac number) of the omitted node.  For the
orthogonal and symplectic groups that are our main examples, this
merely means that $S$ is of order 2, since the relevant labels equal 2.

\ifig\ExtendedDynkinDiagrams{Extended Dynkin diagrams for semisimple Lie algebras
with Coxeter labels $a_i$ (we set $a_0 = 1$).}
{\epsfxsize4.8in\epsfbox{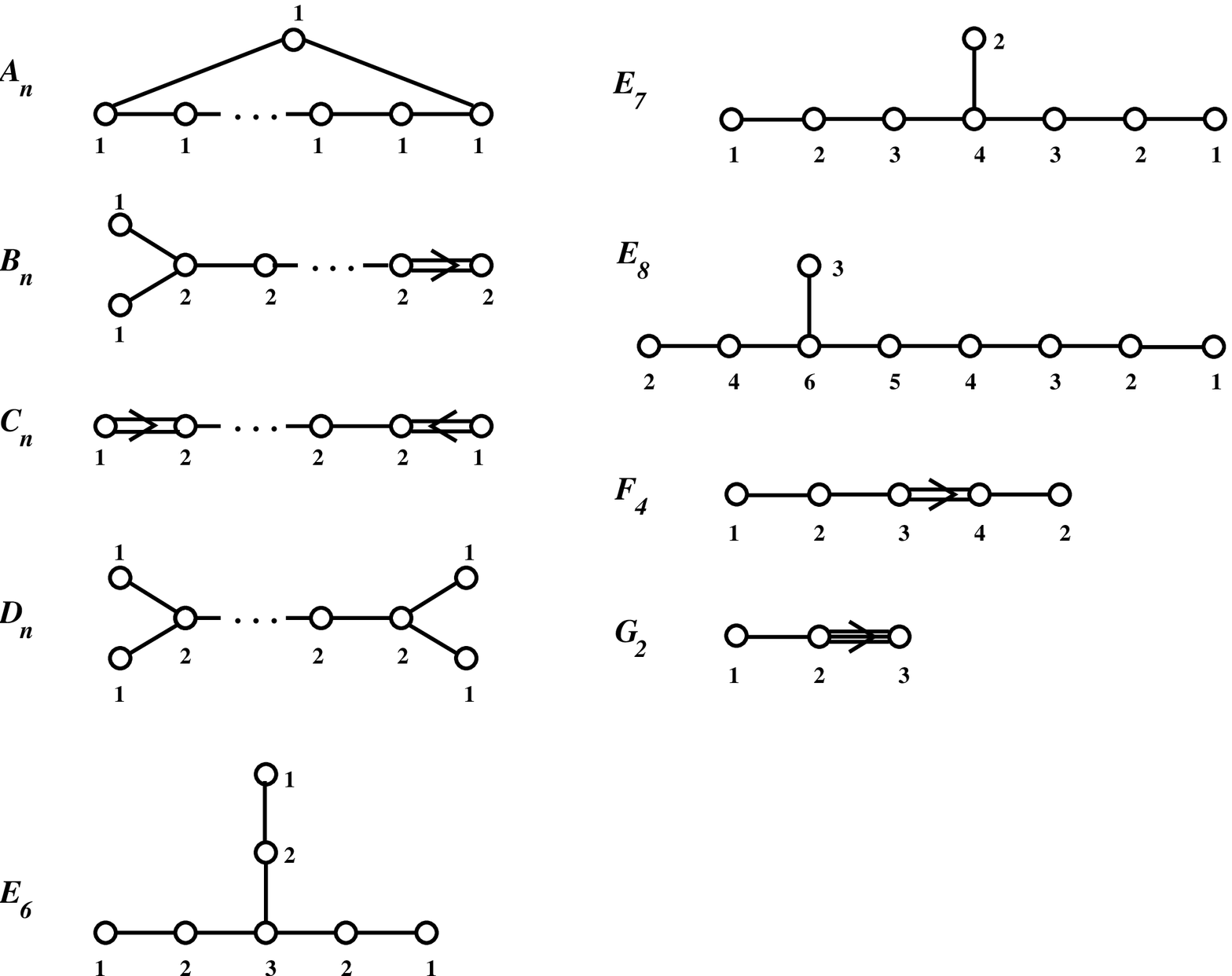}}

In the general theory of  rigid semisimple orbits, it is shown that
all of these statements hold for any $G$. Indeed, let $\Lambda_{\rm
rt}$ be the root lattice of $G$, $\Lambda_{\rm rt}^+ \subset
\Lambda_{\rm rt}$ the set of positive roots, and $\Delta = \{ \a_1 ,
\ldots, \a_r \} \subset \Lambda_{\rm rt}^+$ the corresponding set of
simple roots. Furthermore, let
\eqn\highestrt{ \th = \sum_{i=1}^r a_i \a_r }
be the highest root in $\Lambda_{\rm rt}^+$. The coefficients $a_i$
are the Coxeter labels (or Kac numbers). We denote $\a_0 = - \th$
and $\bar \Delta = \Delta \cup \a_0$. A proper subset of simple
roots, $\Theta \subset \bar \Delta$, defines a parabolic subgroup
$\eusm P (\Theta) \subset G_{\C}$ with the Levi subgroup $\Bbb{L} (\Theta)$,
which will be identified with the centralizer $\G^S$
of a semisimple element $S$ in $G$.
We remind that every parabolic subalgebra $\frak p$ has a direct sum
decomposition
\eqn\levi{ \frak p = \frak l \oplus \frak n }
called Levi decomposition, where $\frak l$ is the Levi factor and
$\frak n$ the nilpotent radical of $\frak p$.
Specifically, in our case, the parabolic subalgebra $\frak p (\Theta)$
associated with the subset of simple roots $\Theta$ is generated
by $\frak t$ and all the root spaces $\frak g_{\a}$ such that
$\a \in \bar \Delta$ or $-\a \in \Theta$.
Similarly, the Levi subalgebra $\frak l (\Theta)$ corresponding to
$\Bbb{L} (\Theta)$ is
\eqn\ltheta{ \frak l (\Theta)
= \frak t \oplus \sum_{\a \in \Lambda_{\Theta}}
\frak g_{\a} }
where $\Lambda_{\Theta}$ denotes the subroot system generated by $\Theta$.
We note that, since elements of $\bar \Delta$ correspond to nodes of
the extended Dynkin diagram of $G$, we can think of $\Theta$ as a subset
of nodes of the extended Dynkin diagram.

Since we are interested in rigid surface operators, the centralizer
$$
\G^S = \Bbb{L} (\Theta)
$$
must be of the same rank as $G$.
In other words, $\Theta$ must be a proper subset of $\bar \Delta$
obtained by removing a single node;
we denote such subsets $\Theta_i$, $i=1, \ldots, r$,
\eqn\thetai{ \Theta_i = \bar \Delta \setminus \{ \a_i \} }
Every such subset of simple roots $\Theta_i \subset \bar \Delta$,
corresponds to (the conjugacy class of) a rigid semisimple element
$S_i$ in the simply-connected form of $G$ (that is $G = \Gsc$).
Generalization to other forms of $G$ ({\it e.g.} to the adjoint form $G = \Gad$)
will be discussed below.

Specifically, we define $S_0 = 1$ and $S_i$ for $i=1, \ldots, r$
as follows (see {\it e.g.} \dCKac)
\eqn\rigidsselt{ S_i = \exp (2\pi i \omega_i^{\vee} / a_i) }
where $\omega^{\vee}_i \in \TT$ are the fundamental coweights
defined by
\eqn\omegajdef{ \langle \a_i , \omega_j^{\vee} \rangle = \delta_{ij} }
One important consequence of the fact that rigid semisimple
elements are of finite order (the order being a divisor of one of
the Coxeter labels) is that any rigid semisimple element of $G_\C$
can actually be conjugated to the compact group $G$.  This is
important in the context of ${\cal N}=4$ super Yang-Mills theory,
since only $G$ and not $G_\C$ is a group of gauge symmetries in this
theory.

We described rigid semisimple elements assuming that $G$ is
simply-connected. Now we wish to relax this assumption and, in
particular, to describe rigid semisimple elements when $G$ is of the
adjoint type. Note, that in the construction of rigid semisimple
elements in the simply-connected form of $G$ we found rigid elements
$S_i$ for every choice of the proper subset $\Theta_i \subset \bar
\Delta$, where index $i$ runs from $0$ to $r$, not taking account
symmetries of the Dynkin diagram.


\vskip 0.8cm \vbox{ \centerline{\vbox{
\hbox{\vbox{\offinterlineskip
\def\tablespace{height7pt&\omit&&\omit&&\omit&&\omit&&\omit&&\omit&&\omit&\cr}
\def\tablerule{\tablespace\noalign{\hrule}\tablespace}

\hrule\halign{&\vrule#&\strut\hskip0.2cm\hfill
#\hfill\hskip0.2cm\cr
\tablespace
& Type && $A_N$ && $B_N$, $C_N$, $E_7$ && $D_{2N}$ && $D_{2N+1}$ && $E_6$ && $E_8$, $F_4$, $G_2$ &\cr
\tablerule
& $\CZ (\Gsc)$ && $\Z_{N+1}$ && $\Z_2$ && $\Z_2 \times \Z_2$ && $\Z_4$ && $\Z_3$ && $1$ &\cr
\tablespace}\hrule}}}}
%
%
} \vskip 0.5cm

The center $\CZ (\Gsc)$ of the universal cover $\Gsc$ acts on the
extended Dynkin diagram, therefore, relating some of the nodes
which give rise to the same conjugacy classes of rigid semisimple
elements. Hence, if we wish to consider, say, the adjoint form of
$G$, we need to divide by the action of $\CZ (\Gsc)$ and to take
only one conjugacy class for every orbit of the $\CZ
(\Gsc)$-action on the nodes of the Dynkin diagram. This leads to a
similar classification of rigid semisimple elements (and their
conjugacy classes) for the adjoint form of $G$, except that now
the index $i$ that labels proper subsets $\Theta_i \subset \bar
\Delta$ runs only over the subset of the nodes of the Dynkin
diagram not identified by $\CZ (\Gsc)$:
\eqn\irange{\eqalign{
(A_N): \quad\quad\quad\quad\quad~~ & i=0 \cr
(B_N): \quad\quad\quad\quad\quad~~ & 0 \le i \le N-1 \cr
(C_N): \quad\quad\quad\quad\quad~~ & 0 \le i \le \Big[ {N \over 2} \Big] \cr
(D_N): \quad\quad\quad\quad\quad~~ & 0 \le i \le \Big[ {N \over 2} \Big]-1 \cr
(E_6): \quad\quad\quad\quad\quad~~ & 0 \le i \le 2 \cr
(E_7): \quad\quad\quad\quad\quad~~ & 0 \le i \le 4 \cr
(E_8,~F_4,~G_2): \quad\quad & 0 \le i \le r
}}
For example, in type $A$ the center $\CZ (\Gsc)$ acts by ``rotating''
the nodes of the extended Dynkin diagram, in types $B$ and $C$ it acts
by ``reflection'' with respect to the horizontal (resp. vertical) axis, {\it etc.}


\bigskip\noindent{\it Rigid Semisimple Surface Operators}

Now we will explain why rigid semisimple surface operators are
relevant for our purposes.

We will describe a gauge theory singularity in real codimension 2
associated with a rigid semisimple element of $G$. In the notation
of section \review, we  take the singularity to be at $x^2=x^3=0$,
and we use polar coordinates $x^2+ix^3=re^{i\theta}$.

In the absence of any singularity, an adjoint-valued field on the
$x^2-x^3$ plane (for fixed values of the other coordinates
$x^0,x^1$, which we suppress) can be represented by an
adjoint-valued function $\Phi(r,\theta)$ that obeys
$\Phi(r,\theta+2\pi)=\Phi(r,\theta)$. If $S$ is any element of the
gauge group $G$, we can modify this condition to
\eqn\modcon{\Phi(r,\theta+2\pi)=S\Phi(r,\theta)S^{-1}.} Since $G$ is
a symmetry group of ${\cal N}=4$ super Yang-Mills theory, it makes
sense to formulate ${\cal N}=4$ super Yang-Mills theory for fields
that have this sort of behavior, near a codimension two surface $D$
in spacetime.

Of course, if we impose this condition, then along $D$, we should
divide only by gauge transformations that commute with $S$.  This
recipe gives a surface operator that makes sense for any $S\in G$.
It varies smoothly as long as the centralizer $\G^S$ of $S$ in $G$
does not change.
To get a rigid surface operator, we must pick $S$ to be rigid,
meaning that $\G^S$ jumps if $S$ is changed at all.

Let us compare the surface operator obtained in this description to
the type of surface operator that we considered in \Ramified. There,
as in eqn. \norto, we considered a gauge singularity of the form
$A=\alpha\,d\theta$.  (For the present purposes, we set
$\beta=\gamma=\eta=0$.)  One quantizes ${\cal N}=4$ super Yang-Mills
theory for fields with this type of singularity, dividing by gauge
transformations that at $z=0$ are valued in $\G^\alpha$, the
centralizer of $\alpha$ in $G$.  Let us call this type of surface
operator a generic one.

A generic surface operator behaves well as $\alpha$ is varied as
long as the centralizer of $\alpha$ is the same as the centralizer
of the monodromy $S=\exp(-2 \pi \alpha)$.  We are precisely in the
situation in which this is not the case, for if $S$ is strongly
rigid (and noncentral) then the centralizer of $S$ is strictly
larger than the centralizer of any $\alpha\in \frak g$ such that
$S=\exp(-2\pi\alpha)$.  (Likewise, if $S$ is rigid, then invariance
under $\G^S$ does not allow the introduction of any continuous
parameters analogous to $\beta,\gamma,\eta.$)

The rigid surface operator with monodromy a rigid semisimple
element $S\in G$ is therefore not quite a special case of the
generic construction in \Ramified.  But it is a close cousin,
somewhat similar to the construction in \Ramified\ of
surface operators associated with Levi subgroups $\Bbb{L}$
that are strictly larger than $\Bbb{T}$.


\subsec{Combining The Two Constructions}\subseclab\combining

So far we have constructed rigid surface operators whose monodromy
is a  rigid element of $G_\C$ that is either unipotent or
semisimple.   The former construction used Nahm's equations and the
latter one was done by fiat in eqn. \modcon. Actually, we can
combine the two constructions and construct a rigid surface operator
whose monodromy is in any rigid conjugacy class of $G_\C$, not
necessarily semisimple or unipotent.

We need to know a few facts.  To being with, any element $V \in
G_\C$ can be written as $V=SU$, where $S$ is semisimple, $U$ is
unipotent, and $S$ commutes with $U$.  Moreover, let $\G_{\C}^S$ be
the centralizer of $S$ in $G_\C$, so $U \in \G_{\C}^S$.  Then the
condition for $V=SU$ to be  rigid (or strongly rigid) in $G_\C$ is
that $S$ must be rigid (or strongly rigid) in $G_\C$ and $U$ must be
 rigid in $\G^S_{\C}$.

To construct a surface operator with monodromy $V=SU$, we combine
the two constructions as follows.  First we require that near $r=0$,
all fields of ${\cal N}=4$ super Yang-Mills theory obey
$\Phi(r,\theta+2\pi)=S\Phi(r,\theta)S^{-1}$, as in \modcon. Second,
we also pick a homomorphism $\rho:\frak{su}(2)\to\frak{g}^S$
(here $\frak{g}^S$ is the Lie algebra of $\G^S$)
and we require that the fields have a singularity near $r=0$
that is given by the familiar solution \nahmsolut\ of Nahm's equations:
\eqn\nahmsolt{\eqalign{A&={t_1\,d\theta\over \ln r}+\dots \cr
\phi&={t_2\,dr\over r\ln r}-{t_3\,d\theta\over \ln r}+\dots,}}
where the ellipses denote terms that are less singular at $r=0$.
Because $\rho$ commutes with $S$, this condition on the fields is
compatible with \modcon. The combined condition defines a surface
operator with the monodromy
\eqn\vviasu{V=SU.}
There is no need here for $V$ to be rigid.  For every conjugacy
class in $G_\C$, a construction along these lines gives a surface
operator of monodromy $V$. However, for generic $V$, this surface
operator is simply equivalent to a special case of the generic
surface operators constructed in \Ramified\ and reviewed in section
\review. For certain well-chosen $V$, the construction gives
something new. The case that is most novel, or at least most
different from what was already considered in \Ramified, is the case
that $V$ is rigid.

As we explained in the previous subsection,
strongly rigid semisimple elements correspond to proper
subsets of simple roots.
For every such subset $\Theta_i \subset \bar \Delta$,
the corresponding Levi subgroup $\Bbb{L} (\Theta_i)$
is precisely the centralizer $\G^{S_i}$ of the semisimple
element $S_i$. In the case of orthogonal and symplectic groups,
the Levi subgroup $\Bbb{L} (\Theta_i)$ is always a product of two factors,
\eqn\levii{ \Bbb{L} (\Theta_i) = \Bbb{L}' \times \Bbb{L}''}
where to avoid having to specify exceptions we allow the case that
$\Bbb{L}'$ or $\Bbb{L}''$ is trivial and the other is equal to $G$.
(This happens if $\Theta_0$ is obtained by omitting the extended
root, leaving the original Dynkin diagram of $G$. Thus $\Theta_0 =
\Delta$ and $\Bbb{L} (\Theta_0) = G$. We think of the Dynkin diagram
of $G$ as the union of itself with an empty Dynkin diagram. We
simply include this case in our notation as the case that the
product of groups is $\Bbb{L}(\Theta_0)=1\times G$.  This is
precisely the case of a unipotent conjugacy class.) We denote by
$\frak l (\Theta_i) = \frak l' \oplus \frak l''$ the Lie algebra of
$\Bbb{L}(\Theta_i)$.

After picking $S$, the construction of strongly rigid surface
operators with monodromy $V=SU$ also requires a choice of a rigid
unipotent $U \in \G^{S_i}_{\C}$ or, in view of eqn. \levii, a pair
of rigid nilpotent orbits $\frak c'$ and $\frak c''$ in $\frak
l_{\C}'$ and $\frak l_{\C}''$, respectively. A complete
classification of rigid nilpotent orbits for classical groups was
described in section \limit.  For such groups, nilpotent orbits are
labeled by partitions. Therefore, in such cases we can use a pair of
partitions $(\la', \la'')$ to label nilpotent orbits in $\frak
l_{\C}' \oplus \frak l_{\C}''$. To summarize, strongly rigid
conjugacy classes in $G_{\C}$ are labeled by the choice of a root
system $\Theta_i \subset \bar \Delta$ and a  rigid nilpotent orbit
in each factor of $\frak l (\Theta_i)$; in classical types $A$, $B$,
$C$, and $D$ we can naturally label such rigid conjugacy classes by
a pair of partitions,
\eqn\rigidclass{ \frak C^{\Theta_i}_{(\la',\la'')} \subset G_{\C} }

For instance, in section \dualfor\ we consider many examples of dual
pairs of rigid surface operators in theories with gauge groups $G =
Sp(2N)$ and $\LG = SO(2N+1)$. Strongly rigid semisimple conjugacy
classes in these theories are labeled by a choice of node
$i=0,1,\ldots,N$ that defines the root system $\Theta_i \subset \bar
\Delta$ and a pair of partitions $(\la', \la'')$. Omitting the
$i$-th node from the extended Dynkin diagram of $B_N$ gives the root
system of type
\eqn\bnsplit{ D_i \times B_{N-i} }
that we already described explicitly in eqn. \delf.
Hence, in the case of $B_N$ both partitions $\la'$ and $\la''$ are orthogonal.
Similarly, omitting the $i$-th node from the extended Dynkin diagram of $C_N$
gives the root system of type
\eqn\cnsplit{ C_i \times C_{N-i} }
and the corresponding partitions $\la'$ and $\la''$ are symplectic.
One of the factors can be absent, in which case the corresponding
partition is empty. This is the case of a rigid unipotent conjugacy class.


\newsec{Alternative Point of View}\seclab\additional

We begin this section by proposing an alternative point of view
about the surface operators constructed in \Ramified\ and reviewed in
section \rigidsurf. In fact, we will take an ``electric'' viewpoint
in which a surface operator is constructed not by postulating a
singularity (the ``magnetic'' point of view) but by introducing
additional variables.

With this as our starting point, we will then describe some
constructions of rigid surface operators that are more delicate
than those of section \rigidsurf.  This will also lead to our
first duality conjecture.

\subsec{Coupling To Sigma Models}\subseclab\alternative

We simply couple four-dimensional super Yang-Mills theory to
hypermultiplets that are supported on a two-manifold $D$ that is to
be the support of our surface operator.  The hypermultiplets
parametrize a hyper-Kahler manifold $\CQ$ with $G$ action,
so that the supersymmetric sigma model with target $\CQ$ can be
coupled to supersymmetric Yang-Mills theory with gauge group $G$.
Of course, the gauge theory is defined on all of $\R^4$,
with coordinates $x^0,x^1,x^2,x^3$, while the sigma model
is defined on the two-dimensional subspace $x^2=x^3=0$.

Hitchin's equations
\eqn\hitchinteqs{\eqalign{ & F_A - \phi \wedge
\phi = 0 \cr & d_A \phi = 0,\quad d_A \star \phi = 0 }}
assert the vanishing of the moment map for the fields $A,\phi$
(regarded as in section \review\ as hypermultiplets in a two-dimensional sense).
This being so, it is straightforward to include a $\CQ$-valued
hypermultiplet supported at the origin of the $x^2-x^3$ plane.
Let $\vec\mu=(\mu_l,\mu_2,\mu_3)$ be the hyper-Kahler moment map
for the action of $G$ on the hyper-Kahler manifold $\CQ$.
Then the components of $\vec \mu$ appear as delta function
contributions in Hitchin's equations, which can be written in the form
%
\eqn\tyeon{\eqalign{
F_{z \bar z}-[\varphi,\bar \varphi] & = 2 \pi\delta^2(\vec x)\mu_1 \cr
\bar \partial_A \varphi & = \pi\delta^2(\vec x)(\mu_2 + i \mu_3).}}
where $\phi = \varphi dz + \bar \varphi d \bar z$, and
$\delta^2(\vec x)$ is a delta function supported at $z = x^2 +i
x^3=0$; the precise numerical factors multiplying the delta
functions on the right hand side depends on a choice of
normalization of the hyper-Kahler metric on $\CQ$. The labeling of
the components of $\vec \mu$ as $(\mu_1,\mu_2,\mu_3)$ depends on a
choice\foot{The space of pairs $A(x^2,x^3)$, $\phi(x^2,x^3)$ is an
infinite-dimensional hyper-Kahler manifold $\cal W$, with three
independent complex structures.  Likewise, $\CQ$ is a
finite-dimensional hyper-Kahler manifold.  In constructing the
coupling of $\N=4$ super Yang-Mills theory to the sigma model with
target $\CQ$, one can use any hyper-Kahler structure on the product
${\cal W}\times \CQ$.  To endow this product with a hyper-Kahler
structure, one needs to pick a way of ``aligning'' the complex
structures on the two factors.  A choice of such an alignment is
equivalent to a choice of what we mean by the components
$\mu_1,\mu_2,\mu_3$ of $\vec \mu$.} that is made when the
supersymmetric sigma model of target space $\CQ$ is coupled to the
gauge theory.

To decide whether this construction makes sense, we will explore the
solutions of Hitchin's equations with the indicated delta function
source terms.   If there are no reasonable classical solutions, we
surmise that  the sigma model with target $\CQ$ cannot be coupled to
the four-dimensional gauge theory.  We will see that for suitable
$\CQ$, there are reasonable classical solutions.

Since the delta function  ``source'' term in eqn. \tyeon\ is
rotation-invariant, it is reasonable to look for a
rotation-invariant solution.  So away from $r=0$, we simply make the
familiar rotation-invariant ansatz
\eqn\satz{\eqalign{
A & = a(r) d \th + f(r) {dr \over r} \cr
\phi & = b(r) {dr \over r} - c (r) d \th. }}
And, just as in section \review, this leads to Nahm's equations \nahmeqs:
\eqn\ahmeqs{\eqalign{ & {da \over ds} = [b,c]
\cr & {db \over ds} = [c,a] \cr & {dc \over ds} = [a,b] }}
away from $r=0$.  What do the delta functions in eqn. \tyeon\ mean?
Suppose that  $a(r)$ has a limit for $r\to 0$ and call this limit $\alpha$.
Then $A\sim \alpha\,d\theta$ for $r\to 0$.  A connection of this
form is flat away from $r=0$, but has delta function curvature $2\pi
\alpha\delta^2(x)$.  Similarly, if the functions $b(r)$ or $c(r)$
have nonzero limits for $r\to 0$, this gives delta function
contributions in Hitchin's equations.  So it is reasonable to
interpret the delta functions source terms in Hitchin's equations to
mean that we want a solution of Nahm's equations that has the
property that the functions $a,b,c$ have limits for $r\to 0$, and
moreover \eqn\tyono{\lim_{r\to 0}(a,b,c)=(\mu_1,\mu_2,\mu_3).}


This is a very strong condition for the following reason.  First
of all, $r\to 0$ corresponds to $s\to\infty$, so in taking this
limit, we need to solve Nahm's equations on an infinite interval.
For $a,b,c$ to have limits for $s\to\infty$, their derivatives
with respect to $s$ must vanish in this limit, and then Nahm's
equations imply that the limiting values of $a,b,c$ must commute.
On the other hand, for a generic $\CQ$ and a generic point $p\in \CQ$,
the components $\mu_1,\mu_2,\mu_3$ are completely generic elements
of the Lie algebra $\frak g$, and there is no reason at all for
them to commute.

We conclude that the coupling of the sigma model with target $\CQ$
to the gauge theory only makes sense if there are points in $\CQ$
such that the components of $\vec \mu$ commute.  Moreover, in a
sense, these are the only allowed points in the combined system.
Actually, this statement will eventually need some refinement.

\subsec{An Example}\subseclab\anexampl

To test whether this is the right point of view, we would like to
give some interesting examples of hyper-Kahler manifolds that
according to this criterion {\it can} be coupled in the
above sense to ${\cal N}=4$ super Yang-Mills theory.

The half-BPS surface operators constructed in \Ramified\ and
reviewed in section \review\ depend on the choice of a commuting
triple $\alpha,\beta,\gamma\in \frak t\subset \frak g$.  We would
like to reinterpret these surface operators as arising from the
coupling of ${\cal N}=4$ super Yang-Mills theory to a sigma model
with some hyper-Kahler target manifold $\CQ_{\alpha,\beta,\gamma}$.
Such a statement was considered as an approximation in \Ramified,
but we will re-interpret it here as an exact statement.
(The parameter $\eta$ will then further arise as a theta-like angle in
the sigma model with this target, rather as in section 6 of \Ramified.)

We would like $\CQ_{\alpha,\beta,\gamma}$ to have the following two
properties:

(1) For any triple $(a,b,c)=g(\alpha,\beta,\gamma)g^{-1}$ that is
conjugate to $(\alpha,\beta,\gamma)$, with $g$ an element of $G$,
there is a point in $\CQ_{\alpha,\beta,\gamma}$ with $\vec \mu=(a,b,c)$.

(2) Conversely, any point $p\in \CQ_{\alpha,\beta,\gamma}$ such that
the components of $\vec\mu(p)$ commute is of this type, for some $g\in G$.

Remarkably, Kronheimer \Kronheimerii\ has constructed a family of
hyper-Kahler manifolds $\CQ_{\alpha,\beta,\gamma}$ with precisely
these properties. These manifolds are constructed as solution
spaces of Nahm's equations on the half-line $s\geq 0$ for three
$\frak g$-valued fields $X_1,X_2,X_3$:
\eqn\todo{ {dX_i\over ds}+[X_{i+1},X_{i-1}]=0, ~~i=1,2,3.}
The equations are to be solved  on the half-line $s\geq 0$
with the condition that for $s\to\infty$,
$\vec X(s)$ has a limit which is conjugate to $(\alpha,\beta,\gamma)$.
Moreover, the hyper-Kahler moment map turns out to be
\eqn\nodo{\vec\mu=(X_1(0),X_2(0),X_3(0)).}
Given these facts, it is almost a tautology to see that properties
(1) and (2) are satisfied. If $(a,b,c)$ is any commuting triple that
is conjugate to $(\alpha,\beta,\gamma)$, then the constant
solution of Nahm's equations with $(X_1(s),X_2(s),X_3(s))=(a,b,c)$
obeys Kronheimer's boundary conditions and defines a point
$p\in \CQ_{\alpha,\beta,\gamma}$.
In view of \nodo, this point obeys $\vec\mu(p)=(a,b,c)$,
as required to satisfy condition (1).
In condition (2), we are given a point $p$ such that the components
of $\vec\mu$ commute. According to \nodo, it follows that the
initial data $X_1(0)$, $X_2(0)$, $X_3(0)$ in Nahm's equations commute.
For such commutative initial data, the solution of Nahm's equations
is independent of $s$ (the solution is unique, since the equations
are first order, and an $s$-independent set of commuting matrices
does obey the equations). The boundary conditions for $s\to\infty$
are then  obeyed only if $X_1(0),X_2(0),X_3(0)$ are conjugate to $\alpha,\beta,\gamma$.
This demonstrates that condition (2) is satisfied.

We conclude that the coupling of four-dimensional super Yang-Mills
theory to a sigma model with target $\CQ_{\alpha,\beta,\gamma}$
gives the same singular behavior for the two-dimensional  fields
$A,\phi$ as the surface operator constructed in \Ramified\ with
the same parameters.  So we claim that the surface operator can be
obtained by coupling the gauge theory to the sigma model. As we
have seen, this statement depends crucially on the fact that the
coupling of the gauge theory to a hyper-Kahler manifold $\CQ$
singles out the ``good'' points in $\CQ$ at which the components of
$\vec\mu$ commute.

The example that we have just analyzed is related to orbits of
semisimple elements in complex Lie algebras.
Indeed, Kronheimer shows \Kronheimerii\ that, in one of its complex structures,
$\CQ_{\alpha,\beta,\gamma}$ is the orbit in $\frak g_\C$ of
$\xi=\alpha-i\gamma$, assuming that $\xi$ is regular.  (Otherwise,
$\CQ_{\alpha,\beta,\gamma}$ is in general a blowup of this orbit,
depending on $\beta$.)  Our next example is similarly related to
nilpotent orbits in complex Lie algebras, but we will approach it in
a more direct way.


\subsec{Another Example}\subseclab\anotherex

We take $G=SU(2)$, and we will consider a very simple example of a
hyper-Kahler manifold with the action of $G$.  In fact, we will
consider a pair of closely related examples.   One example is
$Y=\R^4$, a flat hyper-Kahler manifold with a natural action of
$SU(2)$.  We can think of this example as consisting of a single
hypermultiplet\foot{Sometimes this object is called a
half-hypermultiplet.} in the representation of $SU(2)$ that has
complex dimension two.  The second example is $Y'=\R^4/\Z_2$.

$Y$ is such a simple hyper-Kahler manifold that one might hope
that the coupling to $Y$ will make sense if any couplings of
${\cal N}=4$ super Yang-Mills to a two-dimensional hypermultiplet
make sense. The sigma model with target $Y'$ is an orbifold of the
sigma model with target $Y$, and so its coupling should make sense
if that of the first one does.

$Y=\R^4$ is completely rigid as a hyper-Kahler manifold;  $\R^4$ has
no hyper-Kahler moduli that preserve its flat structure at infinity.
So if the coupling to $Y$ makes sense, this really should give us a
rigid surface operator.

By contrast, $Y'$ has hyper-Kahler moduli -- it has a singularity at
the origin that can be deformed or resolved.  So coupling to $Y'$
should {\it not} give a rigid surface operator.

It is useful to parametrize $Y$ by four fields $y^{a\dot a}$,
$a,\dot a=1,2$, where the $SU(2)$ gauge group acts on the first
index and a second $SU(2)$, which rotates the three complex
structures of $Y$, acts on the second. We call the second group
$SU(2)'$. The reality condition obeyed by $y^{a\dot a}$ is $\bar
y_{a\dot a}=\epsilon_{ab}\epsilon_{\dot a \dot b} y^{b\dot b}$. An
expectation value of $y$ breaks $SU(2)\times SU(2)'$ to a diagonal
subgroup that we call $SU(2)''$.  The moment map at a given value of
$y$ is, of course, $SU(2)''$-invariant.  This ensures that, up to
conjugation by the original $SU(2)$ group of gauge transformations,
we can write the moment map as \eqn\tyto{\vec\mu=h\vec t,} where
$h=|y|^2$ and $\vec t$ are the $2\times 2$ Pauli sigma matrices.
(The point is that this formula is invariant under conjugation by an
element of $SU(2)$ together with an $SU(2)'$ rotation acting on the
vector $\vec\mu$.  A possible multiplicative constant in \tyto\ has
been eliminated by a choice of normalization of the flat
hyper-Kahler metric of $Y$.)

The components of $\vec t$ do not commute with each other, so if
we take literally the idea that we must restrict to points in $Y$
or $Y'$  for which the components of $\vec \mu$ commute with each
other, we must set $h=0$. This implies that $y^{a\dot
a}=\vec\mu=0$, so it means that the solutions of Hitchin's
equations have no singularity at $r=0$, and are the same as if
there were no surface operator at all.

This conclusion does not seem sensible; it seems that including
the hypermultiplet should give a surface operator that is
different from the trivial one without the hypermultiplet.  What
we think is wrong is that although the condition that the
components of $\vec \mu$ should commute with each other is only
satisfied at $y^{a\dot a}=0$, this is a singular point on the
moduli space (of points at which the components commute) since the
components of $\vec \mu$ are all quadratic in $y$.  Because of
this singularity, a proper analysis is more delicate, and we will
only give a heuristic argument.

In trying to obey the conditions that $(a,b,c)\to \vec\mu$ for
$s\to\infty$ and that $a,b,c$ should commute with each other, we
are driven to take $a,b,c$ to zero for large $s$.  If $a,b,c$ had
nonzero limits for $s\to\infty$, we would get a solution of
Hitchin's equations with a $1/r$ singularity at $r=0$.  The fact
that $a,b,c$ are driven to zero means that instead the singularity
is milder than $1/r$.  We have already encountered this situation
in section \limit\ (it is described much  more fully in \Ramified,
section 3.3), and in that context the right answer is the
following solution of Nahm's equations in which $a,b,c$ vanish for
$s\to \infty$:
\eqn\nahmsolx{ a = - {t_1 \over s+1/ f} \quad,\quad
b = - {t_2 \over s+1/f} \quad,\quad c = - {t_3 \over s+1/f }.}
We see that $a,b,c$ vanish for $s\to\infty$, but in fact they are
proportional to a multiple of the matrices $\vec t$; the multiple
vanishes for $s\to\infty$.

Our proposal is that the coupling to the hypermultiplet $Y$ or
$Y'$ leads to this type of solution of Hitchin's equation, up to
conjugation.  Thus, instead of simply claiming that $a,b,c$ vanish
at $s\to\infty$, so as to commute and equal the hyper-Kahler
moment map of $Y$ or $Y'$, we claim that generically they vanish
in this logarithmic fashion, and are proportional to an
(asymptotically vanishing) multiple of the matrices $\vec t$.

So far it does not matter very much if the hypermultiplet
parametrizes $Y$ or $Y'$.  Now let us consider $Y'$ more carefully.
In section \limit, we analyzed a surface operator described by the
singularity in \nahmsolx.  We showed that its monodromy is
generically an element of the regular unipotent conjugacy class
$\frak C'$ of $SL(2,\C)$, and is always an element of the closure
$\bar\frak C'$ of this class.  As explained in \zor, $\bar\frak C'$
is defined by the equation $\Tr\,U=2$, for $U\in SL(2,\C)$.
Explicitly, to obey $\Tr\,U=2$, we write
\eqn\trofo{U=\left(\matrix{1+a&b\cr c&1-a}\right),}
and then the condition $\det \,U=1$ (for $U\in SL(2,\C)$) gives
\eqn\rofo{a^2 + bc=0.}
This is a standard description of the ${\rm
A}_1$ singularity, and defines the complex manifold $\C^2/\Z_2$.
This complex manifold can of course be endowed with an
$SU(2)$-invariant hyper-Kahler structure, whereupon it becomes
$Y'=\R^4/\Z_2$.

It is possibly better to carry out this analysis for a nilpotent
vector $n\in \frak{sl}(2)$, rather than a unipotent element $U\in SL(2,\C)$.
If we define $n$ to be a traceless $2\times 2$ matrix
\eqn\trnok{n=\left(\matrix{a&b\cr c&-a}\right),}
then the condition $\det\, n=0$ (ensuring that $n$ is nilpotent),
gives again $a^2+bc=0$. Of course, one can map this to the conclusion
of the last paragraph by setting  $U=\exp(n)=1+n$.

Our proposal for interpreting this result is the following.
Let $\frak C$ be a unipotent conjugacy class in $G_\C$
(equivalently, a nilpotent orbit in $\frak g_\C$)
and let $\bar\frak C$ be its closure.  From this data, we
can proceed in either of two ways to define a surface operator.  To
the given unipotent orbit, we can associate an $\frak{su}(2)$
embedding $\rho:\frak{su}(2)\to\frak g$, and to this we associate a
unipotent surface operator as in section \limit. Alternatively,
following \Kronheimeri, we can use Nahm's equations to endow
$\bar\frak C$ with a hyper-Kahler structure.  Then we can define a
surface operator by coupling ${\cal N}=4$ super Yang-Mills theory to
a sigma model with target $\bar\frak C$.  Our proposal is that the
two surface operators made in this way are equivalent.

At least for our example with $SU(2)$, we can justify this
conclusion as follows.   For $G=SU(2)$, the hyper-Kahler manifold
$\CQ_{\alpha,\beta,\gamma}$ is the Eguchi-Hansen ALE manifold,
asymptotic at infinity to $\R^4/\Z_2$. For $\alpha,\beta,\gamma\to
0$, one gets the blowdown of the Eguchi-Hansen manifold, namely
$Y'=\R^4/\Z_2$. We have already proposed that coupling to the
sigma model of $\CQ_{\alpha,\beta, \gamma}$ introduces the surface
operator characterized by given $\alpha,\beta,\gamma$.
So coupling to the sigma model of $Y'$ should give the limit for
$\alpha, \beta,\gamma\to 0$ of the surface operator of generic
$\alpha,\beta,\gamma$. As we explained in section \limit, this
limit is the surface operator associated to the regular unipotent
conjugacy class.


\bigskip\noindent{\it Cover Of A Unipotent Orbit}

The surface operator associated with $Y'$ is not rigid because
$Y'$ has hyper-Kahler moduli.  By the same token, the surface
operator associated to $Y=\R^4$ should be rigid, since $Y$ has no
moduli.

We can formulate what is happening as follows.  $Y'$ is associated
to the regular unipotent conjugacy class $\frak C'$, which
topologically is $\R^4/\Z_2$ with the origin omitted.  (The origin
corresponds to $U=1$, or $n=0$.)  So $\frak C'$ has fundamental
group $\Z_2$.  And $Y'$ has the same fundamental group in the
orbifold sense.  As a result, it is possible to take a double cover
of $\frak C'$ or (after taking the closure) $Y'$, giving us
$Y=\R^4$.

$Y'$ can also be deformed to regular semisimple conjugacy classes
in $\frak g_\C$, and these are simply-connected. For $G=SU(2)$, we
can be very explicit about this.  Going back to \trnok, if we
deform the nilpotent orbit $\det\,n=0$ to a semisimple orbit
$\det\,n=\mu$, we get the equation $a^2+bc=\mu$, which defines a
smooth and  simply-connected manifold. That manifold is a
deformation of the ${\rm A}_1$ singularity; as a hyper-Kahler
deformation of $Y'$, it is the Eguchi-Hansen ALE hyper-Kahler
manifold. More generally, a regular semisimple conjugacy class in
$G_\C$ is simply-connected for any $G$. So although $\frak C'$ can
be deformed to neighboring conjugacy classes and therefore is not
rigid, the fact that $\frak C'$ has a fundamental group that the
neighboring conjugacy classes lack means that $\frak C'$ has a
cover that actually is rigid.

{}From our viewpoint of section \rigidsurf, with surface operators
constructed via singularities, it is not immediately apparent that
covers of conjugacy classes give new surface operators.  From our
present viewpoint in which unipotent surface operators are derived
by coupling to sigma models, this does seem obvious.

\bigskip\noindent{\it Semisimple Surface Operators}

However, this reasoning does {\it not} apply directly to
semisimple surface operators (or any surface operators whose
monodromy is not unipotent). The reason for this is simply that
the conjugacy class $\frak C_S$ of a semisimple element of $G_\C$
is typically {\it not} hyper-Kahler, or even complex
symplectic.\foot{This can be illustrated by considering the rigid
conjugacy classes that were important in section \rigidsemisimple.
For example, in $SO(2n+1,\C)$, consider the orbit of the element
$S={\rm diag}(1,-1,-1,\dots,-1)$, with precisely a single 1.  Such
an element is a reflection with respect to a unit vector $b$,
which is uniquely determined up to sign; the space $\frak C$ of
such $b$'s is a complexification of $\Bbb{RP}^{2n}$, equivalent
topologically to $T^*\Bbb{RP}^{2n}$.  This space does not admit an
$SO(2n+1,\C)$-invariant complex symplectic structure, as one can
show by considering the stabilizer of a point in $\frak C$ and its
action on the tangent space. So there is certainly no
$SO(2n+1,\C)$-invariant hyper-Kahler structure.} So the semisimple
surface operators of section \rigidsemisimple\ {\it cannot} be
constructed by coupling the four-dimensional gauge theory to a
sigma model. Nevertheless, it is possible to construct surface
operators associated with covers of orbits.  We explain an example
in section \rigdual.

Though rigid semisimple conjugacy classes are typically not
hyper-Kahler, surface operators associated with them rather
magically preserve all supersymmetry and therefore preserve the
hyper-Kahler nature of the moduli space of solutions of Hitchin's
equations.


\subsec{Rigid Surface Operator For The Dual Group}\subseclab\rigdual

Coupling to the sigma model with target space $Y=\R^4$ gives an
example of a rigid surface operator for $G=SU(2)$. It is the only
one we know of if one requires some minimality (see section \minimal).
Interestingly, this surface operator does not make sense for the
dual group $^L\neg G=SO(3)$, since the center of $SU(2)$ acts
nontrivially on $Y$, as a result of which $Y$ cannot be regarded
as a space with $SO(3)$ action.

Therefore, we must find a rigid surface operator for $\LG = SO(3)$
that has no simple counterpart for $SU(2)$.  Happily, we can find one
by thinking carefully about an example that arose in section
\rigidsemisimple. Let $S$ be the element of $SO(3)$
\eqn\telgox{S=\left(\matrix{1&0&0\cr 0&-1&0\cr 0&0&-1\cr}\right).}
The centralizer of $S$ is the group $O(2)$, generated by an
$SO(2)$ subgroup, whose typical element is
\eqn\elgox{\tilde S=\left(\matrix{1&0&0\cr 0&a&b\cr0& -b&a\cr}\right),
~~a^2+b^2=1,}
together with
\eqn\belgox{R=\left(\matrix{-1&0&0\cr 0&-1&0\cr 0&0&1\cr}\right).}
The orbit of $S$ is not strongly
rigid, since $S$ can be deformed to a nearby element, namely
$\tilde S$ (with $b$ and $a-1$ small) whose centralizer has the
same dimension. However, the centralizer of $\tilde S$ is a proper
subgroup of that of $S$.  The centralizer of $\tilde S$ is
$SO(2)$, since $\tilde S$ does not commute with $R$, while the
centralizer of $S$ is $O(2)$.

The result of this is that the orbit of $S$ is rigid in a weaker
sense.  It cannot be deformed to a nearby orbit, such as the orbit
of $\tilde S$. One would have to replace the orbit of $S$ by a
double cover before it could be so deformed.

Of course, we can describe the orbits explicitly.  The orbit of $S$
in the compact group $SO(3)$ is $SO(3)/O(2)=\Bbb{RP}^2$.  This has
fundamental group $\Bbb{Z}_2$ (reflecting the fact that $O(2)$ has
two components) and its double cover ${\bf S}^2=SO(3)/SO(2)$ is the
orbit of $\tilde S$.  The orbits in the complex Lie group $SO(3,\C)$
are topologically the cotangent bundles of $\Bbb{RP}^2$ and
${\bf{S}}^2$, respectively.

We can lift $S$ to $SU(2)$; it becomes
\eqn\troy{S'=\pm\left(\matrix{0 & i \cr -i & 0 \cr}\right).} The
centralizer of $S'$ is $U(1)$, which is the same as the
centralizer of a generic semisimple element of $SL(2,\C)$, so a
surface operator with monodromy $S'$ is not rigid.  That is why
this construction gives a rigid surface operator for $SO(3)$ that
has no counterpart for $SU(2)$.

\bigskip\noindent{\it A Duality Conjecture}

We can now state our first duality conjecture. We propose that the
two rigid surface operators that we have found are dual to each
other.\foot{They may be the only non-trivial rigid surface operators for
$SU(2)$ and $SO(3)$ that are minimal in a certain sense; see
section \minimal.} For $SU(2)$, we have the rigid surface operator
associated with $Y$, the double cover of a regular unipotent
orbit, and for $SO(3)$, we have the rigid surface operator
associated to the orbit of $S$.

The two surface operators are candidates for being dual to each
other because the two orbits are both of complex dimension 2. The
significance of this is as follows. When ${\cal N}=4$ super
Yang-Mills is compactified on a product of Riemann surfaces
$\Sigma\times C$, the moduli space of solutions ${\cal M}_H(C)$ of
solutions of Hitchin's equations on $C$ becomes the target space
of an effective sigma model defined on $\Sigma$. As we explained
in eqn. \zofer, including a surface operator associated with an
orbit of complex dimension $n$ (the support of the surface
operator being $\Sigma\times p$, for $p$ a point in $C$) has the
effect of increasing the dimension of ${\cal M}_H(C)$ by $n$. In
sigma models with hyper-Kahler target space, the dimension of the
target space is an invariant under all dualities. (For example, it
is proportional to the central charge.) So the number $n$ must be
duality-invariant.

The proposed dual pair for $SU(2)$ and $SO(3)$ are related in the
following elegant way.  The rigid surface operator for $SU(2)$ has
been rigidified by taking the double cover of an orbit associated
to a non-rigid surface operator.  Conversely, for $SO(3)$ the
rigid surface operator can be ``derigidified'' by taking a double
cover, as we explain next.

\bigskip\noindent{\it Taking A Cover Of A Semisimple Orbit}

A basic idea in section \anotherex\ was  that it is possible to
define a surface operator associated with a cover of a unipotent
orbit. Now we would like to explain how to define a surface
operator associated to a cover of a semisimple orbit. (We cannot
use the same approach as before because semisimple surface
operators do not arise by coupling to sigma models.) We illustrate
the idea in the context of the orbit of the element $S\in SO(3)$.

We recall that the basic idea of the construction of a semisimple
surface operator with monodromy $S$ is simply that near a
two-manifold $D$, all fields have a monodromy
\eqn\telf{\Phi(r,\theta+2\pi)=S\Phi(r,\theta)S^{-1}.} Now we
simply modify the definition by saying that we are given the
following additional data along $D$: a normalized eigenvector $v$
of the monodromy, that is a vector $v$ in the three-dimensional
representation of $SO(3)$ that obeys $Sv=v$ and is normalized to
$(v,v)=1$.  For given $S$, there are two choices of $v$.  The two
choices are exchanged by the $SO(3)$ element $R$, so making a
choice eliminates the invariance under $R$.  This has the effect
of replacing the orbit of $S$ with its double cover.

Including the choice of $v$ as part of the definition of a surface
operator gives us a new surface operator.  In general, this may
give new rigid surface operators; we will see examples in section
\dualfor.  In the present case, however, taking the double cover
gives a non-rigid surface operator.  It eliminates the discrete
symmetry that prevents us from deforming $S$ to the more generic
element $\tilde S$ of eqn. \elgox.  The surface operator
associated with the double cover of the orbit of $S$ is the limit
as $\tilde S\to S$ of a surface operator with monodromy $\tilde
S$.  Equivalently, it can be obtained from a generic surface
operator with parameters $\alpha,\beta,\gamma$ by taking
$\beta,\gamma\to 0$ and taking $\alpha$ to a value such that $S$
is equal to the monodromy $\exp(-2\pi\alpha)$.

We have explained this procedure in a special case, but the
procedure is  general.  Part of the definition of a semisimple
surface operator with monodromy $S$ is that along the support $D$ of
the surface operator, the structure group of the gauge group is
reduced to $\G^S$, the centralizer of $S$.  The case that the orbit
of $S$ is not simply-connected is the case that $\G^S$ is not
connected but has several components. In this case, we can define a
new surface operator by reducing the structure group along $D$ not
to $\G^S$ but to a subgroup of the same dimension that is a union of
some of the components of $\G^S$.  Such subgroups correspond to the
possible covers of the orbit of $S$.


\subsec{Minimal Surface Operators}\subseclab\minimal

Once we construct surface operators as in section \alternative,
by coupling to sigma models defined on a surface, we want to
impose some condition of minimality or the problem becomes too
open-ended.  The reason for this is that there are many
hyper-Kahler manifolds $\CQ$ with $G$ action.  We do not, for
example, want to allow $\CQ\to \CQ \times \CQ'$ where
$\CQ'$ is some hyper-Kahler manifold with a trivial action of $G$.

For another example, the hyper-Kahler manifolds
$\CQ_{\alpha,\beta,\gamma}$ that were described in section
\alternative\ deserve to be considered minimal because the locus
of ``good'' points at which the components of $\vec\mu$ commute is
a homogeneous space for the compact gauge group $G$.  This led in
our analysis to a surface operator with definite values of the
surface operator parameters $\alpha,\beta,$ and $\gamma$.  If one
replaces $\CQ_{\alpha,\beta,\gamma}$ with a generic hyper-Kahler
manifold with $G$ action, the locus of ``good'' points will not be
a homogeneous space for $G$ and $\alpha,\beta,\gamma$ will not
have definite values.  The effect of this will be somewhat like
promoting $\alpha,\beta,\gamma$ from coupling constants to fields.
Although this might give an interesting model, it is not what we
want to study in the present paper. For studying gauge theory with
gauge group $G$, it is reasonable to think that the basic case is
the ``irreducible surface operator'' or  ``surface
eigen-operator'' in which $\alpha,\beta,\gamma$ take definite
values.

In some sense, we want our surface operators to obey a condition of
minimality.  We do not exactly know the right technical notion of
minimality, but an approximation to it is that a minimal surface
operator has the following property. Let $x$ be a point in the
support $D$ of a surface operator. Then any chiral operator ${\cal
O}(x)$ that can be defined at $x$ is the limit of a ``bulk'' chiral
operator ${\cal O}(x')$ ($x'$ is a point not in $D$) for $x'\to x$.
The notion of a ``chiral operator'' depends on the choice of a
subalgebra of the supersymmetry algebra of ${\cal N}=4$ super
Yang-Mills theory, and the condition makes sense for any choice.

The aim is to ensure that, for a minimal surface operator
associated with a hyper-Kahler manifold $\CQ$,  $\CQ$ is related to a
single orbit of $G$.  For example, if $G=SU(2)$ and $\CQ$ is
parametrized by a single hypermultiplet in the two-dimensional
representation, then the associated surface operator is minimal by
the above criterion.  But if $\CQ$ is parametrized by two or more
such hypermultiplets, then one can construct\foot{If one
parametrizes the hypermultiplet in one of its complex structures
by a pair of chiral superfields $C^a$, $a=1,2$, then given also a
second such hypermultiplet $\tilde C^a$, one can form the chiral
operator $\epsilon_{ab}C^a\tilde C^b$.}  chiral operators
supported only on $D$, so the associated surface operator is not
minimal by the above criterion.

However, the condition that we have stated may be too strong.  It
is only intended as an approximation to a good notion of
minimality.  The criterion that we stated does have the virtue of
being duality-invariant.


\newsec{Fingerprints of Surface Operators}\seclab\invariants

Our main goal in the rest of this paper is to learn how rigid
surface operators transform under S-duality of the $\CN=4$ super
Yang-Mills theory. To this end, in this section we discuss several
characteristics of the corresponding conjugacy classes which are
expected to be duality invariant.

\subsec{Invariant Polynomials}

To start with, we consider the set of gauge-invariant polynomials
$P(\varphi(x))$ of the Higgs field $\varphi(x)$.  Such polynomials
are half-BPS local operators in the $\CN=4$ super Yang-Mills
theory.  If $P$ is homogeneous of degree $d$, the corresponding
operator is a superconformal primary of dimension $d$.

Once one picks an invariant quadratic form on the Lie algebra $\frak
g$ (corresponding physically to a choice of the gauge coupling),
there is a natural map from a $G$-invariant polynomial $P:\frak g\to
\Bbb{C}$ to an $^LG$-invariant polynomial $\tilde P:{}^L\neg\frak
g\to \Bbb{C}$.  If $G$ is simply-laced, the two Lie algebras
coincide and this map is the trivial one.  To define the map in
general, one uses the fact that $G$-invariant polynomials on $\frak
g$ correspond naturally to Weyl-invariant polynomials on $\frak t$,
the Lie algebra of a maximal torus $\TT \subset G$.  But $G$ and
$^L\neg G$ have the same Weyl group $\Weyl$, and one can define a
Weyl-invariant map from $\frak t$ to $^L\neg \frak t$, taking a
short coroot to a long coroot and a long coroot to a multiple of a
short coroot. (For example, see the appendix to \Ramified\ for
further detail. The map from $\frak t$ to $^L\neg\frak t$ is unique
up to a constant factor that depends on the gauge coupling and is
not relevant for what follows.)

In ${\cal N}=4$ super Yang-Mills theory, $S$-duality sends a local
operator $P(\varphi)$ to the corresponding local operator  $\tilde
P(\varphi)$. This can be demonstrated by first showing explicitly
that it is true in the abelian case and then by reducing the general
case to the abelian case by Higgsing. (For this, one studies the
theory on its Higgs or Coulomb branch, in a vacuum in which the
gauge group $G$ is spontaneously broken to the abelian subgroup
$\TT$. Then the appropriate transformation of half-BPS operators in
the underlying $G$ theory can be determined by computing what
happens in the effective theory with gauge group $\TT$.)


\bigskip\noindent{\it The Dimension}

If a surface operator is related to a conjugacy class $\frak
C\subset G_\C$ (as are all surface operators considered in the
present paper),  then the most basic invariant of this surface
operator is the dimension of $\frak C$.  Indeed, as we explained
in section \rigdual\ (see also eqn. \zofer), in compactification to two
dimensions, including a surface operator associated with $\frak C$
has the effect of increasing the dimension of the moduli space of
solutions of Hitchin's equations by ${\rm dim}\,\frak C$.
 Since the dimension of the target space of a sigma model
 is a duality invariant, it follows that $\dim (\frak C)$  is a duality invariant.

For unipotent surface operators constructed via $\frak{su}(2)$
embeddings $\rho: \frak{su}(2) \to \frak{g}$, the dimension of the
corresponding nilpotent orbit $\frak c$ can be computed as in
section \searching\ by decomposing $\frak g$ into irreducible
representations of $\frak{su}(2)$. Then the total number of
irreducible summands in the decomposition \tofo\ gives the
dimension $s$ of the centralizer $\G^n_{\C}$ of a nilpotent
element $n \in \frak c$, and the dimension of the nilpotent orbit
$\frak c$ follows from the formula $\dim \frak c = \dim ( G_{\C} )
- s$.  This also gives the dimension of the conjugacy class $\frak
C$ containing $U=\exp(n)$.

As we explained in section \limit, for classical groups of type $A$,
$B$, $C$, and $D$, homomorphisms $\rho:\frak{su}(2)\to \frak g$ or,
equivalently, nilpotent orbits $\frak c \subset \frak g_{\C}$ are
labeled by partitions $\la = [\la_1, \la_2, \ldots, \la_k]$, where
each part $\la_i$ represents the size of the $i$-th Jordan block.
(The general result is summarized at the end of section \limit.) For
each of these classical groups, it is straightforward to compute the
dimension $s$ of the centralizer $\G^n_{\C}$ of a nilpotent element
$n \in \frak c$ associated with a given partition, using eqn. \tofo.
One obtains a simple formula for the dimension of the nilpotent
orbit $\frak c$ in terms of $\la_i$ (see \CMcGovern, section 6.1):
\eqn\cdimabcd{\eqalign{
(A_N):\quad\quad
&\dim ( \frak c_{\la} ) = (N+1)^2 - \sum_i |\{ j \vert \la_j \ge i\}|^2 \cr
(B_N):\quad\quad
&\dim ( \frak c_{\la} ) = 2N^2 + N - {1 \over 2} \sum_i |\{ j \vert \la_j \ge i\}|^2
+ {1 \over 2} \sum_{i {\rm ~odd}} |\{ j \vert \la_j = i\}| \cr
(C_N):\quad\quad
&\dim ( \frak c_{\la} ) = 2N^2 + N - {1 \over 2} \sum_i |\{ j \vert \la_j \ge i\}|^2
- {1 \over 2} \sum_{i {\rm ~odd}} |\{ j \vert \la_j = i\}| \cr
(D_N):\quad\quad
&\dim ( \frak c_{\la} ) = 2N^2 - N - {1 \over 2} \sum_i |\{ j \vert \la_j \ge i\}|^2
+ {1 \over 2} \sum_{i {\rm ~odd}} |\{ j \vert \la_j = i\}|
}}

For example, for $G=SU(2)$ the regular nilpotent orbit discussed in
section \limit\ is an orbit of a nilpotent element $n$ that consists
of a single Jordan block of size 2. As we learned there, the closure
of the corresponding unipotent conjugacy class is the familiar
$\Z_2$ orbifold singularity,
$$
\bar \frak C_{[2]} = \C^2 / \Z_2
$$
It has complex dimension 2, in agreement with \cdimabcd\ for
$G=A_1$ and $\la = [2]$. For $\LG = SO(3)$ the same unipotent
conjugacy class would be labeled by $\la = [3]$, and eqn.
\cdimabcd, now with $G=B_1$, also gives $\dim \frak C = 2$. In
symplectic groups, this example is a special case of a
hyper-Kahler $\Z_2$ orbifold \hoto, the closure of a minimal
nilpotent orbit in $C_N$ labeled by $\la = [2,1^{2N-2}]$. As we
explained in section \searching, it is strongly rigid for $N > 1$,
and has complex dimension $2N$, as is obvious from \hoto. This is
in perfect agreement with \cdimabcd, which gives
$$
\dim \frak C_{[2,1^{2N-2}]} = 2N^2 + N - {1 \over 2} (2N-1)^2
- {1 \over 2} - {1 \over 2} (2N-2) = 2N
$$

The dimension of an orbit is an important invariant, but it is not
enough, since many rigid surface operators may be associated with
orbits of the same dimension.  We will next consider a much more
refined invariant.


\subsec{Polar Polynomials}\subseclab\polar

In the presence of any of the surface operators described in
sections 2 and 3, the Higgs field has a singularity.  For example,
in the presence of a unipotent surface operator at $z=0$, we have
\eqn\higgssing{ \varphi = {n \over z} + \ldots }
where $n$ takes values in a prescribed nilpotent orbit $\frak c$,
and the ellipses refer to regular terms.  There is an
analogous expression near a semisimple surface operator, as we
discuss presently.

Given such a singularity in $\varphi$, the invariant polynomials
$P(\varphi)$ also generally have singularities.  Precisely because
$n$ is nilpotent, the singular terms in $P(\varphi)$ are not
determined\foot{Any coefficient in $P(\varphi)$ that is determined
by the surface operator would be  a variable parameter, and a
surface operator with such a parameter would not be rigid. See the
next footnote for an example.} by the singularity in $\varphi$, but
rather depend on the regular part of $\varphi$, the part indicated
by ellipses in eqn. \higgssing.

The moduli space ${\cal M}_H$ of solutions of Hitchin's equations
has a Hitchin fibration $\pi:{\cal M}_H\to \cmmib B$, where
$\cmmib B$ is parametrized by the values of the invariant
polynomials $P(\varphi)$.  If including a surface operator
increases the dimension of ${\cal M}_H$ by $d={\rm dim}\,\frak C$,
then it must increase the dimension of $\cmmib B$ by $d/2$. This
means that, looking at all possible choices of $P$, there will be
precisely $d/2$ independent complex parameters, characterizing the
singularities of the polynomials $P(\varphi)$, that depend on the
regular terms not written explicitly in eqn. \higgssing. The
pattern of these singularities gives us a rather refined invariant
of a  surface operator that we will call its ``fingerprint.''

Let us illustrate this with a simple example. We consider the
(non-rigid) surface operator associated with the regular nilpotent
orbit of $SL(2,\C)$, that is the orbit $\frak c$ of a nilpotent
element
\eqn\nilsutwo{ n = \pmatrix{ 0 & 1 \cr 0 & 0} }
As explained in \Ramified\ and reviewed in section \limit, this
surface operator can be regarded as a limit of a ``generic'' surface
operator in gauge theory with $G = SU(2)$ as $\a,\b,\g \to 0$.
Supposing that such a surface operator is inserted at $z=0$, the
behavior of $\varphi$ near $z=0$ is
\eqn\ilsut{\varphi=n/z+m+m'z+\dots,} with $m,m'\in \frak{sl}(2)$.
For $G=SU(2)$, the only independent invariant polynomial is
\eqn\ilst{P(\varphi)=\Tr\,\varphi^2={2\,\Tr \,nm\over z}+\dots,}
where the ellipses denote regular terms.  We see that the
singularity of $P(\varphi)$ is characterized by exactly one
coefficient, in agreement with the fact that ${1 \over 2} \dim \frak c = 1$.
Moreover, this coefficient depends on the square-integrable part of
$\varphi$, which is free to fluctuate, so it is not a parameter of
the surface operator; rather, we can regard it as one of the
coordinates that parametrizes the base $\cmmib B$ of the Hitchin
fibration, in the presence of a  surface operator whose definition
depends only on the choice of orbit $\frak c$.\foot{If we deform
this unipotent but non-rigid surface operator to a more general one
with $\varphi=\sigma/z+\dots,$ where now $\sigma=(\beta+i\gamma)/2$
takes values in a prescribed semisimple conjugacy class, we will
get $\Tr\,\varphi^2=a/z^2+b/z+\dots$, where now $a=\Tr\,\sigma^2$ is
a parameter of the surface operator, and $b$ depends on the regular
part of $\varphi$ and is a function on $\cmmib B$.}

We can extract from eqn. \ilst\ the statement that
$\Tr\,\varphi^2$ has a simple pole at $z=0$,
\eqn\higgsaone{ \Tr \varphi^2 \simeq {p \over z} + \dots}
This statement must be invariant under duality.  The assertion
that $p$ can be defined as $2\,\Tr\,nm$, with $n$ and $m$ as
above, has no reason to be preserved by duality.  For $G=SU(2)$,
the ``fingerprint'' of this particular type of surface operator is
simply the statement that the singularity of $\Tr\,\varphi^2$ is a
(variable) simple pole.

The regular unipotent conjugacy class $\frak C_{{\rm reg}}$ of $SL(2,\C)$
is not rigid, however. Closer to the subject of the present paper is
the double cover $\tilde{\frak C}_{{\rm reg}}$, which, as we
explained in section \anotherex, is weakly rigid. Since taking the
cover does not change the invariant polynomials, the weakly rigid
surface operator associated with $\tilde{\frak C}_{{\rm reg}}$ has
the same polar polynomials, namely \higgsaone. In section
\rigdual, we proposed that the dual to this weakly rigid surface
operator is a surface operator associated with a semisimple
conjugacy class of the element \telgox\ in $\LG = SO(3)$:
\eqn\telgoxx{S=\left(\matrix{1&0&0\cr 0&-1&0\cr 0&0&-1\cr}\right).}
The conjugacy class $\frak C_S$ is also weakly rigid since nearby
conjugacy classes have strictly smaller stabilizers.

 As a test of
our duality proposal, we will verify that the surface operator
associated with the weakly rigid semisimple conjugacy class $\frak
C_S$ leads to the same simple pole in $\Tr\,\varphi^2$. Let us
remember the definition of the semisimple surface operator with
monodromy $S$. The fields must all have monodromy $S$ around $z=0$.
In particular, the expansion of the $\frak{so}(3)$-valued field
$\varphi$ looks like (see Appendix B for the Lie algebra
conventions):
\eqn\bonehiggsi{ \varphi = z^{-1/2}
\pmatrix{ 0 & a & b \cr - b & 0 & 0 \cr - a & 0 & 0 }
+ \pmatrix{0&0&0\cr 0&c&0\cr 0&0&-c\cr}, }
where $a,b,$ and $c$ are single-valued functions of $z$ (to get
the right monodromy) and regular at $z=0$ (so that $\varphi$ is
square-integrable). So
\eqn\onehig{\Tr \varphi^2=-{4ab \over z}+\dots}
with the same simple pole as before, though of course
with a different interpretation of the residue of the pole.
This is in perfect agreement with the proposed duality between
the weakly rigid conjugacy classes
$\tilde{\frak C}_{{\rm reg}}$ and $\frak C_S$.

Before describing any systematic theory, we will consider by hand
another, more representative example. This example involves a pair
of strongly rigid orbits in $B_3$ and $C_3$.
As we explained in section \combining, strongly rigid orbits in
classical groups of type $B$ and $C$ can be conveniently labeled
by the proper subset of simple roots $\Theta_i \subset \bar
\Delta$ and a pair of partitions $(\la', \la'')$. In this
notation, in type $B_3$ there is a strongly rigid semisimple
conjugacy class of dimension 6 which corresponds to the root
system $\Theta_3 = D_3$ and the partitions $\la' = [1^6]$ and
$\la'' = [1]$,
\eqn\rigidbthree{  \frak C^{D_3}_{([1^6], [1])}  }
According to the general formula \rigidsselt, in gauge theory it
corresponds to a rigid surface operator, with the holonomy of the
$G_{\C}$-valued gauge connection $\CA = A + i \phi$ conjugate to
the rigid semisimple element
$$
S_3 = {\rm diag} \left( +1,-1,-1,-1,-1,-1,-1 \right)
$$
which breaks the gauge group $G = SO(7)$ down to a subgroup
\eqn\bthreesymmbreaking{ O(6) \subset SO(7), }
as in eqn. \bnsplit.  The conjugacy class of the element $S_3$ is
thus $SO(7)/O(6)$, and so has dimension 6.

On the other hand, in $C_3$ there is also a strongly rigid
conjugacy class of dimension 6, namely the unipotent conjugacy
class labeled by the partition $\la'' = [2,1,1,1,1]$. In gauge
theory, it corresponds to a singularity of the Higgs field with a
single Jordan block of size 2 (corresponding to the part ``$2$''
in the partition $\la'' = [2,1,1,1,1]$),
\eqn\cthreehiggsi{ \varphi = {1 \over z} \pmatrix{ 0 & 1 &0  &0 &0
& 0\cr 0 & 0 &  0& 0 &0  & 0 \cr 0 & 0 & 0 & 0 & 0 & 0 \cr
 0& 0 & 0 & 0 & 0 & 0 \cr
 0& 0 & 0 & 0 & 0 & 0 \cr
 0& 0 & 0 & 0 & 0 & 0 } + \ldots}
In the notation of section \rigidsemisimple, this unipotent
conjugacy class corresponds to $\Theta_0 = \Delta$ and similarly to
\rigidbthree\ can be denoted
\eqn\rigidcthree{  \frak C^{C_3}_{(\emptyset, [2,1^4])}  }
The fact that both conjugacy classes $\frak C^{D_3}_{([1^6], [1])}$
and $\frak C^{C_3}_{(\emptyset, [2,1^4])}$ are rigid and have the same dimension
strongly suggests that they might be related by electric-magnetic duality.

Further evidence for this comes from comparing the fingerprints of
these rigid conjugacy classes.  In both cases, we find the
following behavior of the Higgs field:
\eqn\higgbcthree{\eqalign{ & \Tr \varphi^2 \simeq {p_2 \over z} +
\ldots \cr & \Tr \varphi^4 \simeq {p_4 \over z^2} + {p_1 \over z}
+ \ldots \cr & \Tr \varphi^6 \simeq {p_6 \over z^3} + {p_3 \over
z^2} + {p_5 \over z} + \ldots }}
Since the orbits in question are six-dimensional, the singularity in
the invariant polynomials of $\varphi$ should depend on only $3=6/2$
coefficients.  Accordingly, the 6 parameters $p_1,\dots, p_6$ will
obey 3 relations.  It turns out that one gets the same 3
relations in the two cases\foot{In making this comparison, one needs
a fact explained in Appendix B.  The $S$-duality transformation from
$SO(2N+1)$ gauge theory to $Sp(2N)$ gauge theory maps $\Tr\,
\varphi^{2k}$, with the trace in the $(2N+1)$-dimensional
representation of $\frak{so} (2N+1)$, precisely to $\Tr\,
\varphi^{2k}$ with the trace in the $2N$-dimensional representation
of $\frak{sp} (2N)$.}:
\eqn\pppthree{\eqalign{ & p_4 = {1 \over 2} (p_2)^2 \cr & p_6
= {1 \over 4} (p_2)^3 \cr & p_3 = {3 \over 4} p_1 p_2 }}
The first two relations are manifestly invariant under a
redefinition of the local complex parameter near the surface
operator. It is easy to verify that the last relation is also
invariant, given the first two. Indeed, for $z = w + \epsilon w^2
+ \ldots$, we have ${1 \over z^k} \simeq {1 \over w^k} - {k
\epsilon \over w^{k-1}} + \ldots$, so that
\eqn\ppp{\eqalign{ & p_1 \to p_1 - 2 \epsilon p_4 = p_1 - \epsilon
(p_2)^2 \cr & p_3 \to p_3 - 3 \epsilon p_6 = p_3 - {3 \over 4}
\epsilon (p_2)^3 }}
The first transformation implies $p_1 p_2 \to p_1 p_2 - \epsilon
(p_2)^3$, which together with the second transformation in \ppp\
demonstrates that $p_3 - {3 \over 4} p_1 p_2$ is indeed an
invariant combination.

We will now explain how to find this structure. To compute these
polar polynomials in a theory with gauge group $G = SO(7)$, it is
convenient to pass to a double cover of the $z$-plane
(parametrized by a local coordinate $y = z^{1/2}$) and to describe
this surface operator by a singularity with a generic first-order
pole consistent with the gauge symmetry breaking
\bthreesymmbreaking:
%
%
%
\eqn\bthreehiggsi{
\varphi = y^{-1} \pmatrix{ 0 & a & b \cr - b^t & 0 & 0 \cr - a^t & 0 & 0 } + \ldots}
Here, $B=(a,b)$ is a generic 6-vector (the upper left and lower right
blocks are here $1\times 1$ and $6\times 6$ matrices). From this
starting point, one can compute $\Tr\,\varphi^{2k}$ by adding less
singular terms with the same structure as in \bonehiggsi\ (even
powers of $y$ for diagonal blocks, odd powers for off-diagonal
blocks). On the other hand, in the dual theory with $\LG = Sp(6)$,
the polar polynomials \higgbcthree\ can be computed directly, by
adding a generic regular term to the singularity \cthreehiggsi\
and evaluating $\Tr \varphi^{2k}$.   In either of these two cases,
it is immediate to see that $\Tr\,\varphi^{2k}$ has a pole of
order $k$ at $z=0$, giving the general form in \higgbcthree.  It
is also easy in each case to verify the first two relations in
\pppthree; for this it suffices to compute the coefficient of the
leading singularity $z^{-k}$ in $\Tr\,\varphi^{2k}$, $i=1,2,3$. It
is a little harder to verify the coefficient of $3/4$ in the last
relation in \pppthree.  Actually, as we have observed, this
coefficient is determined by reparametrization invariance.


\bigskip\noindent{\it Kazhdan-Lusztig Map}

It quickly becomes cumbersome to describe detailed relations among
polynomials as in the above example.  An alternative point of view
is useful.
To explain this point of view, we go back to the $SU(2)$ example
of eqn. \ilsut. It is convenient to write
\eqn\telme{\varphi(z)=\pmatrix{ * & z^{-1}+*\cr m & * \cr},}
where all we need to know about the matrix elements
denoted $*$ is that they are regular at $z=0$.
For any non-zero $z$, $\varphi(z)$ is regular semisimple,
and thus can be conjugated to $\frak t_\C$.

We recall that for $G_\C=SL(2,\C)$, $\frak t_\C$ is the abelian Lie algebra
of traceless diagonal $2\times 2$ matrices ${\rm diag}(\xi,-\xi)$.
For $z\not=0$, $\varphi(z)$ is conjugate to
\eqn\zelme{\varphi_D=\pmatrix{\pm \sqrt{ m/z} & 0 \cr 0 & \mp \sqrt{m/z}}.}
(This formula is exact if the matrix elements denoted $*$ in \telme\
are set to zero, and in general is valid near $z=0$.)
Of course, the diagonalization of $\varphi$ is not quite
unique -- it is only unique up to a Weyl transformation,
with the Weyl group $\Weyl$ acting by exchange of
the two eigenvalues or equivalently multiplication by $-1$.
Moreover, we see in eqn. \zelme\ that as $z$ circles around
the origin in the punctured complex $z$-plane, $\varphi_D$
undergoes a monodromy. Inevitably, the monodromy element
is a Weyl transformation, in this case the unique nontrivial
element of the Weyl group of $SU(2)$.

This procedure can be carried out for every surface operator.
Regardless of what singularity or monodromy we require $\varphi$
to have at $z=0$, $\varphi$ is generically regular semisimple for
$z\not=0$.  Hence it can be diagonalized for $z\not=0$, but this
diagonalization is only unique up to a Weyl transformation. Making
a specific choice of how to diagonalize $\varphi$, we get a $\frak
t_\C$-valued ``function'' $\varphi_D(z)$, whose monodromy around
$z=0$ is an element of $\Weyl$.  The element $w\in \Weyl$ that
arises for a generic choice of $\varphi$ is an invariant of the
surface operator.

This invariant is a compact way to describe the ``fingerprint'' of
a surface operator.   Let us explain explicitly for $G=SU(N)$ how
a conjugacy class in $\Weyl$ determines the fingerprint defined in
terms of polar parts of invariant polynomials $P(\varphi)$. The
case of any $G$ is similar.  Consider first the case that $w$ is a
cyclic permutation of the $N$ eigenvalues
$\xi_1,\dots,\xi_N$ of a $\frak t_\C$-valued function
$\varphi_D(z)$. Then $w$-invariance and square-integrability of
$\varphi$ (which implies that the eigenvalues are less singular
than $1/z$) tells us that
\eqn\teefro{\varphi_D(z)=\sum_{m=1}^{N-1} c_m(z) b^{(m)}(z),}
where the functions $c_m(z)$ are regular at $z=0$, and the matrices
$b^{(m)}$ are
\eqn\torty{b^{(m)}(z)=z^{-m/N}{\rm diag}(1,\omega^k,\omega^{2k},\dots,\omega^{(N-1)k}),}
where $\omega=\exp(2\pi i/N)$. (For $N=2$, this reduces to our
example \zelme\ for $G=SU(2)$.   In general, it is invariant under
monodromy in $z$ plus cyclic permutation of eigenvalues.) Clearly,
given this data, we can compute the behavior of the invariant
polynomials $\Tr\,\varphi^k$ for generic functions $c_m(z)$ and thus
describe the ``fingerprint'' of the surface operator.  The extension
to any Weyl conjugacy class is as follows. A general element $w$ of
the Weyl group is obtained from a partition
$N=\la_1+\la_2+\dots+\la_k$. One divides the $N$ eigenvalues of
$\varphi_D$ in $k$ different groups, with $\la_i$ elements in the
$i^{th}$ group; $w$ acts in each group as a cyclic permutation of
the $\la_i$ eigenvalues.   Then $\varphi_D(z)$ is a direct sum of
blocks, the $i^{th}$ block looking just like \teefro, with $N$
replaced by $\la_i$.  Again, this leads to a determination of the
``fingerprint'' starting from a conjugacy class in $\Weyl$.

So instead of describing the fingerprint of a surface operator in
terms of the polar behavior of functions $P(\varphi)$, and
relations among their coefficients, we can summarize this
information by specifying a conjugacy class in $\Weyl$.

Since our surface operators are related to conjugacy classes in
$G_\C$, we really have here a map from  conjugacy classes in
$G_\C$ to conjugacy classes in $\Weyl$. This map is known as the
Kazhdan-Lusztig map. It was originally defined \KLmap\ as a map
that assigns a conjugacy class of $\Weyl$ to a nilpotent orbit in
$\frak g_\C$. (For us, this definition corresponds to the case of
a unipotent surface operator.)  Later, it was extended by
Spaltenstein \SpaltensteinKL\ to a map from nilpotent orbits in
parabolic subalgebras to conjugacy classes in $\Weyl$. This more
general version is exactly what we need for identifying the
fingerprints of a general surface operator with monodromy $V=SU$,
$S$ being semisimple and $U$ unipotent.  We will make use of
these mathematical results in section \families.


\subsec{Center vs. Topology}\subseclab\centertopology

We will describe one other type of invariant of a surface
operator. First we recall some basic observations of \Ramified\
about center, topology, and surface operators.

Let $\CZ$ or $\CZ(G)$ be the center of $G$.  A very elementary but
important example of a surface operator (which moreover is rigid) is
obtained by twisting by an element $\frak z\in \CZ$.  We simply make
the construction of eqn. \modcon, but setting $S=\frak z$.  This
gives a strongly rigid surface operator since the conjugacy class of
a central element, as it consists only of a single point, has
smaller dimension than any noncentral conjugacy class.

Such a surface operator must have a dual, of course. The center of
$G$ is the Pontryagin dual\foot{The Pontryagin dual of a finite
group $F$ is $F^\vee={\rm Hom}(F,U(1))$.} to the fundamental group
of $^L\neg G$, and vice-versa:
\eqn\centerpione{\eqalign{ \pi_1 (\LG) & \cong \CZ (G)^\vee \cr
\CZ (\LG) & \cong \pi_1 (G)^\vee .}}
 In gauge theory with gauge group $^L\neg G$ on a four-manifold
$M$, a basic ingredient is an $^L\neg G$ bundle $^L\neg E\to M$. It
has a characteristic class $\xi\in H^2(M, \pi_1({}^L\neg G))$. (For
example, $\xi$ is the second Stieffel-Whitney class if $^L\neg
G=SO(3)$.) Now let $D$ be a closed two-manifold in $M$, and
\eqn\trut{\hat f: \pi_1({}^L\neg G)\to U(1)} a homomorphism.  We
denote this homomorphism as $\psi\to \exp(i\langle f,\psi\rangle)$,
for $\psi \in \pi_1({}^L\neg G)$ and $f$ in a sense the logarithm of
$\hat f$. We can modify $^LG$ gauge theory by including in the path
integral a factor
\eqn\pefro{\Psi_{\hat f}=\exp\left(i\left\langle f,\int_D\xi\right\rangle\right).}

This modification gives a surface operator for every choice of the
homomorphism $\hat f$ in \trut.  But according to \centerpione, the
homomorphism  $\hat f:\pi_1({}^L\neg G) \to U(1)$ corresponds
naturally to an element of $\CZ(G)$. The proposal in \Ramified\ is
that the surface operator in $G$ gauge theory whose monodromy is a
central element $\frak z$ is dual to the surface operator in $^L\neg
G$ gauge theory associated with the corresponding homomorphism $\hat
f$.

\bigskip\noindent{\it Generalization}

So far nothing is new.  Now let us repeat this story beginning with
a surface operator associated with a conjugacy class $\frak C$ in
gauge theory with gauge group $G$. Let $V \in \frak C$ be the
monodromy around the singularity of the complexified gauge field
$\CA=A + i \phi$.

Let $\frak z$ be an element of $\CZ$. Whatever the original surface
operator may be, we can, using the reasoning of section \rigidsurf,
construct a new one with monodromy $ \frak z V$. Of course, it may
happen that $V$ and $\frak z V$ are conjugate. In this case,
multiplying by $\frak z$ gives nothing new.  This happens not
infrequently if $V$ is semisimple. But if $V$ and $\frak z V$ are
not conjugate
---  which is always the case\foot{Since $\frak z$ is central,
it acts in any irreducible representation $R$ of $G$ as a multiple
of the identity, say $\bar\frak z$.  Pick a representation $R$ for
which $\bar\frak z\not=1$. If $V$ is unipotent, its only eigenvalue
in any representation is 1 (such a $V$ typically is not completely
diagonalizable). Likewise, the only eigenvalue of $\frak z V$ is
$\bar\frak z$.  So $\frak z V$ and $V$ are not conjugate.} if $V$ is
unipotent and $\frak z\not=1$
--- then the twist by $\frak z$ gives a new surface operator. If
this happens for all nontrivial elements of $\CZ$, we say that we
can ``observe the center of $G$'' for this particular surface
operator.

In general, let $\CZ^V$ or $\CZ(G)^V$ be the subgroup of $\CZ$
defined by saying that $\frak z V$ is conjugate to $V$ for $\frak
z\in \CZ^V$. Then we can observe the quotient $\CZ/\CZ^V$, in the
sense that this quotient group parametrizes new surface operators
that we can make by twisting by an element of the center of $G$.

Of course, there is a dual to this, as follows. Along the support
$D$ of a surface operator, say in gauge theory with gauge group
$^L\neg G$, the structure group of the gauge bundle is reduced from
$^L\neg G$ to a group $H$ that is a symmetry of the surface
operator. ($H$ is the subgroup of $^L\neg G$ that commutes with $V$
and with the relevant $\frak{su}(2)$ embedding if $V$ is not
semisimple.) The inclusion $\iota:H\to {}^L\neg G$ determines a
natural homomorphism $\iota:\pi_1(H)\to \pi_1({}^L\neg G)$ and hence
a subgroup $\iota(\pi_1(H))\subset \pi_1({}^L\neg G)$. When
integrated over $D$, the characteristic class $\xi$ of the gauge
bundle actually takes values in this subgroup.  This reflects the
fact that $H$ is the effective gauge group along $D$.

Hence, when we modify the path integral by including the factor
$\Psi_{\hat f}=\exp\left(i\langle f,\int_D\xi\rangle\right),$ so
as to build a new surface operator, some choices of $\hat f$ are
irrelevant. We are not interested in those $\hat f$ that are
trivial when restricted to $\iota(\pi_1(H))$. So in
$\pi_1({}^LG)^\vee$, there is an irrelevant subgroup
$\tilde\pi_1({}^L\neg G)^\vee$, which classifies homomorphisms
that are trivial on $\iota(\pi_1(H))$. We say that we can observe
the topology if $\iota(\pi_1( H))=\pi_1({}^L\neg G)$, so that
$\tilde\pi_1({}^L\neg G)^\vee$ is trivial and a twist by any
factor $\Psi_{\hat f}$ gives a new surface operator.

In general, the surface operators that we can make in $^L\neg G$
gauge theory by twisting by some $\hat f$ are classified by
$\pi_1({}^L\neg G)^\vee/\tilde\pi_1({}^L\neg G)^\vee$. By
contrast, the possible twists in $G$ gauge theory by an element of
the center were classified by $\CZ(G)/\CZ(G)^V$. Since
$\pi_1({}^L\neg G)^\vee=\CZ(G)$ according to \centerpione, we see
that duality requires \eqn\torf{\CZ(G)^V=\tilde\pi_1({}^L\neg
G)^\vee.} One special case is that if on one side we can observe
the full center of $G$ -- that is if $\CZ(G)^V$ is trivial -- then
on the other side we can observe the full topology, meaning that
$\iota(\pi_1(H))=\pi_1({}^L\neg G)$ (so that a homomorphism that
annihilates $\iota(\pi_1(H))$ is trivial). At the other extreme,
the center and topology are completely invisible (for the chosen
pair of dual surface operators) if $\CZ(G)^V=\CZ(G)$, or dually if
$\iota(\pi_1(H))$ is trivial and all homomorphisms annihilate it.

Let us see how this works out for the (noncentral) rigid surface
operators for the pair of groups $SO(3)$ and $SU(2)$.
We recall that for $SU(2)$, the rigid surface operator in question
is associated with the double cover of the nilpotent cone.
It has unipotent monodromy and is associated with an irreducible
$\frak{su}(2)$ embedding.  The dual rigid surface operator for
$SO(3)$ is associated with the conjugacy class of the element
$S={\rm diag}(1,-1,-1)$.  The essential case to consider is
$G=SU(2)$, $^L\neg G=SO(3)$.  (The opposite case $G=SO(3)$, $^L\neg
G=SU(2)$ is of little interest, as $SO(3)$ has trivial center and
$SU(2)$ is simply-connected.)

We can analyze this example as follows. Since the monodromy of the
$G$ surface operator is unipotent, we can observe the center of $G$;
in other words, $\CZ(G)^V$ is trivial for this surface operator.
Dually the group $H$ is the subgroup $O(2)\subset SO(3)$. Any loop
in $SO(3)$ can be deformed to a loop in $SO(2)\subset O(2)$, so
$\iota(\pi_1(H))=\pi_1({}^L\neg G)$; hence we can observe the
topology of ${}^L\neg G$.


\newsec{Duality for Strongly Rigid Surface Operators}\seclab\dualfor

Now, we use the invariants described in the previous section in
order to identify dual pairs of rigid surface operators in theories
with $SO$ and $Sp$ gauge groups, starting with the simplest examples
of small rank. To keep things simple, we will analyze only rigid
surface operators that are associated to strongly rigid conjugacy
classes.  We know that these do not give the whole story, even for
$SU(2)$ and $SO(3)$; in section \rigdual, we described apparently
dual rigid surface operators for these groups that are not
associated to strongly rigid conjugacy classes.

However, strongly rigid conjugacy classes are relatively easy to
analyze, and this analysis will give many examples of what appear to
be dual pairs.  We begin with rank 2, the first case in which
non-trivial strongly rigid conjugacy classes exist.


\subsec{Duality for $G = SO(5)$ and $\LG = Sp(4)$}\subseclab\bctwo

Strongly rigid conjugacy classes give rigid surface operators in
either the adjoint or the simply-connected form of the group. That
being so, the comparison of $B_2$ and $C_2$ may appear trivial,
since these groups are the same. However, as we will see, even in
this case, $S$-duality identifies strongly rigid surface operators
in a nontrivial way. First, let us describe strongly rigid surface
operators in these theories.

In each case, there is only one strongly rigid unipotent surface
operator, which in the $SO(5)$ (resp. $Sp(4)$) theory corresponds to
a rigid unipotent conjugacy class $\frak C_{\la}$ with $\la =
[2,2,1]$ (resp. $\la = [2,1,1]$). These rigid unipotent conjugacy
classes in $B_2$ and $C_2$ are both of dimension 4, so that one
might naively expect that duality maps strongly rigid surface
operators associated with these unipotent conjugacy classes into
each other. However,  by studying the polar polynomials, one can
show that this is not the case.

As we explain in Appendix B, the $S$-duality map between $Sp(2N)$
gauge theory  and $SO(2N+1)$ gauge theory maps $\Tr\, \varphi^{2k}$,
with the trace in the $2N$-dimensional representation of $Sp(2N)$,
to $\Tr \,\varphi^{2k}$ with the trace in the $(2N+1)$-dimensional
representation of $SO(2N+1)$. On the other hand, for the rigid
surface operators associated with unipotent conjugacy classes
labeled by $\la = [2,2,1]$ and $\la = [2,1,1]$ we find that, in both
cases, the invariant polynomials $\Tr\, \varphi^{2k}$ have the same
structure of poles
%
\eqn\higgbctwo{\eqalign{ & \Tr \,\varphi^2 \simeq {p_1 \over z} +
\ldots \cr & \Tr \,\varphi^4 \simeq {p_2 \over z^2} + {p_3 \over z}
+ \ldots }}
but the relations among the coefficients $p_i$ are different. Since
both conjugacy classes in question are four-dimensional, the polar
polynomials should contain only $2=4/2$ independent coefficients,
and there has to be one relation among the 3 parameters $p_1$,
$p_2$, $p_3$. For the conjugacy class $\frak C_{[2,1,1]}$ in
$C_2$, the relation is $p_2 = {1 \over 2} (p_1)^2$. On the other
hand, for the conjugacy class $\frak C_{[2,2,1]}$ in $B_2$, the
relation is different: $p_2 = {1 \over 4} (p_1)^2$.

However, in both $SO(5)$ and $Sp(4)$ theories, there is another
rigid surface operator, this time with  semisimple monodromy.
As explained in section \rigidsemisimple, in each case, the monodromy
preserves a subgroup of the gauge symmetry group whose Dynkin
diagram can be obtained by removing a node from the extended Dynkin
diagram of $B_2$ or $C_2$, respectively. The relevant node is the
middle node, $i=1$.  (Removing the other node $i=0$ leads back to
the rigid unipotent conjugacy class already considered above.) In
the notation of section \combining, these rigid semisimple conjugacy
classes can be denoted by $\frak C^{D_2}_{([1^4],[1])}$ and $\frak
C^{C_1 \times C_1}_{([1^2],[1^2])}$ since they preserve gauge
symmetry groups $O(4) \subset SO(5)$ and $Sp(2) \times Sp(2) \subset
Sp(4)$, respectively, {\it cf.} \bnsplit\ and \cnsplit. Summarizing,
we have the following list of rigid surface operators:
\bigskip
\centerline{\vbox{\offinterlineskip
\def\tablerule{\noalign{\hrule}}
\halign to 2.6truein{\tabskip=1em plus 2em#\hfil&\vrule height16pt
depth5pt#&#\hfil&\vrule height16pt depth5pt#&#\hfil\tabskip=0pt\cr
\hfil $B_2$ \hfil&&\hfil $\dim$ \hfil&&\hfil $C_2$ \hfil\cr
\tablerule ~~$\frak C^{D_2}_{([1^4],[1])}$ && ~~$4$ && $\frak
C^{C_2}_{(\emptyset, [2,1,1])}$ \cr ~~$\frak
C^{B_2}_{(\emptyset,[2,2,1])}$ && ~~$4$ && $\frak C^{C_1 \times
C_{1}}_{([1^2],[1^2])}$ \cr }}}\bigskip \noindent where we put what
we claim to be dual pairs of strongly rigid surface operators on the
same line. In particular, a surface operator associated with the
rigid semisimple conjugacy class $\frak C^{D_2}_{([1^4],[1])}$ has
the same set of polar polynomials \higgbctwo, with $p_2 = {1 \over
2} (p_1)^2$, as the surface operator associated with the unipotent
conjugacy class $\frak C^{C_2}_{(\emptyset, [2,1,1])}$ in $C_2$.
Similarly, surface operators associated with the conjugacy classes
$\frak C^{B_2}_{(\emptyset,[2,2,1])}$ and $\frak C^{C_1 \times
C_{1}}_{([1^2],[1^2])}$ have the polar polynomials \higgbctwo\ with
$p_2 = {1 \over 4} (p_1)^2$, in complete agreement with the duality.

Further evidence for this identification of dual surface operators
comes from comparing discrete invariants discussed in section
\centertopology. Thus, surface operators associated with the rigid
unipotent conjugacy classes $\frak C^{B_2}_{(\emptyset,[2,2,1])}$
and $\frak C^{C_2}_{(\emptyset, [2,1,1])}$ can ``detect'' the center
of the gauge group (in the sense that twisting  $\CZ (G)$ or $\CZ
(\LG)$ gives rise to new surface operators). In fact, as explained
in section \centertopology, the group $\CZ(G)^V$ (resp.
$\CZ(\LG)^V$) is trivial for any unipotent conjugacy class.

On the other hand, surface operators associated with the rigid
unipotent conjugacy classes $\frak C^{B_2}_{(\emptyset,[2,2,1])}$
and $\frak C^{C_2}_{(\emptyset, [2,1,1])}$ can not ``detect''
topology. Thus, in the adjoint form of $G = SO(5)$, the symmetry
group $H$ of the surface operator associated with the rigid
conjugacy classes $\frak C^{B_2}_{(\emptyset,[2,2,1])}$ is a double
cover of $Sp(2)$. It has a trivial fundamental group, $\pi_1 (H) =
1$, which means that the image of the natural map $\iota:\pi_1(H)\to
\pi_1({}^L\neg G)$ is trivial and all homomorphisms \trut\
annihilate $\iota (\pi_1(H))$. Hence, in the notation of section
\centertopology, we have $\tilde \pi (G) = \pi_1 (G)$ and we
conclude that the strongly rigid surface operator associated with
the conjugacy class $\frak C^{B_2}_{(\emptyset,[2,2,1])}$ can not
``detect'' topology. The analysis of topology for the surface
operator associated with a rigid unipotent conjugacy class $\frak
C^{C_2}_{(\emptyset, [2,1,1])}$ is essentially identical since the
adjoint form of $\LG = Sp(4)$ is isomorphic to $G = SO(5)$ and the
symmetry group $H$ is also the same.

The situation is reversed for surface operators associated with
rigid semisimple conjugacy classes $\frak C^{D_2}_{([1^4],[1])}$ and
$\frak C^{C_1 \times C_{1}}_{([1^2],[1^2])}$, which ``detect'' the
fundamental group but not the center of the gauge group.

Let us explain why this is so.  We start with the class $\frak
C^{D_2}_{([1^4],[1])}$ in $B_2$, which corresponds to the element
$S={\rm diag}(1,-1,-1,-1,-1)$ in $SO(5)$, the adjoint form of $B_2$.
This element commutes with $SO(4)$, and the map of fundamental
groups from $SO(4)$ to $SO(5)$ is surjective, so the surface
operator associated with this class detects topology.

Let us explain why this surface operator does not detect the center
of the gauge group.  Let $\frak z$ be the nontrivial element of the
center of $Spin(5)$; of course, $\frak z$ corresponds to a $2\pi$
rotation in space.  Let $T={\rm diag}(-1,-1,1,1,1)\in SO(5)$.  As
elements of $SO(5)$, $S$ and $T$ commute, but when they are lifted
to $Spin(5)$, we have \eqn\tordo{T^{-1}ST=\frak z S,} showing that
$S$ and $\frak z S$ are conjugate in $Spin(5)$.  Thus, a surface
operator with monodromy conjugate to $S$ does not detect topology.

The story is similar for the conjugacy class $\frak C^{C_1 \times
C_{1}}_{([1^2],[1^2])}$ in $C_2$.  This corresponds to an element of
$C_2$ that in $2\times 2$ blocks looks like
\eqn\zodo{S'=\left(\matrix{-1&0\cr 0&1\cr}\right).} In the adjoint
form of the group, which is $Sp(4)/\Z_2$, $S'$ commutes with
$H=(Sp(2)\times Sp(2))/\Z_2$.  The map of fundamental groups from
$H$ to $Sp(4)/\Z_2$ is surjective, so a surface operator with
monodromy $S'$ detects topology.  On the other hand, the center of
$Sp(4)$ is generated by the element $-1$.  $S'$ can be conjugated to
$-S'$ by an element $T'$ which in $2\times 2$ blocks looks like
\eqn\xodo{T'=\left(\matrix{0&1\cr 1&0\cr}\right),} and this shows
that a surface operator with monodromy $S'$ does not detect the
center of $Sp(4)$.

Since duality exchanges the center with the fundamental group, {\it
cf.} \centerpione, the fact that the unipotent surface operators of
this class detect only the center while the semisimple ones detect
only the topology is consistent with the proposal that $S$-duality
exchanges these two kinds of surface operator.  Another indication
of this will emerge in section \quant\ when we study quantization.


\subsec{Duality for $G = SO(8)$}\subseclab\soeight

Another instructive example, in which duality exchanges rigid
surface operators in a far from obvious way, is the self-dual theory
with gauge group $G = SO(8)$. (Of course, $SO(8)$ is self-dual only
to the extent that the difference between the adjoint and
simply-connected forms is not essential.)  In this case, in addition
to the two rigid unipotent conjugacy classes of dimension 10 and 16
that we listed in the end of section \searching, we also have a
16-dimensional conjugacy class of a strongly rigid semisimple
element
$$
S = {\rm diag} (-1,-1,-1,-1,+1,+1,+1,+1)
$$
which corresponds to the gauge symmetry breaking pattern $D_2 \times D_2 \subset D_4$.
We list all of these conjugacy classes in the following table:
\bigskip
\centerline{\vbox{\offinterlineskip
\def\tablerule{\noalign{\hrule}}
\halign to 1.8truein{\tabskip=1em plus 2em#\hfil&\vrule height16pt
depth5pt#&#\hfil\tabskip=0pt\cr \hfil $D_4$ \hfil&&\hfil $\dim$
\hfil\cr \tablerule ~~$\frak C^{D_4}_{(\emptyset,[2^2,1^4])}$ &&
~$10$ \cr ~~$\frak C^{D_4}_{(\emptyset,[3,2,2,1])}$ && ~$16$ \cr
~~$\frak C^{D_2 \times D_2}_{([1^4],[1^4])}$ && ~$16$ \cr
}}}\bigskip \noindent This is the complete list of surface operators
associated with strongly rigid non-central conjugacy classes for
$G=SO(8)$. While the surface operator associated with the
ten-dimensional class is clearly self-dual (if only these strongly
rigid classes are relevant), rigid surface operators associated with
the two 16-dimensional conjugacy classes
 potentially can be mapped into each other. In
particular, they have identical sets of polar polynomials. (The
calculation of polar polynomials is similar to the example of
section \polar.) Moreover, using the techniques of section
\centertopology, we find that the surface operator associated with
the rigid unipotent conjugacy class $\frak
C^{D_4}_{(\emptyset,[3,2,2,1])}$ can detect the center, but not the
topology, of the gauge group. On the other hand, the surface
operator associated with the rigid semisimple conjugacy class $\frak
C^{D_2 \times D_2}_{([1^4],[1^4])}$ can detect the fundamental
group, but not the center, of the gauge group, as expected for the
dual surface operator.

Further evidence for the duality action on the three rigid surface
operators listed here will become clear in the following sections.
Thus, the unipotent conjugacy class labeled by $\la =[2^2,1^4]$ is
{\it special} and, as we explain in section \quant, should map
into a unipotent conjugacy class. On the other hand, the unipotent
conjugacy class labeled by $\la =[3,2,2,1]$ is not special and, in
general, duality maps such conjugacy classes into operators whose
monodromy is not unipotent.


\subsec{Duality for $G = SO(7)$ and $\LG = Sp(6)$}

While two of our previous examples were rather subtle, the duality between
gauge theories with $G = SO(7)$ and $\LG = Sp(6)$ is very simple
in a sense that dual pairs of strongly rigid surface operators in this case
can be identified simply by comparing the most basic invariant,
namely the dimension of the corresponding conjugacy classes.

In both gauge theories, the construction based on $\frak{su} (2)$
embeddings and Nahm's equations gives only one strongly rigid surface operator.
In gauge theory with $G = SO(7)$, this is a strongly rigid surface operators
associated with a rigid nilpotent orbit labeled by $\la = [2,2,1,1,1]$.
Similarly, in the dual theory with $\LG = Sp(6)$, there is one rigid
surface operator associated with a strongly
rigid nilpotent orbit labeled by $\la=[2,1,1,1,1]$ (see table in sec. \searching).
These nilpotent orbits have dimensions 8 and 6, respectively,
which makes it clear that the construction of rigid surface operators
based on $\frak{su} (2)$ embeddings and Nahm's equations is not sufficient
for producing a set of surface operators closed under duality.

This situation is rectified if we include surface operators
which correspond to strongly rigid semisimple conjugacy classes.
In the theory with $G=SO(7)$, there are two such surface operators,
which correspond to rigid semisimple conjugacy classes
$\frak C^{D_3}_{([1^{6}],[1])}$ and
$\frak C^{D_2 \times B_{1}}_{([1^4],[1^{3}])}$
of dimension 6 and 12, respectively.
On the other hand, in the dual theory with $\LG = Sp(6)$,
there are two surface operators
which correspond to rigid semisimple conjugacy classes
$\frak C^{C_1 \times C_{2}}_{([1^2], [1^{4}])}$ and
$\frak C^{C_2 \times C_{1}}_{([2,1,1], [1,1])}$
of dimension 8 and 12, respectively.
Now, the complete list of strongly rigid surface operators has a nice form:
\bigskip
\centerline{\vbox{\offinterlineskip
\def\tablerule{\noalign{\hrule}}
\halign to 2.8truein{\tabskip=1em plus 2em#\hfil&\vrule height24pt
depth5pt#&#\hfil&\vrule height24pt depth5pt#&#\hfil\tabskip=0pt\cr
\hfil $B_3$ \hfil&&\hfil $\dim$ \hfil&&\hfil $C_3$ \hfil\cr
\tablerule ~~$\frak C^{D_3}_{([1^{6}],[1])}$ && ~~$6$ && $\frak
C^{C_3}_{(\emptyset, [2,1^{4}])}$ \cr ~~$\frak
C^{B_3}_{(\emptyset,[2,2,1^{3}])}$ && ~~$8$ && $\frak C^{C_1 \times
C_{2}}_{([1^2], [1^{4}])}$ \cr ~~$\frak C^{D_2 \times
B_{1}}_{([1^4],[1^{3}])}$ && ~~$12$ && $\frak C^{C_2 \times
C_{1}}_{([2,1,1], [1,1])}$ \cr }}}
\bigskip \noindent
In each case, we find three strongly rigid conjugacy classes
of the same dimension, which allows to identify unambiguously
dual pairs of rigid surface operators for $G = SO(7)$ and $\LG = Sp(6)$.
As a strong test of the duality, we have checked that all other
invariants of dual surface operators also match.
In particular, in the previous section we already gave a detailed
comparison of the polar polynomials for the six-dimensional
conjugacy classes $\frak C^{D_3}_{([1^{6}],[1])}$
and $\frak C^{C_3}_{(\emptyset, [2,1^{4}])}$.


\subsec{Duality for $G = SO(9)$ and $\LG = Sp(8)$}

Now we turn to $SO(9)$ and $Sp(8)$.  This turns out to be the first
case in which we do not get a consistent picture in considering only
strongly rigid conjugacy classes.  Possibly, this means that for
these groups, a full duality statement involves also the more
delicate constructions that were described for $SU(2)$ and $SO(3)$
in section \rigdual.  (Of course, these contructions may also be
relevant for a more complete treatment of the groups that we have
just considered.)

As in the previous examples, we start with rigid surface operators
constructed via $\frak{su} (2)$ embeddings $\rho:\frak{su}(2)\to
\frak g$. These surface operators correspond to rigid unipotent
conjugacy classes which, for $B_4$ and $C_4$, we summarized in the
table in the end of section \searching. Namely, in $B_4$ there are
two rigid unipotent conjugacy classes labeled by $\la = [2,2,1^{5}]$
and $\la = [2^4,1]$. Similarly, in $C_4$ there are also two rigid
unipotent conjugacy classes labeled by $\la = [2,1^6]$ and $\la =
[2^3,1^2]$. However, these rigid unipotent conjugacy classes have
completely different dimensions which, again, makes it clear that
S-duality cannot work unless we enlarge the set of rigid surface
operators at least by including those corresponding to strongly
rigid semisimple conjugacy classes. Once we do this, the list of
strongly rigid surface operators in $B_4$ and $C_4$ becomes
considerably larger, with some obvious matches:
\bigskip
\centerline{\vbox{\offinterlineskip
\def\tablerule{\noalign{\hrule}}
\halign to 3.0truein{\tabskip=1em plus 2em#\hfil&\vrule height24pt depth5pt#&#\hfil&\vrule height24pt depth5pt#&#\hfil\tabskip=0pt\cr
\hfil $B_4$ \hfil&&\hfil $\dim$ \hfil&&\hfil $C_4$ \hfil\cr
\tablerule
~~$\frak C^{D_4}_{([1^{8}],[1])}$ && ~~$8$ && $\frak C^{C_4}_{(\emptyset, [2,1^6])}$ \cr
~~$\frak C^{B_4}_{(\emptyset,[2,2,1^{5}])}$ && ~~$12$ && $\frak C^{C_1 \times C_3}_{([1^2], [1^6])}$ \cr
~~$\frak C^{B_4}_{(\emptyset,[2^4,1])}$ && ~~$16$ && $\frak C^{C_2 \times C_2}_{([1^4], [1^4])}$ \cr
~~$\frak C^{D_3 \times B_1}_{([1^6],[1^3])}$ && ~~$18$ && $\frak C^{C_4}_{(\emptyset,[2^3,1^2])}$ \cr
~~$\frak C^{D_4}_{([2^2,1^4],[1])}$ && ~~$18$ && $\frak C^{C_1 \times C_3}_{([1^2],[2,1^4])}$ \cr
~~$\frak C^{D_2 \times B_2}_{([1^4],[1^5])}$ && ~~$20$ && $\frak C^{C_2 \times C_2}_{([2,1^2], [1^4])}$ \cr
~~$\frak C^{D_4}_{([3,2,2,1],[1])}$ && ~~$24$ && $\frak C^{C_2 \times C_2}_{([2,1^2], [2,1^2])}$ \cr
~~$\frak C^{D_2 \times B_{2}}_{([1^4],[2,2,1])}$ && ~~$24$ && ~~? \cr
}}}\bigskip
\noindent
In particular, strongly rigid surface operators which correspond
to conjugacy classes of dimension 8, 12, 16, and 20
can be identified simply by matching the dimension.
For these surface operators, one can also check that
all other invariants match, in complete agreement with the duality.

Strongly rigid surface operators which correspond to conjugacy
classes of dimension 18 and 24 are more interesting. In dimension
18, there is an ambiguity in the matching that is actually not
resolved by our other invariants. All four surface operators of
dimension 18 listed in the table have the same set of polar
polynomials:
\eqn\higgseighteen{\eqalign{ & \Tr \,\varphi^2 \simeq {p_2 \over z}
+ \ldots \cr & \Tr \,\varphi^4 \simeq {p_4 \over z^2} + {p_1 \over
z} + \ldots \cr & \Tr\, \varphi^6 \simeq {p_6 \over z^3} + {p_3
\over z^2} + {p_5 \over z} + \ldots \cr & \Tr\, \varphi^8 \simeq
{p_8 \over z^4} + {p_7 \over z^3} + {p_{9} \over z^2} + {p_{10}
\over z} + \ldots }}
Since these orbits have dimension 18, we expect the space of polar
polynomials to be 9-dimensional. Therefore, we expect one relation
among the 10  parameters $p_i$. This relation turns out to be
\eqn\higgseighteenrel{ p_8 = {1 \over 48} p_2^4 - {1 \over 4} p_2^2
p_4 + {1 \over 4} p_4^2 + {2 \over 3} p_2 p_6 }
(which one can verify to be invariant under reparametrization of the
local coordinate $z$). Moreover, even the discrete invariants of
section \centertopology\ do not help to resolve the ambiguity in
matching of orbits.
Indeed, in $B_4$, both rigid surface operators associated with
the 18-dimensional conjugacy classes
$\frak C^{D_3 \times B_1}_{([1^6],[1^3])}$
and $\frak C^{D_4}_{([2^2,1^4],[1])}$
can detect the fundamental group $\pi_1 (G)$,
but not the center $\CZ (G)$.
On the other hand, in $C_4$, both rigid surface operators
associated with the 18-dimensional conjugacy classes
$\frak C^{C_4}_{(\emptyset,[2^3,1^2])}$
and $\frak C^{C_1 \times C_3}_{([1^2],[2,1^4])}$
can detect the center $\CZ (\LG)$,
but not the fundamental group $\pi_1 (\LG)$.
This is consistent with the duality, but one needs finer invariants
in order to say more precisely how these 18-dimensional conjugacy
classes are paired up.

In dimension 24, we have a worse problem: there are two strongly
rigid conjugacy classes in $B_4$ and only one in $C_4$ so there is
no hope of matching them. Perhaps the inclusion of more delicate
constructions of surface operators --- such as those of section
\rigdual\ --- is needed for resolving this problem.

Still, it is attractive that all but one strongly rigid surface
operators in our table does appear to have a dual.


\newsec{More Examples}\seclab\families

Although we do not know the general mapping from rigid surface
operators in a theory with gauge group $G$ to similar operators in
the dual theory with gauge group $\LG$, in this section we make a
duality conjecture for certain infinite families of surface
operators.  The proposal generalizes examples seen in the previous
section. Again, our main tools for identifying dual pairs will be
invariants described in section \invariants.


\subsec{Special Rigid Orbits}\subseclab\specialsurf

As we have already mentioned, there is no bijection between
nilpotent orbits (rigid or not) for the dual groups $G_{\C}$ and
$\LG_{\C}$. There is, however, a nice bijection between a certain
subset of nilpotent orbits called {\it special}
orbits\foot{Special nilpotent orbits  include Richardson orbits,
which correspond in the following sense to surface operators
studied in \Ramified\ and reviewed in section \review.  Let
$\Bbb{L}$ be a Levi subgroup of $G$, with $\Bbb{L}$-regular
parameters $\alpha,\beta,\gamma$.  In the limit that these
parameters are all taken to zero, the monodromy of the surface
operator generically takes values in the Richardson unipotent
orbit associated to $\Bbb{L}$. For more on this see section 3.3 of \Ramified.}
which has been studied in the mathematical literature
\refs{\Lusztigiv,\Lusztigii} (see also \LusztigS).
This bijection is defined by
considering representations of the Weyl group associated to an
orbit by the Springer correspondence, while we are interested in a
duality map that preserves the invariants of section \invariants.
The most important invariant in what follows is the conjugacy
class in the Weyl group associated to an orbit by the
Kazhdan-Lusztig map; this is somewhat analogous to the invariant
associated with the Springer correspondence.

Many special orbits are not rigid; however, some of them are.
And even though explaining how generic special orbits transform
under duality involves a rather sophisticated combinatorics,
the case of surface operators associated with special rigid
orbits is considerably easier.
First, we will give an idea of what special orbits look like,
and then specialize to the rigid ones.

There are several ways to define special orbits. For example, one
definition, related to quantization, will be mentioned in  section
\quant. Here, we present another, equivalent definition which is
helpful for better understanding of the set of nilpotent orbits (or
unipotent conjugacy classes) as a whole. We have seen in section
\limit, for the example of $G=SU(2)$, that it is possible for one
nilpotent orbit (the orbit of the zero element of $\frak{sl}(2)$) to
be in the closure of another (the orbit of a regular nilpotent
element). Exploiting this idea, we get a natural partial order on
the set of nilpotent orbits. Let $\frak c_\lambda$ and $\frak c_\mu$
be two nilpotent orbits. One says that $\frak c_{\lambda} \le \frak
c_{\mu}$ if $\frak c_{\lambda}$ is contained in the closure of
$\frak c_\mu$, that is if $ \frak c_{\lambda} \subset \bar \frak
c_{\mu}$.  For classical groups, we can think of $\lambda $ and
$\mu$ as partitions, $N=\lambda_1+\dots+\lambda_n=\mu_1+\mu_2+\dots
+\mu_{n'}$ (where we take $\lambda_i\geq \lambda_j,$ $\mu_i\geq
\mu_j$ for $j>i$). We say that $\lambda\leq\mu$ if
$$
\sum_{i=1}^k \lambda_i \le \sum_{i=1}^k \mu_i
$$
for all $k$. This condition is equivalent  with one
exception\foot{The exception arises for $D_n$ in the case of a
very even partition, that is a partition such that the $\lambda_i$
are all even.  Such a partition corresponds to two distinct
nilpotent orbits, neither of which is in the closure of the
other.} that will not concern us to the condition $\frak
c_\lambda\leq \frak c_\mu$.
 For example, the closure ordering of nilpotent
orbits in $B_3$ and $C_3$ can be summarized in the diagram below.
\ifig\hassebcthree{Hasse diagram for $B_3$ and $C_3$.  If
$\lambda\leq \mu$, then $\lambda$ is shown below $\mu$ in the
diagram.  Nilpotent orbits which are not special are shown in red
and labeled by an asterisk; omitting such orbits gives a diagram
with an order-reversing involution.}
{\epsfxsize3.8in\epsfbox{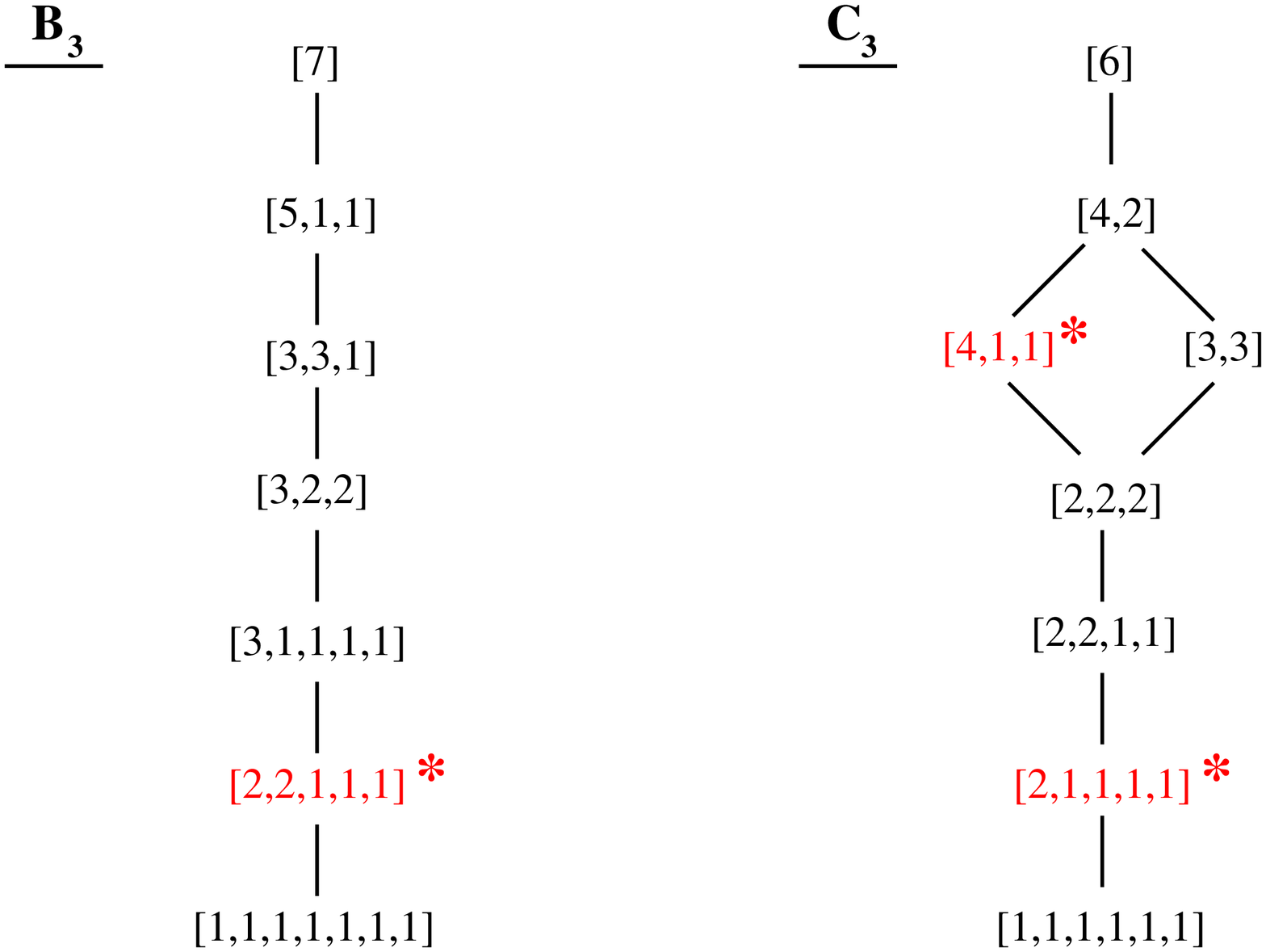}}

There is a natural order-reversing involution on the set of
nilpotent orbits, which in type $A$ corresponds to a map $\la \to
\la^t$, where $\la^t$ is the transpose partition of $\la$, see
{\it e.g.} \CMcGovern. The transpose partition is described as
follows.  We relate a partition to a Young tableau by turning
every ``part'' $\lambda_i$ into a column of height $\lambda_i$.
For example, for $N=11$, to the partition $N=3+3+2+2+1$, we
associate the Young tableau \eqn\geft{\tableau{2 4 5}.} To make
the transpose partition, we just take the transpose of the
picture. Thus, for the example that we just gave, the transpose
operation is
\eqn\jurf{ \la = \tableau{2 4 5} \quad\quad \leadsto
\quad\quad \la^t = \tableau{1 2 2 3 3},}
so the transpose of the partition $[3^2,2^2,1]$ is $[5,4,2]$.

For a group of type $A$, the transpose operation makes sense for
any partition.  It can be shown that it reverses order, meaning
that if $\lambda\leq \mu$, then $\mu^t\leq\lambda^t$.

For other classical groups $B,C,$ and $D$, the transpose operation
does not make sense for an arbitrary partition.   For these
groups, nilpotent orbits are associated with partitions that obey
certain constraints (the constraints are stated at the end of
section \limit).  These constraints are not invariant under the
transpose operation.  For example, in the $B$ case, the constraint
is that if $\lambda_i$ is even, it must occur with even multiplicity.
In the example of eqn. \jurf, we see that $\lambda=[3^2,2^2,1]$
obeys this constraint, and $\lambda^t=[5,4,2]$ does not.

For groups of type $B,$ and $C$, we say that a partition $\lambda$
is special (or the corresponding orbit $\frak c_\lambda$ is special)
if the transpose $\lambda^t$ obeys the relevant conditions (for the same group).
Thus, in the above example, $\lambda$ is not special.
On the other hand, for $SO(9)$ the partition $[3,2^2,1^2]$
is special since its transpose, which is $[5,3,1]$, has no even parts:
\eqn\zefto{  \tableau{1 3 5} \quad\quad \leadsto \quad\quad
\tableau{1 1 2 2 3}.}

For type $A$, all nilpotent orbits are special since $\la$
is subject to no constraint.
For groups of type $D$, a nilpotent orbit $\frak c_\lambda$
labeled by an orthogonal partition $\lambda$ is special
if and only if the transpose partition $\lambda^t$ is symplectic.
For example, the orbit labeled by $\la = [3,2,2,1]$ in $D_4$
that we discussed in section \soeight\ is not special
since the transpose partition $\lambda^t = [4,3,1]$ is not symplectic:
\eqn\dfoursp{  \la = \tableau{1 3 4} \quad\quad \leadsto \quad\quad
\la^t = \tableau{1 2 2 3}.}
This definition  is rather surprising, but we will not describe it
here, as we will not go into any depth concerning groups of type
$D$.

Now we focus on groups of type $B$ and $C$. Clearly, with the
above definition, the operation $\lambda\to\lambda^t$ makes sense
for special partitions. Since we are interested in rigid surface
operators, our next task is to single out those special partitions
$\lambda$ for which $\frak c_\lambda$ is rigid. First, let us
consider $B_N$. As usual, we label nilpotent orbits by partitions
\eqn\nnnpart{ \la = [k^{n_k}(k-1)^{n_{k-1}} \ldots  2^{n_2}1^{n_1}]
}
where we assume that the multiplicities $n_i $ are positive
precisely if $i\leq k$. (This is one of the criteria for rigidness
that were described in section \searching.) Of course, we also have
$n_{2i}$ even since we are considering orbits in $B_N$. Finally, the
other criterion for rigidness is that $n_{2i+1} \ne 2$ for all $i$.
Imposing an extra condition that the orbit is special, one finds
that $n_k$ is odd and $n_i$ is even for all $i < k$. For example,
since $n_{k-1}\not=0$, we find that the transpose partition has for
its smallest part the  number $n_k$ occurring with multiplicity 1,
as in this example: \eqn\znnpart{\tableau{2 4 5}
\quad\quad\leadsto\quad\quad\tableau{1 2 2 3 3}.} Hence, if $n_k$ is
even (as in this example) then $\lambda^t$ is not orthogonal. A
similar argument shows that $n_i$ must be even for $i<k$.

 Therefore, we conclude that special rigid orbits in $B_N$ are
 associated with partitions of the form
\eqn\nnnpartbn{ \la = [k^{2m_k+1} (k-1)^{2m_{k-1}}\ldots
2^{m_2}1^{2m_1}] }
with $k$ odd.

Similarly, a rigid orbit in $C_N$ is labeled by the partition in the
form \nnnpart\ with $n_{2i+1}$ even and $n_{2i}$ not equal to 2. In
addition, such an orbit is special if all $n_i$ are even. Hence, we
label special rigid orbits in $C_N$ by
\eqn\nnnpartcn{ \la = [k^{2m_k}(k-1)^{2m_{k-1}} \ldots  2^{2m_2}
1^{2m_1}] }
Notice that in the case of $C_N$ we do not need to assume that $k$
is odd.

Now, let us identify dual pairs of special rigid surface operators.
Starting with a special rigid orbit labeled by the symplectic
partition \nnnpartcn\ of $2N$, we need to describe the orthogonal
partition of $2N+1$ that labels the dual orbit. It can be
constructed by the following simple rule:\foot{This rule was found
by comparing to some constructions of Lusztig \Lusztigii\ as well as
to examples in the last section.} in the symplectic partition, every
block $l^{n_l}$ with $l$ odd remains invariant, while in every block
$l^{n_l}$ with $l$ even one of the parts is replaced by $l+1$ and
one other part is replaced by $l-1$:
$$
l^{n_l} \mapsto \cases{(l+1) l^{n_l-2} (l-1) & $l$ even \cr l^{n_l} & $l$ odd}
$$
Notice that this operation does not change the net sum of all the
parts. Hence, we also add ``$1$'' to the resulting partition in
order to obtain a partition of $2N+1$ (instead of a partition of
$2N$). In the end, we obtain the following map\foot{Here $(k+1)^1$
refers to a part $k+1$ that appears with multiplicity 1.}
\eqn\specrigiddual{ [k^{2m_k} \ldots 3^{2m_3} 2^{2m_2} 1^{2m_1}]
\mapsto \cases{ [(k+1)^1 k^{2m_k-2} \ldots 3^{2m_3+2} 2^{2m_2-2}
1^{2m_1+2}] & $k$ even \cr [k^{2m_k+1} (k-1)^{2m_{k-1}-2}\ldots
3^{2m_3+2} 2^{2m_2-2} 1^{2m_1+2}] & $k$ odd} }
where $m_j > 1$ if $j$ is even, and $m_j > 0$ in general.

This map  preserves all the invariants of surface operators
introduced in section \invariants. For example, using \cdimabcd\
we find that special rigid orbits labeled by the partitions
\specrigiddual\ have the same dimension, given by
$$
\dim (\frak c_{\la}) = 2N^2 + N - 2 \sum_i \big( \sum_{j \ge i} n_j \big)^2
- \sum_{i {\rm ~odd}} n_i
$$

Similarly, one can identify the corresponding polar polynomials or,
equivalently, the conjugacy class in the Weyl group $\Weyl$ under
the Kazhdan-Lusztig map. For groups of type $B_N$ and $C_N$ that we
are considering here, conjugacy classes in $\Weyl$ are indexed by
pairs of partitions $(\nu_+ ,\nu_-)$ such that $|\nu_+| + |\nu_-| =
N$, see {\it e.g.} \SpaltensteinKL. In particular, $(\nu_+,\nu_-) =
([1,1,\ldots,1] , \emptyset)$ corresponds to the class of the
identity in $\Weyl$, while $(\nu_+,\nu_-) = (\emptyset, [N])$
corresponds to the Coxeter class (the class which contains a cyclic
permutation of order $N$). After a somewhat lengthy calculation,
using formulas in \SpaltensteinKL, one finds that under the
Kazhdan-Lusztig map, dual orbits identified in \specrigiddual\ map
to the same conjugacy class in $\Weyl$, namely:
$$
([ \ldots 5^{n_5} 3^{2n_3} 1^{n_1}],[\ldots 3^{2n_6} 2^{2n_4} 1^{2n_2}])
$$
Therefore, we conclude that special rigid orbits identified
by \specrigiddual\ have the same fingerprints.

Finally, we consider the center and topology of $G$ and $^L\neg G$
as follows.  For a unipotent surface operator with gauge group
$G$, one can always observe the center of $G$, as explained in
section \centertopology. Dually, consider a surface operator in a
theory with gauge group $^L\neg G$ and let $H$ be the automorphism
group of this surface operator.  Then one can observe the topology
of $^L\neg G$ if the natural map $\iota:\pi_1(H)\to \pi_1({}^L\neg
G)$ is surjective.  For $\LG$ of type $B_N$, this will be so if
$H$ contains a factor $SO(n)$, $n\geq 2$, since the map of
fundamental groups $SO(2)\to SO(2N+1)$ is surjective.  In turn,
for a surface operator associated with a partition $\lambda$, $H$
has such a factor if one of the odd parts in $\lambda$ has
multiplicity at least 2. According to \nnnpartbn, this is so
whenever $\lambda$ is rigid and special. For $\LG$ of type $C_N$,
the condition we need is that the central element $-1$ of $\LG$
should be connected to the identity in the subgroup $H$ that
commutes with the embedding $\rho:\frak{sl}(2)\to {}^L\neg\frak g$.
(A path from 1 to $-1$ in $C_N=Sp(2N)$ projects in
$Sp(2N)/\Z_2$ to a generator of the fundamental group.)
$H$ is a product of factors $H_{\la^*}$, associated respectively with the
parts of size $\la^*$.  If a part $\la^*$ appears with
multiplicity $m$, then $H_{\la^*}$ is $SO(m)$ or $Sp(m)$
(depending on whether $\la^*$ is even or odd). The element $-1$
is connected to the identity in $SO(m)$ or $Sp(m)$ if $m$ is even,
which is always true in the rigid special case, according to eqn. \nnnpartcn.


\subsec{Dualities Involving Rigid Semisimple Orbits}\subseclab\ssfamilies

Now let us consider dual pairs that involve rigid semisimple
conjugacy classes on at least one side. Simple  classes of this type
were described in section \rigidsemisimple.

We start with the minimal unipotent orbit in $C_N$ (whose dual turns
out to be semisimple). We already discussed this orbit in detail in
section \searching,  eqns. \gret\ - \hoto. In the notation of
section \combining, this orbit corresponds to $\Theta_0 = \Delta$
with $\la' = \emptyset$ and $\la'' = [2,1^{2N-2}]$. It has dimension
$2N$ and via the Kazhdan-Lusztig map is identified with the
following conjugacy class in the Weyl group:
$$
([1^{N-1}],[1])
$$
In type $B_N$, there is also a $2N$-dimensional conjugacy class
$\frak C^{D_N}_{([1^{2N}],[1])}$ of the semisimple element
$$
S = {\rm diag} (1,-1,-1,\ldots,-1)
$$
which has the same behavior of the Higgs field and discrete invariants
introduced in section \centertopology.
This is a strong hint that the corresponding surface operators are  dual.

For our next example we take the strongly rigid unipotent conjugacy
class in $B_N$ corresponding to the partition $\la'' = [2,2,1^{2N-3}]$.
This conjugacy class was also discussed in section \searching.
It has dimension $4(N-1)$ and via the Kazhdan-Lusztig map
is identified with the following conjugacy class in the Weyl group:
$$
([2,1^{N-2}],\emptyset)
$$
One finds the same behavior of the Higgs field for the strongly rigid
conjugacy class of the semisimple element
$$
S = {\rm diag} (1,1,-1,-1,\ldots,-1)
$$
in $C_N$ associated with $\Theta_1 = C_1 \times C_{N-1}$
and $(\la', \la'') = ([1^2], [1^{2N-2}])$.

Our next example involves dual pairs of surface operators, both of
which correspond to strongly rigid semisimple conjugacy classes. In
type $B_N$, we consider the conjugacy class associated to $\Theta_2
= D_2 \times B_{N-2}$ and $(\la', \la'') = ([1^4], [1^{2N-3}])$. On
the other hand, in type $C_N$ the candidate for the dual conjugacy
class has $\Theta_2 = C_2 \times C_{N-2}$ and $(\la', \la'') =
([2,1,1], [1^{2N-4}])$. It is possible to check that both of these
conjugacy classes have equal polar polynomials and their image under
the Kazhdan-Lusztig map is the same conjugacy class in the Weyl
group:
$$
(\emptyset, [1^N])
$$

Summarizing, we find the following families of dual pairs of rigid
semisimple surface operators:
\bigskip
\centerline{\vbox{\offinterlineskip
\def\tablerule{\noalign{\hrule}}
\halign to 2.5truein{\tabskip=1em plus 2em#\hfil&\vrule height28pt depth5pt#&#\hfil\tabskip=0pt\cr
\hfil $B_N$\hfil&&\hfil $C_N$ \hfil\cr
\tablerule
$\frak C^{D_N}_{([1^{2N}],[1])}$ && $\frak C^{C_N}_{(\emptyset, [2,1^{2N-2}])}$ \cr
$\frak C^{B_N}_{(\emptyset,[2,2,1^{2N-3}])}$ && $\frak C^{C_1 \times C_{N-1}}_{([1^2], [1^{2N-2}])}$ \cr
$\frak C^{D_2 \times B_{N-2}}_{([1^4],[1^{2N-3}])}$ && $\frak C^{C_2 \times C_{N-2}}_{([2,1,1], [1^{2N-4}])}$ \cr
$\ldots$ && $\ldots$ \cr
}}}\bigskip

\noindent We note that these examples completely cover all dual
pairs of strongly rigid conjugacy classes in small rank $N \le 3$.
In particular, for $N=2$ we recover a somewhat subtle duality
between rigid surface operators in $B_2$ and $C_2$ discussed in
section \bctwo.

Lusztig in \Lusztigiii, section 13.3, generalizes the
correspondence between special unipotent classes in $G_\C$ and
$^L\neg G_\C$ to a surjective map from special conjugacy classes
in $G_\C$ that are not necessarily unipotent to unipotent classes
in $^L\neg G_\C$ that are not necessarily special.  The first two
entries in the above table appear to be special cases of this
definition, though the third is not.  The table is hopefully an
approximation to a more complete table that would describe a
bijection between suitable objects on the two sides.


\newsec{Duality and Quantization}\seclab\quant

In this paper, we have mainly studied ``static'' half-BPS surface
operators supported on $D = \R^2$ in the space-time manifold $M =
\R^4$. This problem admits various generalizations; in particular,
one can consider more general space-time manifolds $M$ and embedded
surfaces $D \subset M$, including those with boundary. A simple
example of such a generalization is obtained by taking
$$
M = \R^3 \times [0,1]
$$
with supersymmetric boundary conditions $\CB_{\pm}$ at $W_- = \R^3
\times \{ 0 \}$ and $W_+ = \R^3 \times \{ 1 \}$, and with a
``static'' surface operator on $D = \R \times [0,1]$, where
$\R \subset \R^3$ stands for the ``time'' direction, parametrized by $x^0$.

\ifig\WWWFig{A time zero slice of a time-independent configuration
on $M = \R^3 \times [0,1]$ with boundary conditions $\CB_+$ and $\CB_-$.
The support $D$ of a surface operator intersects the time zero slice
on the interval $I=[0,1]$, parametrized by $x^1$.}
{\epsfxsize3.5in\epsfbox{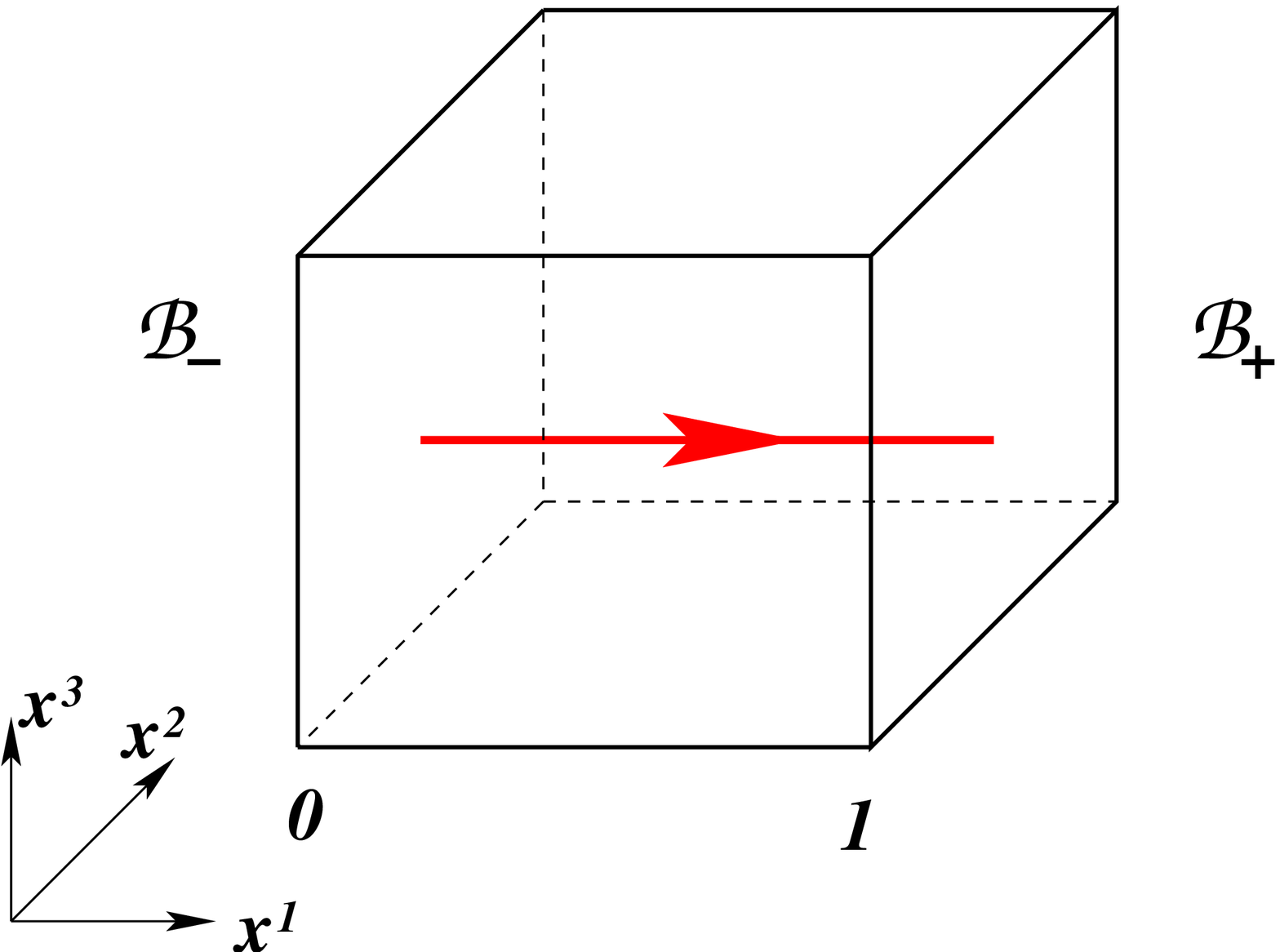}}

Quantization of this theory gives a Hilbert space, $\CH$, which
depends on all the choices involved, in particular, on the surface
operator as well as on the boundary conditions $\CB_+$ and $\CB_-$.
As will be explained in more detail elsewhere \toappear, for a
suitable choice of boundary conditions $\CB_{\pm}$ and a surface
operator on $D = \R \times [0,1]$,
the space $\CH$ is a
representation of a real form $G_{\R}$ of the complexified gauge
group $G_{\C}$.

In this construction, we consider a surface operator associated
with a unipotent conjugacy class $\frak C$ or with a semisimple
conjugacy class obtained by a deformation of $\frak C$.
Furthermore, the real form $G_{\R}$ is determined by
one of the boundary conditions, say $\CB_-$,
while the other boundary condition, $\CB_+$, is universal.
In compactification to two dimensions, $\CB_+$ corresponds
to the so-called canonical coisotropic brane;
see \KWitten, section 12.4 for a detailed description
of this boundary condition in four-dimensional gauge theory.
In particular, $\CB_+$ includes mixed Neumann-Dirichlet boundary
conditions for the gauge field $A$ and the Higgs field $\phi$:
\eqn\ccbrane{\eqalign{
D_0 \phi_2 + \p_1 A_2 & = 0 \cr
D_0 \phi_3 + \p_1 A_3 & = 0 \cr
F_{02} - \p_1 \phi_2 & = 0 \cr
F_{03} - \p_1 \phi_3 & = 0 }}
where $x^1$ is the coordinate on the interval $[0,1]$.

For applications to the present paper, the details of the second
boundary condition $\CB_-$ are not important, as long as it
preserves the same supersymmetry\foot{In fact, it is really only
necessary to preserve part of their common supersymmetry.  The
important part is the supersymmetry of the relevant $A$-model.} as
$\CB_+$. A particularly nice class of boundary conditions $\CB_-$
corresponds to imposing Dirichlet boundary conditions on half of the
fields $(A,\phi)$.   This can be done so that the boundary
conditions, upon reduction to the sigma-model, define a Lagrangian
brane supported on a $G_{\R}$-conjugacy class $\frak C_{\R} \subset
\frak C$, for some real form $G_{\R}$ of $G_{\C}$. For example, it
is easy to see that the Dirichlet boundary condition
\eqn\zerophi{ \CB_-: \quad \phi \vert_{W_-} = 0 }
restricts the monodromy $V$ of the connection $\CA = A + i \phi$ to
be in a conjugacy class of the compact group $G$. More generally,
one can define a boundary condition $\CB_-$ associated with a
$G_{\R}$-conjugacy class $\frak C_{\R}$ for some real form $G_{\R}$
of $G_{\C}$, such that
\eqn\vinreal{ V \in \frak C_{\R} }

For our purposes, all we need to know from \toappear\ is that the
central character $\zeta$ of the representation $\CH$ attached to
$\frak C_{\R}$ depends only on the surface operator (that is, on the
corresponding conjugacy class $\frak C$) involved in this
construction and not on the particular choice of the boundary
condition $\CB_{-}$ (which, among other things, determines the real
form $G_{\R}$ and $\frak C_{\R} \subset \frak C$). In particular,
for a surface operator \norto\ labeled by a Levi subgroup
${\Bbb{L}}$ and continuous parameters $(\alpha,\beta,\gamma,\eta)$,
the central character is given by \toappear:
\eqn\cchara{ \zeta = \eta + i \g }
More generally, in all examples that we have checked of surface
operators constructed via $\frak{su} (2)$ embeddings and solutions
to Nahm's equations, it turns out that the central character $\zeta$
of the $G_{\R}$-representation attached to $\frak C_{\R} \subset \frak C$
is related to the semisimple part $\dual{S}$ of the conjugacy class
$\dual{\frak C}$ associated with the dual surface operator,
\eqn\sviacchar{ \dual{S} = \exp (2\pi \zeta) }
This motivates the following conjecture:

\medskip\noindent
{\bf Conjecture:} {\it Let $\frak C$ be a unipotent conjugacy class
(or a semisimple conjugacy class obtained by a deformation of $\frak C$).
Then, the parameter ${1 \over 2\pi} \log \dual{S}$ of the dual conjugacy class $\dual{\frak C}$
is equal to the central character $\zeta$ of (any) $G_{\R}$-representation
attached to $\frak C_{\R} \subset \frak C$.}

\medskip\noindent
This conjecture can be verified for many surface operators. In
particular, it holds for all the surface operators constructed in
\Ramified\ and reviewed in section \review. Indeed, let us consider
a surface operator \norto\ labeled by a Levi subgroup ${\Bbb{L}}$
and continuous parameters $(\alpha,\beta,\gamma,\eta)$. As explained
in \toappear\ and summarized in eqn. \cchara, in this case the
central character is $\zeta = \eta + i \g$. Since $\eta$ is a
``quantum'' parameter, it is convenient to use the duality
transformation of the parameters of such surface operators
\Ramified,
\eqn\surfduality{\eqalign{
(\alpha, \eta) & \to (\eta, - \alpha)
}}
to write $\zeta$
in terms of ``classical'' parameters in the dual theory:
\eqn\ccharb{ \zeta = - \dual{\a} + i \dual{\g} }
This indeed equals the semisimple part $\dual{S}$
of the dual conjugacy class $\dual{\frak C}$,
thus justifying \sviacchar\ for surface operators
associated with Richardson conjugacy classes and their
semisimple deformations.
However, one can verify the above conjecture for more
general surface operators, including rigid unipotent
surface operators studied in this paper.

A simple class of  surface operators that are rigid (and, therefore,
not included in those of \Ramified) can be constructed in a theory
with gauge group $G = Sp(2N)$ via $\frak{su} (2)$ embeddings labeled
by $\la = [2,1^{2N-2}]$. We already considered such rigid surface
operators in the previous sections; they correspond to minimal
orbits $\frak c_{{\rm min}}$ in $C_N$. The minimal orbit $\frak
c_{{\rm min}}$ in $C_N$ has dimension $2N$, and the representation
attached to this orbit is the familiar Weyl representation obtained
by quantizing the phase space of $N$ decoupled harmonic oscillators.
(For mathematical literature on quantization of the minimal orbit
$\frak c_{{\rm min}}$, see {\it e.g.}
\refs{\Brylinski,\BravermanJ}.) The central character of this
representation gives a rigid semisimple element
$$
\dual{S} = \exp (2\pi \zeta) = {\rm diag} (+1,-1,-1,\ldots,-1)
$$
in the dual group, $\LG = SO(2N+1)$. It corresponds to the root
system of type $D_{N}$, and its conjugacy class $\frak C_{\dual{S}}$
also has dimension $2N$. In fact, $\frak C_{\dual{S}} = \frak
C^{D_N}_{([1^{2N}],[1])}$ is precisely the rigid semisimple
conjugacy class of a surface operator in the $\LG = SO(2N+1)$ theory
that, by matching invariants, we proposed in section \ssfamilies\
as the dual to the rigid surface operator associated with
the minimal nilpotent orbit $\frak c_{{\rm min}}$ in the $G = Sp(2N)$ theory.
In particular, this analysis implies that the minimal orbit in $C_N$ is dual
to a rigid semisimple orbit in $B_N$, thus explaining a somewhat subtle duality
in the case of $B_2$ and $C_2$ discussed in section \bctwo.

Suppose that a surface operator associated with a  nilpotent orbit
maps  under duality to a surface operator associated with a
nilpotent orbit of the dual group. The conjecture implies that
$G_{\R}$-representations obtained by quantizing such an orbit have
trivial (or integral) central character $\zeta$. These are precisely
the special nilpotent orbits (which were described in section
\specialsurf).  Because of this property, they are sometimes called
``quantizable'' orbits in the mathematical literature, {\it cf.}
\refs{\BBMacPherson,\Lusztigiii}. In particular, it follows that the
pairs of rigid unipotent conjugacy classes in $G_{\C}$ and
$\LG_{\C}$ which are related to each other by $S$-duality are
precisely the special ones. We have checked this prediction for all
strongly rigid surface operators in $B_N$ and $C_N$ up to rank
$N=10$, assuming that only strongly rigid conjugacy classes need to
be considered in the dual group, and using the invariants we know
for surface operators to partly constrain the duality map.



\newsec{Stringy Constructions of Rigid Surface Operators}\seclab\holog

\subsec{Holographic Description}

For gauge groups of classical types $A$, $B$, $C$, or $D$, the large
$N$ limit of the $\CN=4$ super-Yang-Mills theory
is believed to be equivalent to type IIB string theory
in space-time AdS$_5 \times Q$, where the ``horizon'' $Q$ equals
$\S^5$ if $G = SU(N)$ and or $\RP^5$ if $G$ is an orthogonal or
symplectic group \refs{\Maldacena,\Wbaryons}.
In particular, under this duality, the
superconformal symmetry group $PSU(2,2 \vert 4)$ of the $\CN=4$
gauge theory is identified with the isometry group of the
super-geometry whose bosonic reduction is ${\rm AdS}_5\times Q$. In
orientifold models with $Q = \RP^5$, different choices of the gauge
group $G$ correspond to different values of the discrete torsion for
the 2-form fields $B_{NS}$ and $B_{RR}$.
Following \Wbaryons, we introduce discrete holonomies
\eqn\thnsrr{
\th_{NS} = \int_{\RP^2} {B_{NS} \over 2\pi}
\quad,\quad
\th_{RR} = \int_{\RP^2} {B_{RR} \over 2\pi}
}
which can take two values, $0$ and ${1 \over 2}$, since $\RP^2
\subset \RP^5$ is a two-torsion element, generating $H_2 (\RP^5,
\tilde \Z) = \Z_2$. ($\tilde \Z$ is a twisted version of the
constant sheaf of integers.)

A rigid surface operator of a type studied in this paper breaks the
four-dimensional conformal group $SU(2,2) \cong SO(2,4)$ down to a
subgroup $SO(2,2) \times SO(2)$. Moreover, just like half-BPS
surface operators in \Ramified, it introduces a singularity for two
components of the Higgs field and, therefore, breaks the
$R$-symmetry group $SO(6)_{\CR}$ down to a subgroup
\eqn\rsymbreaking{
SO(6)_{\CR} \to SO(4) \times SO(2) }
The remaining symmetry group
\eqn\bosonicsymm{ SO(2,2) \times SO(2) \times SO(4) \times SO(2) }
is precisely the part of the isometry of AdS$_5 \times \S^5$
preserved by a ``probe'' D3$'$-brane embedded in AdS$_5 \times Q$ as
\eqn\dthreebrane{\matrix{
{\rm space-time:}~~~~~~ & {\rm AdS}_5 & \times & Q \cr
& \cup & & \cup \cr
{\rm D3}'{\rm -brane:}~~~~~~ & {\rm AdS}_3 & \times & \ell } }
where $\ell \subset Q$ is ``equator'' of $Q$.

The identification of the parameters is similar to the $SU(N)$ case
considered in \Ramified. Thus, if we denote the gauge field on the
D3$'$-brane by $A'$, the parameter $\a$ is simply the holonomy of
$A'$, while $\eta$ is identified with the holonomy of the dual
photon $\tilde A'$,
\eqn\aaa{
\a = {\rm Hol}_{\ell} (A')
\quad,\quad
\eta = {\rm Hol}_{\ell} (\tilde A')
}
Similarly, in this description, $\b$ and $\g$ determine the
asymptotic behavior of the Higgs field $\varphi'$ on the
D3$'$-brane. Namely, it has the familiar form,
$$
\varphi' = {1 \over 2z} (\b + i\g) + \ldots
$$
where $z$ is a complex variable on the D3$'$-brane world-volume.
It is convenient to introduce coordinates $(y_1,y_2,y_3,\chi)$
on the D3$'$-brane world-volume, AdS$_3 \times \ell$,
such that the metric takes the standard form
\eqn\adssone{
ds^2 = {1 \over y_3^2} (dy_1^2 + dy_2^2 + dy_3^2) + d \chi^2 }
In terms of these coordinates, we have $z = y_3 e^{i \chi}$.

In general, the number of D3$'$-branes determines the number of
independent parameters $\a$, as well as $\b$, $\g$, and $\eta$.
(This is clear from the intersecting brane model, which will be
discussed below.) In other words, the number of D3$'$-branes is
equal to the number of abelian factors in the Levi subgroup $\LL$.
For example, in a theory with gauge group $G=SU(N)$, a single
D3$'$-brane corresponds to a surface operator with maximal Levi
subgroup $\LL = SU(N-1) \times U(1)$. In general, larger number of
D3$'$-branes corresponds to surface operators with smaller Levi
subgroups and larger orbits $\frak c \subset \frak g_{\C}$.

The identification of the parameters allows to see how surface
operators in this holographic description transform under S-duality.
Indeed, since S-duality in the D3- and D3$'$-brane theory correspond
to S-duality in type IIB string theory, it easily follows that $\a$
and $\eta$ transform as
\eqn\aetasdual{S: \quad (\a,\eta) \to (\eta, -\a) }
while $\b$ and $\g$ are essentially invariant under S-duality.


\subsec{Application: $SO(2N)$ Gauge Theory}

In section \quant, we made a general proposal on how
S-duality should act on surface operators associated
with (rigid) unipotent conjugacy classes.
Here, we will go in the opposite direction and
use the holographic description to study the action
of S-duality on surface operators associated with
rigid semisimple conjugacy classes.

Let us consider a simple case of $SO(2N)$ gauge theory,
whose holographic dual is given by AdS$_5 \times \RP^5$
with no discrete torsion, $(\th_{NS},\th_{RR}) = (0,0)$.
A particular class of rigid surface operators which is easy
to describe in this holographic setup consists of strongly rigid surface
operators with semisimple holonomy $V = S$ of the form \sssimple,
\eqn\sssdn{ S = {\rm diag} \big( +1,+1 \ldots,+1,
\underbrace{-1,-1,\dots,-1,-1}_{2k} \big),
\quad\quad 1< k \le \Big[ {N \over 2} \Big]. }
This surface operator corresponds to $k$ ${\rm D}3'$-branes with non-trivial
holonomy $\a={1 \over 2}$ and with $\b = \g = \eta = 0$. An element $S$ of
the form \sssdn\ breaks the gauge group $G = SO(2N)$ down to a subgroup,
$$
G \to S \left( O(2k) \times O(2N-2k) \right)
$$
so that we label this surface operator by the conjugacy class,
\eqn\sscdn{
\frak C = \frak C^{D_k \times D_{N-k}}_{([1^{2k}],[1^{2N-2k}])} }

Now, let us consider what happens under duality. Since in the
D3$'$-brane theory, $S$-duality exchanges $\a$ and $\eta$, just as
in \aetasdual, it follows that strongly rigid surface operators
associated with $S \ne 1$ given by \sssdn\ are mapped to rigid
surface operators with $\dual{\a}=\dual{\b}=\dual{\g}=0$ and
$\dual{\eta} \ne 0$. In particular, such surface operators should
correspond to rigid unipotent conjugacy classes $\dual{\frak C}
\subset \LG_{\C}$ since they have $\dual{\a}=\dual{\g}=0$ and,
hence, $\dual{S}=1$, {\it cf.} \doof. In other words, the
holographic description of these surface operators suggests that,
under duality, they are mapped to strongly rigid surface operators
labeled by  rigid unipotent conjugacy classes $\dual{\frak C}$,
\eqn\ssdndual{\matrix{
S:~~~ & {\rm rigid~semisimple} & \longrightarrow & {\rm rigid~unipotent} \cr
& (S \ne 1,~U=1) & & (\dual{S}=1,~\dual{U} \ne 1) }}
It is not yet clear how to deduce from this holographic description
the dictionary between values of $\dual{\eta}$ and the corresponding
unipotent conjugacy classes $\dual{\frak C}$.  However, we can
determine what the answer must be by comparing invariants described
in section \invariants. We find that rigid semisimple conjugacy
classes \sscdn\ are dual to rigid unipotent conjugacy classes
labeled by the partition
\eqn\sscdndual{ \la = [3,2^{2k-2},1^{2N-4k+1}] }
It is easy to see that the unipotent conjugacy class labeled by
this partition is indeed rigid.
Namely, it has no gaps --- which, according to section \searching,
is one of the criteria for rigidness --- as long as
$1< k \le \big[ {N \over 2} \big]$.
The second condition for rigidness says that no odd part of $\la$
should occur exactly twice. This condition automatically holds
true for the partition \sscdndual\ since the only odd parts are
``$3$'' and ``$1$'', and their multiplicities are always odd.

We note that the duality we arrived at, which relates rigid
semisimple surface operators with the monodromy \sssdn\ and rigid
unipotent surface operators associated with the conjugacy class
labeled by \sscdndual, is consistent with the general proposal of
section \quant. Indeed, from the relation with quantization
discussed in section \quant\ it follows that the only rigid
unipotent surface operators which under duality are mapped to rigid
unipotent operators are those associated with special conjugacy
classes. Therefore, as a consistency check, we should verify that
the unipotent conjugacy class labeled by the partition \sscdndual\
is not special. This is easy to do using the criterion described in
section \specialsurf. According to this criterion, a unipotent
conjugacy class $\frak C_{\la}$ in $D_N$ is special if and only if
the transpose partition $\la^t$ is symplectic. The transpose of the
partition \sscdndual\ is
\eqn\sscdndualtransp{ \la^t = [2N-2k, 2k-1, 1] }
Clearly, this $\la^t$ is not symplectic since odd parts ``$(2k-1)$''
and ``$1$'' have odd multiplicity.
Hence, we conclude that the unipotent conjugacy class labeled by
the partition \sscdndual\ is not special and, therefore, according
to the general proposal in section \quant\ under duality should
transform into a rigid semisimple conjugacy class.
This is precisely what we find in this section, from the holographic
description of the corresponding rigid surface operators.

Finally, we remark that the duality between surface operators in $SO(8)$
gauge theory studied in section \soeight\ is a special case of the duality
found in this section; it corresponds to $N=4$ and $k=2$.


\subsec{Intersecting Brane Models}

Now we will reconsider the same subject from the point of view of
intersecting brane models.  (The holographic models just considered
arise from the near-horizon limit of the intersecting brane models.)

In intersecting brane models, gauge theories with symplectic and
orthogonal gauge groups can be engineered by introducing orientifold
$p$-planes ($Op$-planes). Therefore, let us start by recalling a few
basic facts about $Op$-planes, see {\it e.g.} \BergmanGS. In type II
string theory, an orientifold $p$-plane is defined using a $\Z_2$
projection that combines world-sheet orientation symmetry $\Omega$
with a space-time involution $\CI_{9-p}$ and, possibly, the action
of $(-1)^{F_L}$ on fermions,
$$
Op : \quad \R^{1,p} \times \R^{9-p} / \CI_{9-p} \Omega \cdot
\cases{1 & $p=0,1$ mod 4 \cr (-1)^{F_L} & $p=2,3$ mod 4.}
$$
The action of the orientifold on anti-symmetric tensor fields is given by
\eqn\oponforms{\eqalign{
& B_{NS} \to - B_{NS} \cr
& C_{p'} \to + C_{p'} \quad p'=p+1 {\rm ~~mod~~} 4 \cr
& C_{p'} \to - C_{p'} \quad p'=p+3 {\rm ~~mod~~} 4
}}

For $2 \le p \le 5$, there are four types of orientifold planes,
labeled by the discrete torsion $(\th_{NS},\th_{RR})$
of the NS-NS 3-form field $H$ and of the $(6-p)$-form field $G_{6-p}$
in the R-R sector, {\it cf.} \thnsrr.
We summarize these $Op$-planes, their charges, and the corresponding
gauge groups in the table below.


\vskip 0.8cm \vbox{ \centerline{\vbox{ \hbox{\vbox{\offinterlineskip
\def\tablespace{height7pt&\omit&&\omit&&\omit&&\omit&\cr}
\def\tablerule{\tablespace\noalign{\hrule}\tablespace}

\hrule\halign{&\vrule#&\strut\hskip0.2cm\hfill #\hfill\hskip0.2cm\cr
\tablespace & $(\th_{NS},\th_{RR})$ && orientifold plane && $G$ && charge &\cr
\tablerule & $(0,0)$ && $Op^{-}$ && $SO(2N)$ && $-2^{p-5}$ &\cr
\tablerule & $({1 \over 2}, 0)$ && $Op^{+}$ && $Sp(2N)$ && $+2^{p-5}$ &\cr
\tablerule & $(0, {1 \over 2})$ && $\tilde{Op}^{-}$ && $SO(2N+1)$ && $-2^{p-5} + {1 \over 2}$ &\cr
\tablerule & $({1 \over 2} , {1 \over 2})$ && $\tilde{Op}^{+}$ && $Sp(2N)$ && $+2^{p-5}$ &\cr
\tablespace}\hrule}}}}
%
%
} \vskip 0.5cm

Maximally supersymmetric $\CN=4$ gauge theory with $G=U(N)$ can be
realized on the world-volume of $N$ D3-branes in type IIB string
theory. In this realization, half-BPS surface operators can be
obtained by introducing $k$ extra D3$'$-branes, which intersect
D3-branes over the surface $D \subset M$.

\ifig\branefig{A D-brane realization of surface operators in $U(N)$ gauge
theory. $N$ D3-branes (shown horizontally) intersect $k$ extra D3$'$-branes
(shown vertically) over a two-dimensional subspace $D \subset M$.}
{\epsfxsize4.0in\epsfbox{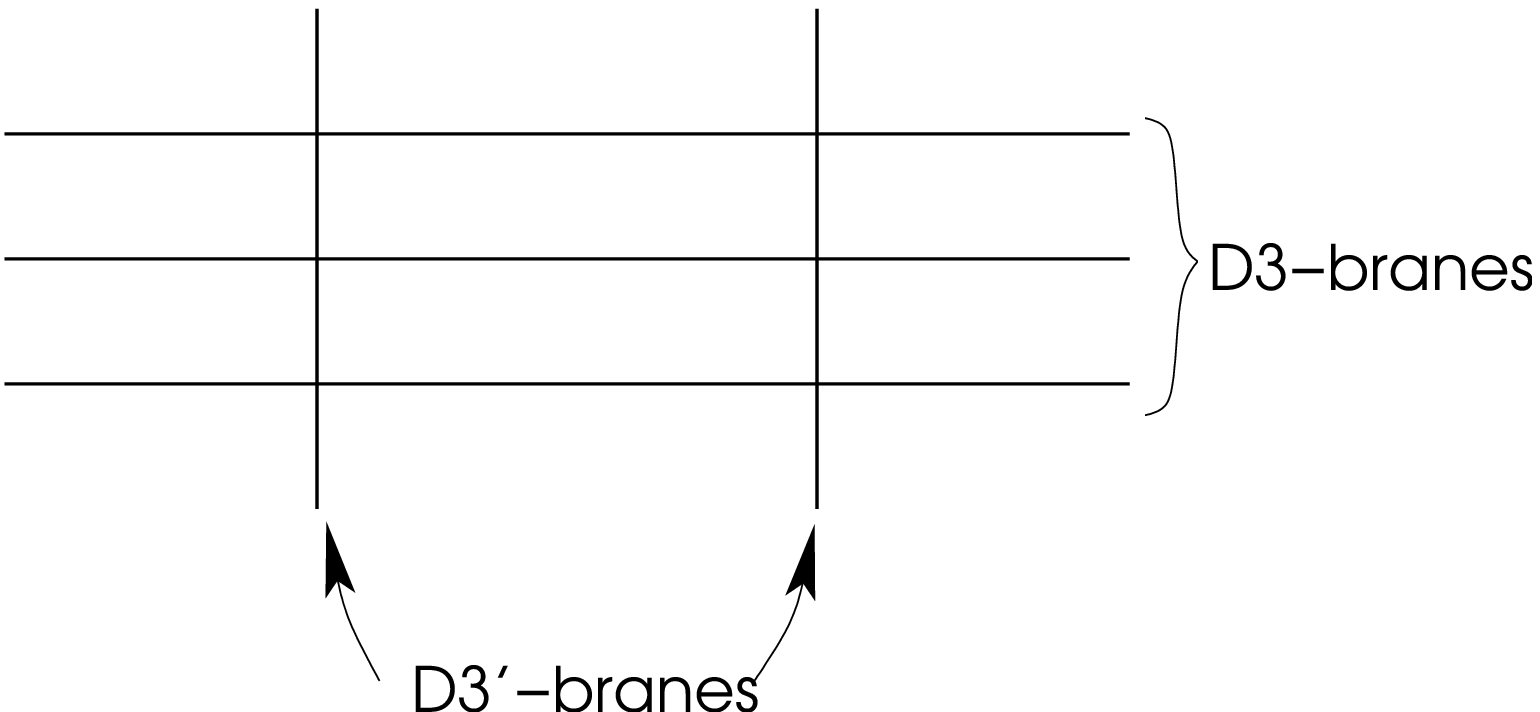}}

Introducing orientifold 3-planes on top of the D3-branes leads to
stringy realizations of gauge theories with orthogonal and
symplectic gauge groups, where the gauge group is determined by the
particular type of the O3-plane, according to the above table. On
the other hand, the gauge group $G'$ on the D3$'$-branes is a $\Z_2$
extension of $U(k)$, which we call $\bar{U(k)}$,
\eqn\uext{
1 \to U(k) \to \bar{U(k)} \to \Z_2 \to 1 }
Specifically, $\bar{U(k)}$ is generated by elements $g_i$ of $U(k)$
and the generator $\epsilon$ of $\Z_2$, with the commutation relations
\eqn\uextrels{\eqalign{ & (g_i,\epsilon)\cdot(g_j,\epsilon) = (g_i \bar g_j ,1) \cr
& (g_i,\epsilon)\cdot(g_j,1) = (g_i \bar g_j ,\epsilon) \cr
& (g_i,1)\cdot(g_j,\epsilon) = (g_i g_j ,\epsilon) }}
The holonomy $V' = \epsilon$ in the D3$'$-brane theory breaks
$G'=\bar{U(k)}$ down to a subgroup, which is the centralizer of $V'$ in $G'$.
For example, $V'=\epsilon$ breaks $G'$ down to a subgroup $O(k) \times \Z_2$.
Indeed, it consists of the elements $(g,s) \in G'$, such that
$$
(1,\epsilon) \cdot (g,s) = (g,s) \cdot (1,\epsilon)
$$
which implies $g = \bar g \in O(k)$.


\subsec{Bubbling Geometries}

So far, we discussed stringy description of rigid surface operators
in terms D3$'$-branes realizing $\CN=4$ gauge theory either on the
world-volume of D3-branes or via its holographic dual. However,
there is yet another, equivalent description, in which D3$'$-branes
are also replaced by a dual geometry. Following \refs{\LLMi,\LLMii},
we call these {\it bubbling geometries}.
Conformally invariant half-BPS surface operators in $\CN=4$ gauge
theory with gauge group $G=SU(N)$ can be obtained \GomisM\
by analytic continuation of the LLM solutions \refs{\LLMi,\LLMii}.
These solutions are asymptotic to AdS$_5 \times \S^5$.

\ifig\BubblingFig{Bubbling geometries are specified by point ``charges''
in the base space $X = \R_+ \times \R^2$.
Semi-infinite dashed lines represent disks $D_i$.}
{\epsfxsize3.0in\epsfbox{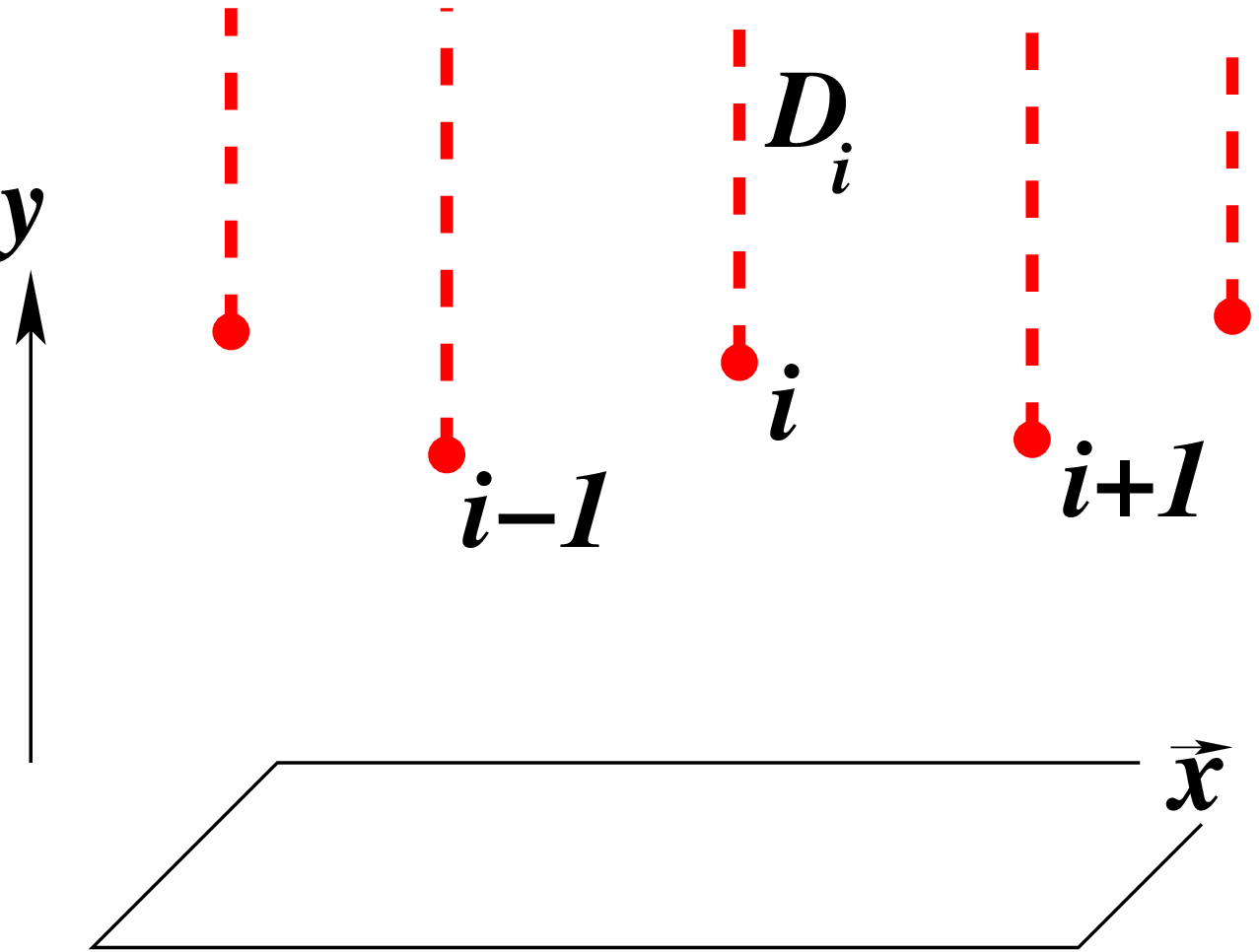}}

In order to make the symmetry group \bosonicsymm\ manifest,
it is convenient to construct the bubbling geometries as
AdS$_3 \times \S^3 \times \S^1$ fibrations over
the three-dimensional base space $X = \R_+ \times \R^2$
(for more detail, see \GomisM).
In the case of $SU(N)$ gauge theory,
every half-BPS geometry is parametrized by positions $(\vec x_i, y_i)$
of point ``charges'' in $X$ of total charge $N$,
where $y_i \in \R_{\ge 0}$ and $\vec x_i \in \R^2$.
The coordinate $y_i$ is related to the value of each charge, $N_i$, as
$$
N_i = {y_i^2 \over 4 \pi l_p^4}.
$$
whereas the coordinate $\vec x$ is related to the (eigen-)values
of $\b$ and $\g$.
Namely, we have\foot{in the conventions where $\ell_s = 1$}
$$
\vec x_i = (\b_i,\g_i).
$$
In order to describe the geometric interpretation of $\a$ and $\eta$,
we note that $\S^1$ degenerates at every point $(\vec x_i, y_i)$
(location of the $i$-th charge) and $\S^3$ degenerates at the plane $y=0$.
Therefore, every bubbling geometry contains some number of
5-spheres (one for every point charge in $X$) represented by
a Hopf-like fibration of $\S^3 \times \S^1$ over the interval $y \in [0,y_i]$,
and some number of disks (also, one for every point charge in $X$)
ending on the asymptotic boundary,
$$
D_i = \{ (y,\chi) ~\vert~ y \in [y_i,\infty), \chi \in \S^1 \}.
$$
Here, $\chi$ is the variable parametrizing the $\S^1$, as in \adssone.
The (eigen-)values of $\a$ and $\eta$ are holonimies of the NS and RR
2-form fields \GomisM:
\eqn\aetaviabb{ \a_i = - \int_{D_i} {B_{NS} \over 2\pi} \quad,\quad
\eta_i = \int_{D_i} {B_{RR} \over 2\pi}. }
The $S$-duality of type IIB string theory exchanges $B_{NS}$ and
$B_{RR}$, thus, providing another evidence for \aetasdual\ --
\ssdndual.

For example, the bubbling geometry corresponding to a single
charge at $(\vec x_0, y_0)$, is the familiar space AdS$_5 \times \S^5$,
with the usual metric
\eqn\adsbubbling{ ds^2 = y_0 \big[ (\cosh^2 u ~ds^2_{AdS_3} + du^2 +
\sinh^2 u ~d\psi^2  ) + (\cos^2 \th d \Omega_3 + d\th^2 + \sin^2 \th
d \phi^2) \big], }
where the variables are
\eqn\adsbubblingvars{\eqalign{ & x^1 - x^1_0 + i (x^2 - x^2_0) = r
e^{i (\phi - \psi)} \cr & r = y_0 \sinh u \sin \th \cr & y = y_0
\cosh u \cos \th \cr & \chi = {1 \over 2} (\phi + \psi) .}}

Now we can extend this construction to describe bubbling geometries
representing conformally invariant half-BPS surface operators in
$\CN=4$ gauge theory with symplectic and orthogonal gauge groups.
As usual, this can be achieved by introducing a $\Z_2$ orientifold
projection, such that the corresponding quotient
of the AdS$_3 \times \S^3 \times \S^1$ fibrations over $X$
is asymptotic to AdS$_5 \times \RP^5$.
Since the $\Z_2$ involution $\CI$ acts trivially on AdS$_5$
and as the antipodal map on $\S^5$,
it follows from \adsbubbling\ -- \adsbubblingvars\
that it acts as
\eqn\bubblinginv{ \CI: \quad\quad \eqalign{ \S^3 & \to \S^3 / \Z_2
\cr \chi & \to \chi + {\pi \over 2} \cr \vec x & \to - \vec x }}
Notice that $\CI$ has no fixed points. Moreover, as usual, the
orientifold projection acts non-trivially on the 2-form fields
$B_{NS}$ and $B_{RR}$, {\it cf.} \oponforms.

The generic surface operator which has deformation parameters
$(\a,\b,\g,\eta)$ and corresponds to the regular conjugacy class
$\frak C_{{\rm reg}}$ is represented by $N$ pairs of charges
at $\pm \vec x_i$, that is $N$ charges and their ``mirror images''.

On the other hand, surface operator associated with the rigid
semisimple conjugacy class \sssdn\ -- \sscdn\ is described by
the ``rigid'' configuration with two charges, $N_1 = 2k$ and $N_2 = 2N-2k$,
located at $\vec x = 0$.


\vskip 30pt

\centerline{\bf Acknowledgments}

We would like to thank R.~Bezrukavnikov, A.~Braverman,
A.~Elashvili, D.~Gaiotto, V.~Kac, G.~Lusztig, C.~Vafa, and
especially D.~Kazhdan for valuable discussions and correspondence.
Research of SG is supported in part by NSF Grant DMS-0635607,
in part by RFBR grant 07-02-00645,
and in part by the Alfred P. Sloan Foundation.
Research of EW is partly supported by NSF Grant PHY-0503584.
Conclusions reported here are those of the authors
and not of funding agencies.

\vfill
\eject

\appendix{A}{Rigid Nilpotent Orbits for Exceptional Groups}

Here we describe rigid nilpotent orbits in exceptional cases. In
such cases, the appropriate language to classify nilpotent orbits is
based on Bala-Carter theory which we summarize below. According to
Bala and Carter, nilpotent orbits in $\frak g_{\C}$ are in
one-to-one correspondence with pairs $(\frak l , \frak p_{\frak
l})$, where $\frak l \subset \frak g$ is a Levi subalgebra, and
$\frak p_{\frak l}$ is a distinguished\foot{A nilpotent orbit in
$\frak g_{\C}$ is called distinguished if its centralizer contains
no semisimple elements which are not in the center of $\frak
g_{\C}$. In type $A$, the only distinguished orbit is a principal
orbit. In types $B$, $C$, or $D$, an orbit is distinguished if and
only if its partition has no repeated parts. Thus, the partition of
a distinguished orbit in type $B$ and $D$ has only odd parts, each
occuring once, while the partition of a distinguished orbit in type
$C$ has only even parts, also occuring only once.} parabolic
subalgebra of the semisimple algebra $[\frak l , \frak l]$. Such
pairs can be conveniently labeled as $X_N (a_i)$ where $X_N$ is the
Cartan type of the semisimple part of $\frak l$ and $i$ is the
number of simple roots in any Levi subalgebra of $\frak p_{\frak
l}$. If $i=0$ one simply writes $X_N$, and if a simple component of
a Levi subalgebra $\frak l$ involves short roots (when $\frak g$ has
two root lengths) then one labels its Cartan type with a tilde.
Using this notation, below we list rigid nilpotent orbits in $G_2$:
\bigskip
\centerline{\vbox{\offinterlineskip
\def\tablerule{\noalign{\hrule}}
\halign to 2.2truein{\tabskip=1em plus 2em#\hfil&\vrule height12pt depth5pt#&#\hfil&\vrule height12pt depth5pt#&#\hfil\tabskip=0pt\cr
\hfil orbit $\frak c$ \hfil&&\hfil $\dim (\frak c)$ \hfil&&\hfil $\pi_1 (\frak c)$ \hfil\cr
\tablerule
~~$A_1$ && $6$ && $1$ \cr
~~$\tilde A_1$ && $8$ && $1$ \cr
}}}\bigskip
\noindent
These are the only nilpotent orbits in $G_2$ which are not special.
As usual, we omit the trivial orbit, and in the last column
we also list the $\Gsc$-equivariant fundamental group of $\frak c$
(defined as $\pi_1 (\frak c) = \Gsc (\frak c) / \Gsc (\frak c)^o$,
where $\Gsc (\frak c)$ is the centralizer of $\frak c$ in the simply-connected form of $G$).
The $G_{{\rm ad}}$-equivariant fundamental group, usually denoted $A(\frak c)$,
is the same as $\pi_1 (\frak c)$ in types $G_2$, $F_4$, and $E_8$.

In the following table we list rigid nilpotent orbits in $F_4$:
\bigskip
\centerline{\vbox{\offinterlineskip
\def\tablerule{\noalign{\hrule}}
\halign to 2.5truein{\tabskip=1em plus 2em#\hfil&\vrule height12pt depth5pt#&#\hfil&\vrule height12pt depth5pt#&#\hfil\tabskip=0pt\cr
\hfil orbit $\frak c$ \hfil&&\hfil $\dim (\frak c)$ \hfil&&\hfil $\pi_1 (\frak c)$ \hfil\cr
\tablerule
~~$A_1$ && $16$ && $1$ \cr
~~$\tilde A_1$ && $22$ && $S_2$ \cr
~~$A_1 + \tilde A_1$ && $28$ && $1$ \cr
~~$A_2 + \tilde A_1$ && $34$ && $1$ \cr
~~$\tilde A_2 + A_1$ && $36$ && $1$ \cr
}}}\bigskip
\noindent
All of these orbits, except for $\tilde A_1$ and $A_1 + \tilde A_1$, are not special.

In type $E_6$, rigid nilpotent orbits are the following:
\bigskip
\centerline{\vbox{\offinterlineskip
\def\tablerule{\noalign{\hrule}}
\halign to 2.5truein{\tabskip=1em plus 2em#\hfil&\vrule height12pt depth5pt#&#\hfil&\vrule height12pt depth5pt#&#\hfil\tabskip=0pt\cr
\hfil orbit $\frak c$ \hfil&&\hfil $\dim (\frak c)$ \hfil&&\hfil $\pi_1 (\frak c)$ \hfil\cr
\tablerule
~~$A_1$ && $22$ && $1$ \cr
~~$3A_1$ && $40$ && $1$ \cr
~~$2A_2 + A_1$ && $54$ && $\Z_3$ \cr
}}}\bigskip
\noindent
The orbit $A_1$ is special, while $3A_1$ and $2A_2 + A_1$ are not.
The group $A (\frak c)$ is trivial for all of these rigid orbits.

In type $E_7$, rigid nilpotent orbits are the following:
\bigskip
\centerline{\vbox{\offinterlineskip
\def\tablerule{\noalign{\hrule}}
\halign to 2.5truein{\tabskip=1em plus 2em#\hfil&\vrule height12pt depth5pt#&#\hfil&\vrule height12pt depth5pt#&#\hfil\tabskip=0pt\cr
\hfil orbit $\frak c$ \hfil&&\hfil $\dim (\frak c)$ \hfil&&\hfil $\pi_1 (\frak c)$ \hfil\cr
\tablerule
~~$A_1$ && $34$ && $1$ \cr
~~$2A_1$ && $52$ && $1$ \cr
~~$(3A_1)'$ && $64$ && $1$ \cr
~~$4A_1$ && $70$ && $1$ \cr
~~$A_2 + 2A_1$ && $82$ && $1$ \cr
~~$2A_2 + A_1$ && $90$ && $1$ \cr
~~$(A_3 + A_1)'$ && $92$ && $1$ \cr
}}}\bigskip
\noindent
All of these orbits have $A (\frak c) =1$.
Among these, the orbits $A_1$, $2A_1$, and $A_2 + 2A_1$ are special.

Finally, in the following table we list rigid nilpotent orbits in $E_8$:
\bigskip
\centerline{\vbox{\offinterlineskip
\def\tablerule{\noalign{\hrule}}
\halign to 2.8truein{\tabskip=1em plus 2em#\hfil&\vrule height12pt depth5pt#&#\hfil&\vrule height12pt depth5pt#&#\hfil\tabskip=0pt\cr
\hfil orbit $\frak c$ \hfil&&\hfil $\dim (\frak c)$ \hfil&&\hfil $\pi_1 (\frak c)$ \hfil\cr
\tablerule
~~$A_1$ && $58$ && $1$ \cr
~~$2A_1$ && $92$ && $1$ \cr
~~$3A_1$ && $112$ && $1$ \cr
~~$4A_1$ && $128$ && $1$ \cr
~~$A_2 + A_1$ && $136$ && $S_2$ \cr
~~$A_2 + 2A_1$ && $146$ && $1$ \cr
~~$A_2 + 3A_1$ && $154$ && $1$ \cr
~~$2A_2 + A_1$ && $162$ && $1$ \cr
~~$A_3 + A_1$ && $164$ && $1$ \cr
~~$2A_2 + 2A_1$ && $168$ && $1$ \cr
~~$A_3 + 2A_1$ && $172$ && $1$ \cr
~~$D_4 (a_1) + A_1$ && $176$ && $S_3$ \cr
~~$A_3 + A_2 + A_1$ && $182$ && $1$ \cr
~~$2A_3$ && $188$ && $1$ \cr
~~$A_4 + A_3$ && $200$ && $1$ \cr
~~$A_5 + A_1$ && $202$ && $1$ \cr
~~$D_5 (a_1) + A_2$ && $202$ && $1$ \cr
}}}\bigskip
\noindent
The only special orbits in this list are $A_1$, $2A_1$, $A_2 + A_1$, $A_2 + 2A_1$, $D_4 (a_1) + A_1$.

\vfill\eject


\appendix{B}{Orthogonal and Symplectic Lie Algebras and Duality}

In this appendix, we recall the root systems of the Lie algebras
$\frak{so} (2N+1)$ and $\frak{sp} (2N)$.
In particular, we describe a convenient matrix realization
that leads to a simple identification of the invariant polynomials
$\Tr \varphi^k$ in the corresponding fundamental representations.


We begin with the symplectic group $Sp(2N)$.
It consists of $(2N) \times (2N)$ matrices $A$ that satisfy
\eqn\cjcond{ A^t J A = J }
where
$$
J = \pmatrix{0 & I_N \cr -I_N & 0}
$$
The matrix form of the corresponding Lie algebra, $\frak{sp} (2N)$,
can be obtained by writing $A = \exp (X) \simeq I + X$ in terms of
$N \times N$ matrices $X_i$,
\eqn\xmatrix{ X = \pmatrix{ X_1 & X_2 \cr X_3 & X_4} }
Then, the condition \cjcond\ implies
\eqn\spmatrix{
X_1^t = - X_4,
\quad\quad
X_2^t = X_2,
\quad\quad
X_3^t = X_3 }
The Lie algebra, $\frak t$, of the maximal torus of $Sp(2N)$ can be
represented by diagonal matrices of the form
\eqn\ccartan{ X = \pmatrix{D & 0 \cr 0 & -D} }
where $D = {\rm diag} (x_1, x_2, \ldots, x_N)$.

Now, in this $2N$-dimensional representation,
let us define the root system of $\frak{sp} (2N)$
$$
\Lambda_{{\rm rt}} = \{ \pm (e_i \pm e_j),~ 1 \le i < j \le N \}
\cup \{ \pm 2e_i,~ i = 1, \ldots, N \}
$$
the set of positive roots
$$
\Lambda_{{\rm rt}}^+ = \{ e_i \pm e_j,~ 1 \le i < j \le N \}
\cup \{ 2e_i,~ i = 1, \ldots, N \}
$$
and the set of simple roots
$$
\Delta = \{ e_i - e_{i+1} ,~ 1 \le i < N \} \cup \{ 2e_N \}
$$
Here, $e_i$ denote basis elements of $\frak t^* \cong \IR^N$.
The $2(N^2-N)$ short roots $\pm e_i \pm e_j$ can be represented by matrices
(see {\it e.g.} \CMcGovern):
\eqn\crootmatrixshort{\eqalign{
& X_{e_i - e_j} = E_{i,j} - E_{j+N,i+N} \cr
& X_{e_i + e_j} = E_{i,j+N} + E_{j,i+N} \cr
& X_{- e_i - e_j} = E_{i+N,j} + E_{j+N,i}
}}
where $E_{i,j}$ is a matrix with $1$ at the position $(i,j)$ and zeros elsewhere.
Similarly, $2N$ long roots $\pm 2 e_i$ are represented by matrices
\eqn\crootmatrixlong{\eqalign{
& X_{2 e_i} = E_{i,i+N} \cr
& X_{-2 e_i} = E_{i+N,i}
}}

Choosing a metric on $\frak t$ defines a natural isomorphism between
$\frak t$ and $\frak t^*$ that we need later. We normalize the metric
so that short coroots (equivalently, long roots) have length squared 2.
With this normalization, in type $C_2$ we have
\eqn\ctworoot{
\a_1 = \sqrt{2} e_1 \quad,\quad \a_2 = {1 \over \sqrt{2}} (e_2 - e_1) }
where $\{ e_1 , e_2 \}$ is an orthonormal basis of $\frak t^* \cong \IR^2$.

\ifig\btwoctwofig{The root systems of type $B_2$ and $C_2$.}
{\epsfxsize3.3in\epsfbox{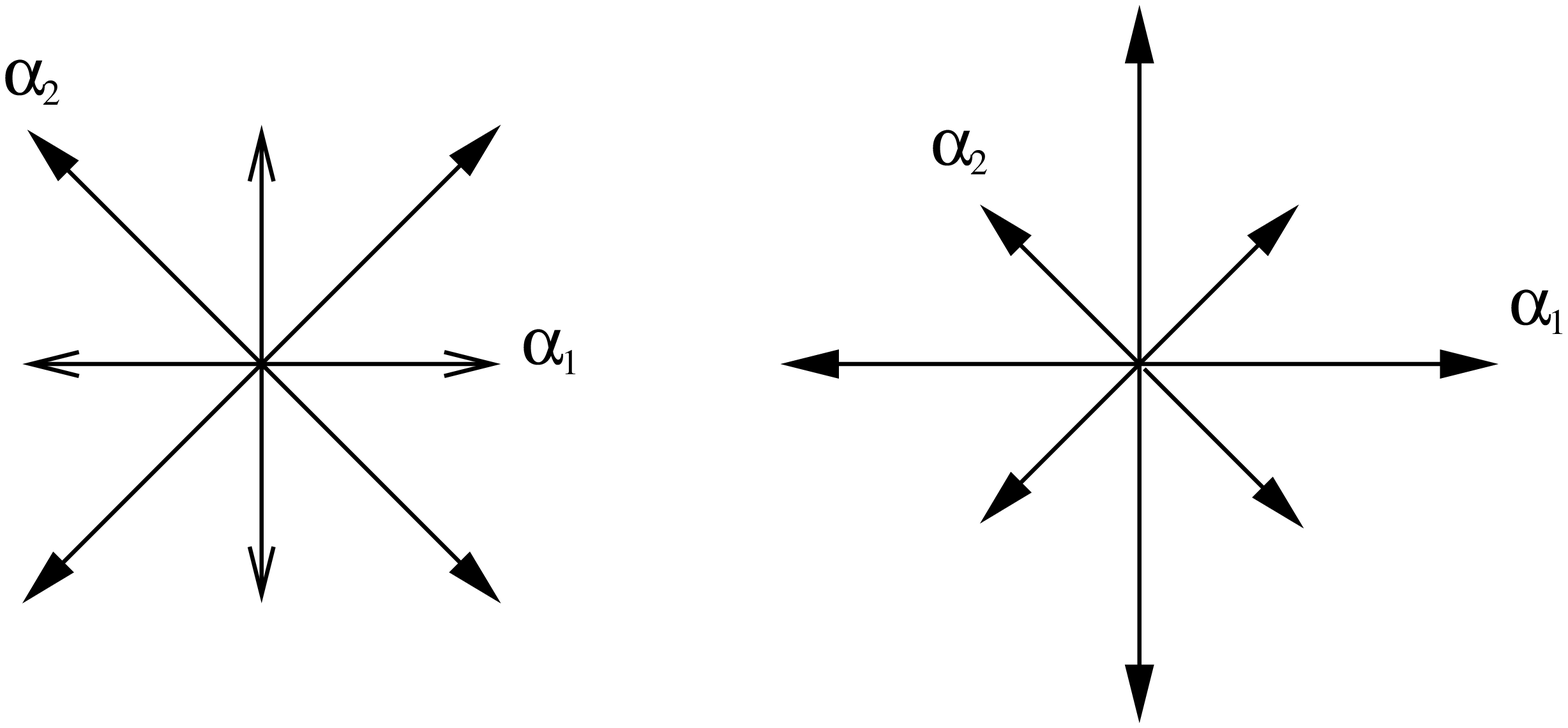}}


Now, let us consider the orthogonal group $SO(2N+1)$.
In the $(2N+1)$-dimensional representation, it is realized
by $(2N+1) \times (2N+1)$ matrices $A$ which satisfy
\eqn\bjcond{ A^t A = I }
In order to obtain the matrix form of the corresponding
Lie algebra $\frak{so} (2N+1)$, we write $A = \exp (X) \simeq I + X$.
Then, the condition \bjcond\ leads to the following
condition on the Lie algebra element $X$,
$$
X + X^t = 0
$$
In particular, this condition implies that all diagonal
elements of $X$ vanish.

Our goal, however, is to describe a matrix realization
of the Lie algebra $\frak{so} (2N+1)$ which would allow
a simple comparison of the invariant polynomials
in the dual Lie algebras $\frak{sp} (2N)$ and $\frak{so} (2N+1)$.
This will be easy to achieve if we can realize the Cartan
subalgebra of $\frak{so} (2N+1)$ by diagonal matrices,
as we did in eqn. \ccartan\ for $\frak{sp} (2N)$.
For this reason, it is convenient to perform
a unitary transformation on matrices $A$,
$$
A = U B U^t
$$
which after substituting to \bjcond\ and writing
$B = \exp (X) \simeq I + X$ gives a condition on
the Lie algebra element $X$,
\eqn\bxcond{ X^t K + KX = 0 }
with $K = U^t U$.

In the $(2N+1)$-dimensional representation that we are considering,
we write matrices $X$ in the block form,
$$
X = \pmatrix{
X_0 & a & b \cr
c & X_1 & X_2 \cr
d & X_3 & X_4}
$$
where the diagonal blocks $X_0$, $X_1$, and $X_4$
have size $1$, $N$, and $N$, respectively.
In this presentation, we choose
$$
U = {1 \over \sqrt{2}} \pmatrix{
\sqrt{2} & 0 & 0 \cr
0 & i I_N & -i I_N \cr
0 & -I_N & -I_N}
$$
which gives
$$
K = U^t U = \pmatrix{
1 & 0 & 0 \cr
0 & 0 & I_N \cr
0 & I_N & 0}
$$
so that the condition \bxcond\ becomes
$$
\eqalign{
& X_0 = 0, \quad\quad X_1^t = - X_4, \cr
& c = - b^t, \quad\quad X_2^t = - X_2, \cr
& d = - a^t, \quad\quad X_3^t = - X_3
}
$$
Therefore, in this representations, we can realize elements
of the Lie algebra $\frak{so} (2N+1)$ by matrices of the form
\eqn\somatrix{
X = \pmatrix{ 0 & a & b \cr - b^t & X_1 & X_2 \cr - a^t & X_3 & - X_1^t } }
where $X_1$ is arbitrary and $X_2$ and $X_3$ are anti-symmetric.
This form is similar to the realization \xmatrix\ - \spmatrix\
of the Lie algebra $\frak{sp} (2N)$. In particular, as in \ccartan\
the Cartan subalgebra of $\frak{so} (2N+1)$ is realized by
diagonal matrices of the form
\eqn\bcartan{ X = \pmatrix{ 0 & 0 & 0 \cr 0 & D & 0 \cr 0 & 0 & - D^t } }

Now, let us describe the root system of $\frak{so} (2N+1)$,
$$
\Lambda_{{\rm rt}} = \{ \pm (e_i \pm e_j),~ 1 \le i < j \le N \}
\cup \{ \pm e_i,~ i = 1, \ldots, N \}
$$
with the standard choice of positive roots
$$
\Lambda_{{\rm rt}}^+ = \{ e_i \pm e_j,~ 1 \le i < j \le N \}
\cup \{ e_i,~ i = 1, \ldots, N \}
$$
and simple roots
$$
\Delta = \{ e_i - e_{i+1} ,~ 1 \le i < N \} \cup \{ e_N \}
$$
In the $(2N+1)$-dimensional representation \somatrix,
$2(N^2-N)$ long roots $\pm e_i \pm e_j$ are represented by matrices
\eqn\brootmatrixlong{\eqalign{
& X_{e_i - e_j} = E_{i+1,j+1} - E_{j+N+1,i+N+1} \cr
& X_{e_i + e_j} = E_{i+1,j+N+1} - E_{j+1,i+N+1} \cr
& X_{- e_i - e_j} = E_{i+N+1,j+1} - E_{j+N+1,i+1}
}}
and $2N$ short roots $\pm e_i$ are represented by matrices
\eqn\brootmatrixshort{\eqalign{
& X_{e_i} = E_{1,i+N+1} - E_{i+1,1} \cr
& X_{- e_i} = E_{1,i+1} - E_{i+N+1,1}
}}
For example, with our choice of normalization, in type $B_2$ we have
\eqn\btworoot{ \a_1 = e_1 \quad,\quad \a_2 = e_2 - e_1 }
Notice, coroots of $B_2$ are the same as roots of $C_2$ scaled by the factor
$\sqrt{ n_{\frak g} } = \sqrt{2}$, and vice versa.

The matrix realizations of the Lie algebras $\frak{sp} (2N)$
and $\frak{so} (2N+1)$ described here have a nice feature that,
in both cases, the Cartan subalgebras are realized by the set
of diagonal matrices, \ccartan\ and \bcartan, respectively.
This defines a natural map from the Cartan subalgebra of
these two Lie algebras, in which we simply identify
the ``eigenvalues'' in eqs. \ccartan\ and \bcartan\
(and add an extra ``$0$'' in the case of $\frak{so} (2N+1)$).

In particular, this map between Cartan subalgebras of
$\frak{sp} (2N)$ and $\frak{so} (2N+1)$ gives rise to a map from
invariant polynomials of $\frak{sp} (2N)$ to invariant polynomials
of $\frak{so} (2N+1)$, with the property that $\Tr \varphi^k$,
with the trace in the $2N$-dimensional representation of $\frak{sp} (2N)$,
maps to $\Tr \varphi^k$, with the trace in the $(2N+1)$-dimensional
representation of $\frak{so} (2N+1)$.

\listrefs
\end